\documentclass[aip,reprint]{revtex4-2}

\usepackage{lipsum}
\usepackage{graphicx}
\usepackage{dcolumn}
\usepackage{bm}
\usepackage{appendix} 
\usepackage{amssymb}
\usepackage{array}
\usepackage{algorithm}
\usepackage{algorithmic}
\usepackage{graphicx}
\usepackage{color}
\usepackage{soul}
\usepackage{amsmath}
\usepackage{physics}
\usepackage[inkscapelatex=false]{svg}
\usepackage{hyperref}
\usepackage{tikz}
\usepackage{amsmath,amssymb}
\usepackage{caption}
\usepackage{subcaption}
\usetikzlibrary{arrows.meta,decorations.pathreplacing,positioning}

\begin{document}


\title{Chirality-sensitive mobility and dissipation of Brownian motion on a helical landscape } 



\author{Debankur Bhattacharyya}
\email{dbhattac@sas.upenn.edu}
\affiliation{Department of Chemistry, University of Pennsylvania, Philadelphia, Pennsylvania 19104, USA}
\author{Abraham Nitzan}
\email{anitzan@sas.upenn.edu}
\affiliation{Department of Chemistry, University of Pennsylvania, Philadelphia, Pennsylvania 19104, USA}
\affiliation{School of Chemistry, Tel Aviv University, Tel Aviv 69978, Israel}


\date{\today}

\begin{abstract}
We study the Brownian dynamics and linear response of a particle with inertia moving in a 2-dimensional helical landscape imprinted on a cylindrical surface. In the harmonic well approximation, the deterministic motion separates into free propagation along the screw direction and harmonic motion in the transverse screw-normal direction. We show that for isotropic damping this simplification survives in the Langevin description, whereas anisotropic damping along the axial and angular directions couples the stochastic dynamics and destroys separability. The resulting anisotropic model is formulated as a linear Ornstein-Uhlenbeck process in phase space with a zero mode associated with diffusion along the screw coordinate, so that in an infinite system the full phase-space dynamics does not relax to a stationary distribution. To treat transport in this setting, we construct the stationary dynamics in the stable subspace obtained after projecting out the zero mode. This leads to a linear response theory for this system and yields closed analytical expressions for stationary time-correlation functions and the dynamical mobility tensor in both the time and frequency domains. The off-diagonal elements of the mobility tensor describe cross-response between axial forcing and angular motion, and between applied torque and axial transport. Consistent with time reversal symmetry, these cross mobilities are equal and provide a direct dynamical signature of the helical geometry. In addition, a simultaneous application of driving in both the axial and angular direction reveals asymmetry in energy dissipation rate due the helical landscape.
\end{abstract}


\maketitle 

\section{Introduction}
\label{Sec:Intro}

The study of particle motion in confined geometries and the interplay between particle transport and the topology of the underlying environment is often encountered as a dominating issue in the behavior of physical, chemical, and biological processes. Helical or chiral landscapes are especially consequential because they can couple drift and rotation, enabling rectification, chiral selectivity, geometry-controlled mobility and interfacial transport. \cite{CISS_review_acs_chemrev,experiment_helical_particle_diffusion_tensor_doi:10.1073/pnas.2220033120,colloidal_particle_in_helical_field_cholestrics_microrheology_PhysRevLett.105.178302,Hydrodynamic_helical_microchannel_chase2022hydrodynamically,Helical_water_flow_zelenovskii2024detection,surface_corrugation_helicity_kurzthaler2024surface,ion_channel_hu2021helical,aristov2013separation-977,meinhardt2012separation-d91,huseby2025helical-5ab}Several theoretical and experimental works have established that helicoidal or general asymmetrical rigid particles show coupling between rotational and translational motion through their hydrodynamic mobility or diffusion tensor\cite{brenner1965coupling-bc6,brenner1967coupling-86e,kraft2013brownian-280,gonzalez2010hydrodynamic,cichocki2012communication-ad2}. Driven optically levitated nano-particles are a natural experimental platforms for stochastic thermodynamics where the particle inertia, geometry of the potential and damping anisotropy are required to be taken into account\cite{e20050326_Gieseler,tebbenjohanns2019cold-fbf,hempston2017force-aa3,zhalenchuck2022optical-358}. Several other works have also explored the consequences of helical geometry on dynamics in processes such as diffusion in a helical tube\cite{theory_helical_tube_diffusion_chavez2018unbiased},
diffusion on a curved surface\cite{theory_surface_diffusion_curved_PhysRevE.107.034801,langevin_on_curved_surface_PhysRevE_2025,castaeda-priego2012brownian-8f0},
         motion of charged particle in a toroidal helix\cite{theory_chaotic_dynamics_toroidal_helix_caharged_particle_PhysRevE.103.052217,theory_driven_toroidal_helix_kapitza_pendulum_PhysRevE.105.054204}, ion scattering in helical confinements\cite{theory_ion_scattering_in_helical_confinement_PhysRevE.111.014140}, curvature induced transport enhancement of colloids in confined environments\cite{curvature_induced_transport_spinning_colloids_RSC_2024} to name a few. Despite this broad relevance, systematic understanding of the dynamics of a structureless Brownian particle in helical landscapes, particularly in the context of the interplay between the helical geometry of the landscape, particle inertia, damping anisotropy, thermal noise and external driving  remains incomplete, which motivates our present study.\par 
    
    In this article we use a simple two-dimensional model to investigate the way by which the motion of a Brownian particle, and its linear response behavior are affected by the symmetry of a helical channel. An important characteristic of motion in a helical landscape is the topology induced correlation between linear and angular momentum. Consequences of this coupling can be explored in two dimensions, leading to a simplified mathematical description. 
    Using the standard Langevin setup of a particle moving in a given potential field under the influence of white noise and damping (constructed to satisfy the fluctuation dissipation relation with a given temperature), and optionally a driving force, we examine the linear response behavior focusing on (a) manifestations of the helical symmetry on the system dynamics as expressed by time-correlation functions and linear transport properties, and (b) the consequence of anisotropic friction (different damping parameters along the axial and angular helical directions) that may characterize a system with such symmetry.  \par 
    This paper is organized as follows: Sec.~\ref{Sec:Model_Details} discusses a two-dimensional model that exhibits the implications of the helical landscape symmetry on a particle motion. This symmetry makes it possible to describe the motion in terms of a screw coordinate along the helical valley, and a transverse coordinate which is locally normal to it (henceforth referred to as the screw-normal(SN) coordinate), along which the helical confinement is set. In terms of these coordinates the dynamics is separable and a further model assumption, setting this confinement as a harmonic potential, makes it possible to describe the motion analytically: free motion along the screw coordinate and harmonic oscillations in the SN direction.   
    The corresponding Langevin dynamics is described in Sec.~\ref{Sec:Brownian_motion_section_main_text} and analyzed further in Sec.~\ref{sec:main_text_header_sec_time_Correlations}. The case of isotropic friction (Sec.~\ref{Sec:istropic_case}) is shown to remain separable, combining a free Brownian particle diffusing in the screw direction with a trapped Brownian particle in harmonic potential in the SN direction. Sec.~\ref{sec:anisotropic_model} explores the more general case, where the frictions along the axial and angular directions are different, which is considerably more involved because the ensuing stochastic process is no longer separable. For such system we further discuss (in Sec.~\ref{Sec:Time_corr_functions}) how time correlation functions of linear observables of phase space variables of the system can be calculated, with a focus on the observables in the stationary subspace of the dynamics and particularly momentum time correlation functions. Sec.~\ref{Sec:Mobilites} focuses on calculation of dynamical mobility of the particle through the helical channel under a time-dependent drive as an exemplary transport property calculation and then analyzes how the interplay of the drive frequency, damping anisotropy and the geometry of the helix creates various transport regimes in both the zero frequency (DC) and the non-zero frequency (AC) cases,and then analytically calculates the lineshape of energy dissipation rate against the frequency of a monochromatic drive.Sec.~\ref{Sec:Conclusion} concludes with presenting  a summary of insights obtained from this study, and potential direction for future works.

\section{Model}
\label{Sec:Model_Details}
\begin{figure}
    \centering
    \includegraphics[width=1\linewidth]{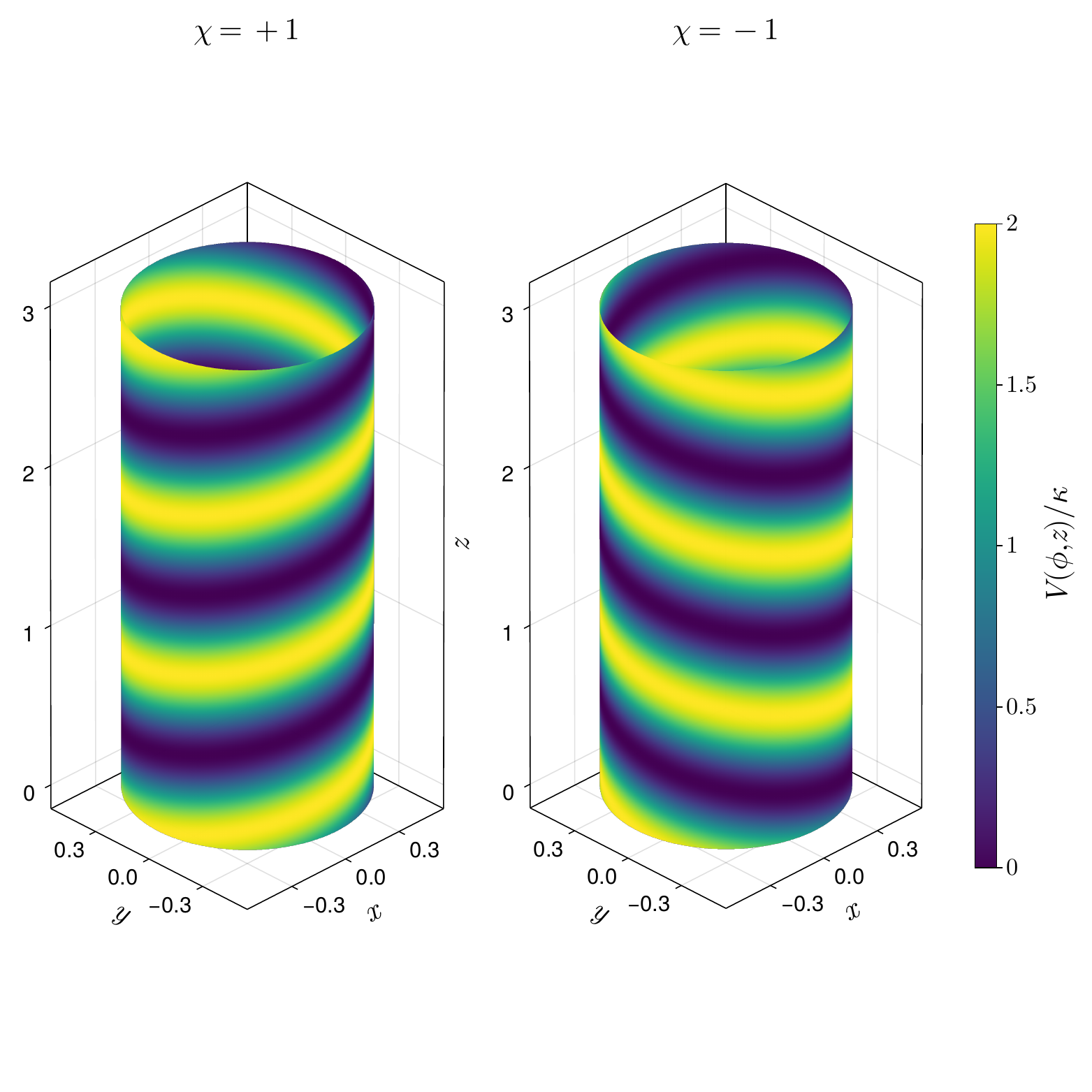}
    \caption{Plotted here for an example a helical potential landscape on a cylindrical manifold: $V(\phi,z;R,\chi)=\kappa(1-\cos\left(\phi-\frac{2\pi\chi z}{b}\right))$, which may be approximated as a quadratic potential near the helical valley. Parameters: $R=1/2,b=1$.}
    \label{fig:helical_pot}
\end{figure}
Our model is defined by a particle moving on the surface of a cylinder of radius R under the 2-dimensional helical potential of the form  
(in cylindrical coordinates $(R,\phi,z)$)
 \begin{equation}
 \begin{split}
     V(\phi,z;R,\chi)&=U\left(\phi-\frac{2\pi\chi}{b}z; R\right), \\ b&>0, \chi=\pm 1, R= \mathrm{const.}
 \end{split}
\end{equation}
with a non-negative potential that satisfies $U(\psi;R)=U(\psi+2\pi;R)$,
so that  $V(\phi, z)=V(\phi+2\pi,z)=V(\phi,z-\chi b)$, and has a minimum along the helical path $\psi - \frac{2\pi \chi z}{b}=0$. $R$ is taken constant so this helical path is confined to the surface of a cylinder of radius $R$. Here $b$ represents the pitch of the helix and $\chi$ represents its handedness: $\chi=-1$  and  $\chi=+1$  correspond to left and right-handed helices, respectively. See Fig.~\ref{fig:helical_pot} for an example of such potential.

It is convenient to recast these equations in a dimensionless form.
Expanding $U$ about the minimum in the form 
\begin{equation}
    \label{eq:potential_full_series}
     U(\psi)= \frac{1}{2}\kappa\psi^2 +\dots \ ,
\end{equation}
we use $\kappa$, $b$ and $\bar{\tau}=\sqrt{mb^2/\kappa}$ as the unit of energy, length and time respectively and redefine 
$(z/b)\rightarrow z$, $(R/b) \to R$\footnote{Henceforth $R$ is dimensionless}, and $(t/\bar{\tau})\to t$, and $(U/\kappa) \to U$. 
Also any external axial force $f_z$ and torque $f_\phi$ (used in Sec.~\ref{Sec:Mobilites}) are scaled accordingly to 
\begin{equation}
    \label{eq:driving_force_dimensions}
    \begin{split}
        \frac{f_z}{mb\bar{\tau}^{-2}}&=\frac{f_z}{\kappa b^{-1}}\to f_z, \\
        \frac{f_\phi}{mb^2\bar{\tau}^{-2}}&=\frac{f_\phi}{\kappa }\to f_\phi.
    \end{split}
\end{equation}
The dimensionless kinetic energy of the particle, along the angular direction is $\frac{1}{2}R^2\dot{\phi}^2$, and along the axial direction is $\frac{1}{2}\dot{z}^2$. Hence, the Lagrangian $L(\phi,\dot{\phi},z,\dot z)$ of the system in dimensionless coordinate is 
\begin{equation}
    \label{eq:Lagarangian_eq}
    \mathrm{L}(\phi,\dot{\phi},z,\dot z;R,\chi)=\frac{1}{2} R^2\dot{\phi}^2 +\frac{1}{2}\dot{z}^2 -U\left(\phi-2\pi\chi z\right).
\end{equation}
Note that $R$ (the ratio between the radius and pitch of the helix) and $\chi$ (the helix handedness) are the only geometric parameters that enters the Lagrangian and determines the dynamics. In this dimensionless form, the momenta conjugate to $z$ and $\phi$ are  $p_z=\frac{\partial \mathrm{L}}{\partial\dot{z}}=\dot{z}$, and $L_\phi=\frac{\partial \mathrm{L}}{\partial\dot{\phi}}=R^2\dot{\phi}$  respectively, and the Hamiltonian is 
\begin{equation}
\begin{split}
    H(\phi,z,L_\phi,p_z;R,\chi)&=\dot{\phi}L_\phi+\dot{z}p_z-\mathrm{L}\\
&=\frac{L_\phi^2}{2R^2}+\frac{p_z^2}{2}+U\left(\phi-2\pi\chi z\right),
\end{split}
    \label{eq:Hamiltonian_scaled_general}
\end{equation}

\subsection{Screw/screw-normal representation}
\label{subsec:coordinate_transform}

\begin{figure*}
    \centering
    \includegraphics[width=1.0\linewidth]{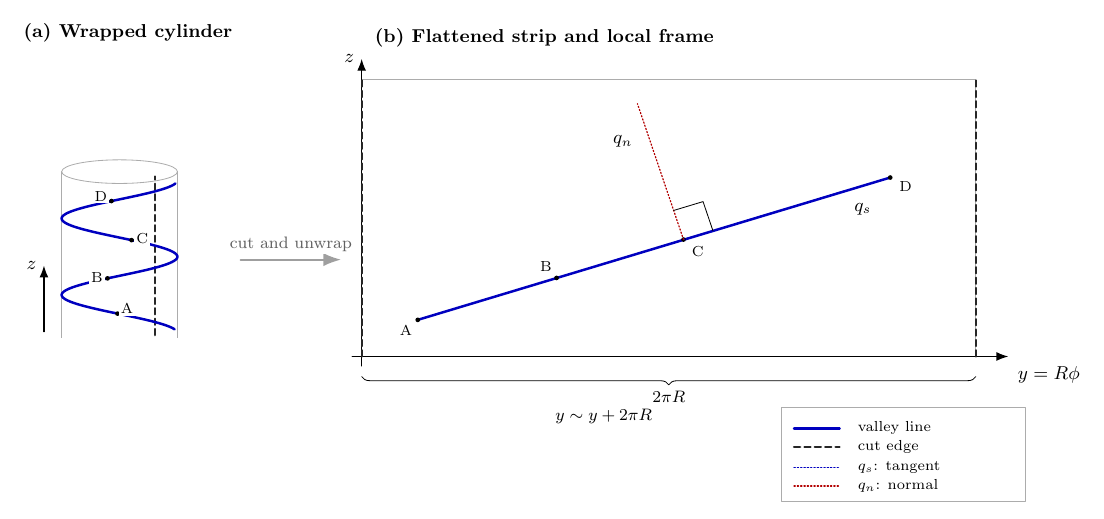}
    \caption{Screw/Screw-Normal coordinate conversion. This figure illustrates how the helical coordinates $(\phi,z)$ are converted to the screw/screw-normal coordinates as given by Eq.~\eqref{eq:coord_transform_yz_to_s-lambda}. Panel (a) shows the cylindrical manifold where the particle is confined in. The blue curve $ABCD$ represent the helical valley along which the minima of the potential exist. The black dotted line shows a cut in the manifold along z direction, which allows us to unwrap the cylindrical manifold in a plane. Panel (b) illustrates the unwrapped manifold. We scale the angular coordinate as $y=\phi R$ which has now the periodicity of $2\pi R$. The same helical valley $ABCD$ is taken to be the direction of the screw coordinate $q_s$. We chose point $C$ as a reference point in the diagram and and draw a perpendicular to the helical valley shown as the red dotted line which represent the screw-normal coordinate $q_n$. $(q_s,q_n)$ forms an orthonormal coordinate system which can be obtained by rotating the $(y,z)$ coordinate.   }
    \label{fig:coordinate_transform}
\end{figure*}

Eqs.~\eqref{eq:deterministic_eoms} are easily solved by transforming to the screw/screw-normal (S/SN) coordinates\cite{petrie2007potential-fc0}  $(q_s,q_n)$, defined as follows (see Fig.~\ref{fig:coordinate_transform}) 

\begin{equation}
\label{eq:coord_transform_yz_to_s-lambda}
    \begin{split}
        q_n&= \frac{1}{\sqrt{\ g(R) }}\left(\frac{y}{2\pi\chi R}- z\right), \\
        q_s&=\frac{1}{\sqrt{\ g(R)}}\left(y+\frac{ z}{2\pi\chi R}\right)
    \end{split}
\end{equation}
with $y=\phi R$, and 
\begin{equation}
    \label{eq:geometric_factor}
    \ g(R)=\left(1+\frac{1}{4\pi^2 R^2}\right).
\end{equation}
Note that $q_s$ and $q_n$ are orthogonal to each other \footnote{Note that Eq.~\eqref{eq:coord_transform_yz_to_s-lambda} can be written in the form $q_s=y\cos\alpha-z\sin\alpha$ and $q_n=y\sin\alpha +z\cos\alpha$, showing the rotation by the angle $\alpha$ with $\tan{\alpha}=2\pi\chi R$} and the linear transformation $(y,z)\leftrightarrow(q_n,q_s)$ is area preserving $\left|\pdv{(q_n,q_s)}{(y,z)}\right|=1$. The corresponding conjugate momenta are given by 
\begin{equation}
\label{eq:momentum_transform_yz_to_s-lambda}
    \begin{split}
        p_n :=\frac{\dd q_n}{\dd t}&=
            \frac{1}{\sqrt{\ g(R)}}\left(\frac{p_y}{2\pi \chi R}- p_z\right)\\&=\frac{1}{\sqrt{\ g(R)}}\left(\frac{L_\phi}{2\pi\chi R^2}- p_z\right),\\
        p_s:=\frac{\dd q_{s}}{\dd t}&=\frac{1}{\sqrt{\ g(R)}}\left(p_y+\frac{p_z}{2\pi \chi R}\right) \\ &=\frac{1}{\sqrt{\ g(R)}}\left(\frac{L_\phi}{R}+\frac{ p_z}{2\pi \chi R}\right) 
    \end{split}
\end{equation}
This is a canonical transformation\cite{goldstein2002classical} that leads to the separable transformed Hamiltonian
\begin{equation}
    \label{eq:modified_hamiltonain_general}
   \bar{H}(q_s,q_n,p_s,p_n)=\bar{H}_s(q_s,p_s)+\bar{H}_n(q_n,p_n),
\end{equation}
with
\begin{equation}
\label{eq:screw_hamiltonian_in_screw_coord}
 \bar{H}_s(q_s,p_s)= \frac{p_s^2}{2}  , 
\end{equation}
\begin{equation}
\label{eq:normal_hamiltonian_innormal_coord}
    \bar{H}_n(q_n,p_n)=\frac{p_n^2}{2}+U\left[\left(2\pi \chi \sqrt{\ g(R)}\right)q_n\right].
\end{equation}
Eq.~\eqref{eq:screw_hamiltonian_in_screw_coord} describes the motion of a free particle carrying momentum $p_s$ along the valley (the screw direction) created by the minima of the helical potential. While, the Hamiltonian Eq.~\eqref{eq:normal_hamiltonian_innormal_coord}
represents the dynamics in the $(q_n,p_n)$ subspace, i.e., in the SN direction, which is a one dimensional motion under the potential $U\left[\left(2\pi \chi \sqrt{\ g(R)}\right)q_n\right]$. Note that the screw momentum $p_s$ is a constant of motion.
\subsection{Harmonic approximation}
For sufficiently small deviations from the minimum along the SN direction $|(\phi-2\pi\chi z)| \ll 1 $ (or equivalently $2\pi\sqrt{\ g(R)}|q_n|\ll1$), the potential $U(\psi)$ can be approximated by disregarding the higher order terms in Eq.~\eqref{eq:potential_full_series}.The Hamiltonian in Eq.~\eqref{eq:Hamiltonian_scaled_general} then takes the form
\begin{equation}
\label{eq:Hamiltonian_scaled} 
\begin{split}
H(\phi,z,L_\phi,p_z;R,\chi)=\frac{L_\phi^2}{2R^2}+\frac{p_z^2}{2}+\frac{1}{2}\left(\phi-2\pi\chi z\right)^2,
\end{split}  
\end{equation}
with the corresponding equations of motions \begin{equation}
\label{eq:deterministic_eoms}
\begin{split}
     \dot{\phi} &= \frac{L_\phi}{R^2},\\
    \dot{L_\phi} &= -(\phi-2\pi \chi z),\\
    \dot{z} &= p_z, \\
    \dot{p}_z&= 2\pi\chi(\phi -2\chi\pi z).
\end{split}
\end{equation}
In the S/SN representation, under this harmonic approximation the Hamiltonian is 
\begin{equation}
\begin{split}
    \bar{H}(q_s,p_s,q_n,p_n;R)&=\frac{p_s^2}{2}+\frac{p_n^2}{2}+\frac{1}{2}\omega_0^2(R)q_n^2, \\ \omega_0(R)&=\sqrt{4\pi^2 \ g(R)},
\end{split}
\label{eq:modified_hamiltonain}
\end{equation}
leading to the equations of motions
\begin{equation}
    \label{eq:SN_coordinate_deterministic_EOM}
    \begin{split}
       \dot{q}_s &=p_s,\\
       \dot{p}_s&=0,\\
       \dot{q}_n&=p_n, \\
       \dot{p}_n&=-\omega_0^2(R)q_n.
    \end{split}
\end{equation}

\section{Brownian dynamics}
\label{Sec:Brownian_motion_section_main_text}
Next, we address Brownian dynamics on this helical landscape by supplementing Eq.~\eqref{eq:deterministic_eoms} by Markovian random noise and damping terms that satisfy standard fluctuation-dissipation (detailed balance) conditions\cite{kubo1966fluctuation-dissipation-d79,hnggi1982stochastic-873,Seifert_2025,chorin_hald2009stochastic,sekimoto_book2010stochastic,kurtJacobs_2010_book}. To account for possible differences between the interaction of the particle and its thermal environment along the axial and angular directions, we allow for different noise and damping parameters  $\gamma_z$ and $\gamma_\phi$ in these directions. In cylindrical coordinates and our dimensionless representation this leads to the following stochastic equations of motion (see Appendix.~\ref{app:derivation_langevin_equations}) 
\begin{equation}
\label{eqn:SDE_scaled_cylinder_coordinates}
\begin{split}
    \dd \phi &= \frac{L_\phi}{R^2}\dd t \\
    \dd L_\phi &= -(\phi-2\pi \chi z)\dd t - \gamma_{\phi} L_{\phi} \dd t + \sqrt{2\gamma_{\phi} R^2\beta^{-1}} \dd W^{\phi}_{t}\\
    \dd z &= p_z  \dd t\\
    \dd p_z&= 2\pi\chi(\phi -2\pi\chi z)\dd t -\gamma_{z} p_z \dd t + \sqrt{2\gamma_{z} \beta^{-1}} \dd W_{t}^{z}
\end{split}
\end{equation}
where $dW_t^\phi$ and $dW_t^z$ are Wiener increments\footnote{informally can be understood as $\sigma\frac{\mathrm{d}W_t}{\mathrm{d}t}\equiv \eta(t)$ with $\langle\eta(t)\eta(s)\rangle=\sigma^2\delta(t-s)$} which have the following properties\footnote{In Eq.~\eqref{eq:Weiner_increment_prop_main_text} the correlation among the Wiener increments are represented in shorthand as $\langle\mathrm{d}W_t^i{d}W_t^j\rangle=\mathrm{d}t \mathrm{d}s \delta(t-s)$ in the main text, and it should be interpreted as follows. For test functions $f(t)$ and $g(t)$, $\mathbb{E}\left[\left((\int f(t)\mathrm{d}W_t^i\right)\left(\int g(s)\mathrm{d}W_s^i\right)\right]=\delta_{ij}\int\mathrm{d}t\int\mathrm{d}s \delta(t-s)f(t)g(s)=\delta_{ij}\int\mathrm{d}t f(t)g(t)$.} \cite{chorin_hald2009stochastic}\cite{kurtJacobs_2010_book}:
\begin{equation}
\label{eq:Weiner_increment_prop_main_text}
\begin{split}  \langle  dW_t^{\phi} \rangle &=0,\\ \langle dW_t^{z} \rangle &=0,\\
    \langle dW_t^{\phi} dW_s^{\phi}\rangle  &=\dd s  \dd t \delta(t-s),\\
    \langle dW_t^{z} dW_s^{z} \rangle &= \dd s \dd t \delta(t-s), \\
    \langle dW_t^{\phi} dW_s^{z} \rangle  &=0,
\end{split}
\end{equation}
and $\beta$ stands for the dimensionless inverse temperature 
\begin{equation}
    \beta=\frac{\kappa}{k_BT}.
\end{equation}
with $T$ being the temperature of the environment and where $k_B$ Boltzmann's constant (note that the dimensionless thermal energy is defined in terms of the energy parameter $\kappa$ introduced in Eq.~\eqref{eq:potential_full_series}). Again in this case, $R$ (the helix radius pitch ratio), and $\chi$ (the handedness parameter) are the only geometric parameters that characterizes this stochastic dynamics. Further notice that the noise strengths along the $\phi$ and $z$ coordinates also differ reflecting their geometric difference: the noise strength in $\phi$ direction explicitly depends on $R$, which captures the inertial effect arising in the particle motion due to confinement in the cylindrical manifold. \footnote{Here we have introduced the model in Eq.~\eqref{eqn:SDE_scaled_cylinder_coordinates} by heuristic arguments, however we can also derive these equations of motions starting from Langevin equations for a system with thermostat in cartesian coordinates and then applying constraint on the particle to move in a cylindrical manifold. Details are given in Appendix~\ref{app:derivation_langevin_equations}.}
Indeed starting from Langevin dynamics in the $(\phi, L_\phi, z, p_z)$ representation and transforming to S/SN coordinates as done in Sec~\ref{Sec:istropic_case} leads to
\begin{equation}
    \label{eq:anisotropic_transverse}
    \begin{split}
        \dd q_n&= p_n \dd t;\\
        \dd p_n 
        &=-\omega_0^2(R) q_n\dd t -\Gamma_n(p_n, p_s) \dd t +\mathrm{d}B_t^n
    \end{split}
\end{equation}
and,
\begin{equation}
    \begin{split}
        \label{eq:anisotropic_screw}
        \dd q_s&= p_s \dd t\\
        \dd p_s 
        &=-\Gamma_s(p_n,p_s) \dd t + \dd B^s_t
    \end{split}
\end{equation}
where the effective damping factors $\Gamma_n,\Gamma_s$ and the noise terms $\mathrm{d}B_t^\lambda,\mathrm{d}B_t^s$ are given
by
\begin{equation}
\label{eq:noise_damping_eqn_screw_transverse_anisotropic}
    \begin{split}
        \Gamma_n(p_n, p_s)&=\frac{1}{\sqrt{\ g(R)}}\left( \frac{\gamma_{\phi}L_\phi(p_n,p_s)}{2\pi \chi R^2}-\gamma_zp_z(p_n,p_s)\right)\\
        \Gamma_s(p_n,p_s)&=\frac{1}{\sqrt{\ g(R)}}\left(\frac{\gamma_{\phi}L_\phi(p_n,p_s)}{R}+\frac{\gamma_{z}p_z(p_n,p_s)}{2\pi\chi R} \right)\\
        \mathrm{d}B_t^n&=\frac{\sqrt{2  \beta^{-1}}}{\sqrt{\ g(R)}}\left[\frac{1}{2 \pi \chi R}\sqrt{\gamma_{\phi}}\dd W_t^\phi -\sqrt{\gamma_{z}}\dd W_t^z\right]\\
         \mathrm{d}B_t^s&=\frac{\sqrt{2  \beta^{-1}}}{\sqrt{\ g(R)}}\left[\frac{1}{2 \pi \chi R}\sqrt{\gamma_{z}}\dd W_t^z +\sqrt{\gamma_{\phi}}\dd W_t^\phi\right].
    \end{split}
\end{equation}
with
\begin{equation}
\begin{split}
    \langle \mathrm{d}B_t^n\mathrm{d}B_{t'}^n\rangle&=\frac{2 \beta^{-1}}{\ g(R)}\left(\frac{\gamma_{\phi}}{4\pi^2 R^2}+\gamma_{z}\right)\delta(t-t')\mathrm{d}t \mathrm{d}t'\\
    \langle \mathrm{d}B_t^s\mathrm{d}B_{t'}^s\rangle &=\frac{2 \beta^{-1}}{\ g(R)}\left(\gamma_{\phi}+\frac{\gamma_{z}}{4\pi^2 R^2}\right)\delta(t-t')\mathrm{d}t \mathrm{d}t'\\
    \langle\mathrm{d}B_t^s\mathrm{d}B_{t'}^n \rangle &=\frac{2 \beta^{-1}}{2\pi\chi R\ g(R)}(\gamma_{\phi}-\gamma_{z})\delta(t-t')\mathrm{d}t \mathrm{d}t'
\end{split}
\end{equation}
It is interesting to note that although $p_s$ is no longer conserved in the presence of thermal interactions, the quantity  
\begin{equation}
\begin{split}
\label{eq:I_s_defn_main}
    I_s(\phi,L_\phi, z,p_z)
    &=p_s(p_z,L_\phi) + \bar\gamma q_s (\phi,z)\\
    &+\frac{\delta\gamma}{\sqrt{\ g(R)}}\left(R \phi-\frac{z}{2\pi\chi R}\right)
\end{split} 
\end{equation}
 with  $\bar{\gamma}=\frac{\gamma_\phi+\gamma_z}{2}$ and $\delta\gamma=\frac{\gamma_\phi-\gamma_z}{2}$, can be shown (Appendix~\ref{app:I_s_conservation_and_related_properties}) to be conserved on the level of ensemble average, \begin{equation}
 \label{eq:I_s_mean_conservation}
     \frac{\mathrm{d}\langle I_s\rangle}{\mathrm{d}t}=0
 \end{equation}
 This holds for arbitrary helical potential $U(\phi -2\pi\chi z)$, not just the harmonic approximation.
 From Eq.~\eqref{eq:I_s_mean_conservation} it follows that
 \begin{equation}
 \begin{split}
     \langle\dot{p}_s\rangle + \bar{\gamma}\langle p_s\rangle&=-\frac{\delta\gamma}{\sqrt{\ g(R)}}\left(\frac{ \langle L_\phi\rangle}{R}-\frac{\langle p_z\rangle}{2\pi\chi R}\right)
 \end{split}
 \end{equation}
 or,
 \begin{equation}
    \begin{split}
        \langle p_s(t)\rangle&= \langle p_s(0)\rangle e^{-\bar\gamma t}\\ &-\frac{\delta\gamma}{\sqrt{\ g(R)}} \int_0^t\mathrm{d}t'e^{-\bar\gamma (t-t')}\left(\frac{ \langle L_\phi(t')\rangle}{R}-\frac{\langle p_z(t')\rangle}{2\pi\chi R}\right).
    \end{split}
 \end{equation}    
 Thus we see that anisotropy $\delta\gamma$ can create memory effects in the dynamics of $\langle p_s\rangle $, which depends on the helical parameters $\chi$ and $R$. This effective memory vanishes in the isotropic case for which $\delta\gamma=0$.
In Appendix~\ref{app:I_s_conservation_and_related_properties} we also show that $I_s$, Eq.~\eqref{eq:I_s_defn_main}, plays the role of an effective random force on the screw coordinate, which satisfies a generalized fluctuation-dissipation relation. We find (again for a general helical potential)
 \begin{equation}
\langle\mathrm{d}I_s(t)\mathrm{d}I_s(t')\rangle=2\beta^{-1}\gamma_{\mathrm{eff}}(R)\delta(t-t')\mathrm{d}t\mathrm{d}t'
 \end{equation}
 with  $\chi$-independent effective friction \footnote{If we define an effective diffusion coefficient for $I_s$ using the relation $\mathrm{Var}\left[I_s(t)\right]=\langle I_s^2(t)\rangle-\langle I_s(t)\rangle^2=2 D^p_\mathrm{eff}(R)t$, we get a momentum space Einstein relation relating the inverse bath temperature $\beta$, effective momentum diffusion constant $D^{p}_\mathrm{eff}(R)$, and effective damping $\gamma_\mathrm{eff}(R)$ as $ \beta D^p_\mathrm{eff}(R)=\gamma_\mathrm{eff}(R)$.} 
 \begin{equation}
     \begin{split}
     \label{eq:gamma_effective}
         \gamma_{\mathrm{eff}}(R)&=\frac{4\pi^2 R^2\gamma_\phi +\gamma_z}{4\pi^2R^2+1}\\
         &=\bar{\gamma}+\left(\frac{4\pi^2 R^2-1}{4\pi^2 R^2+1}\right)\delta\gamma.
     \end{split}
 \end{equation}
We next proceed to examine how the stochastic dynamics described by Eq.~\eqref{eqn:SDE_scaled_cylinder_coordinates} is manifested in the linear response behavior of a particle moving in a helical landscape while interacting with its thermal environment. We start with the simplest isotropic limit $\delta\gamma=0$ then proceed with more general anisotropic case.
\section{Time correlation functions}
\label{sec:main_text_header_sec_time_Correlations}
\subsection{Isotropic thermal interactions}
\label{Sec:istropic_case}
In the isotropic case, $\gamma_{\phi}=\gamma_{z}=\gamma$, the equations of motion in the S/SN coordinates remain separable. Thus Eq.~\eqref{eqn:SDE_scaled_cylinder_coordinates} leads to
\begin{equation}
    \label{eqn:sde_lambda_coord}
    \begin{split}
        \dd q_n&= p_n \dd t\\
        \dd p_n&=-\omega_0^2(R) q_n\dd t - \gamma p_n \dd t + \dd B_t^n
    \end{split}
\end{equation}
and,
\begin{equation}
    \label{eq:screw_sde}
    \begin{split}
        \dd q_s&= p_s \dd t\\
        \dd p_s &= -\gamma p_s \dd t + \dd B^s_t
    \end{split}
\end{equation}
where the noises $dB_t^n$ and $dB_t^s$ are expressed in terms of the original Wiener increments $dW_t^\phi$ and $dW_t^z$ as  
\begin{equation}
    \label{eq:dB_n_dB_s_defn}
    \begin{split}
        \dd B^n_t= \frac{\sqrt{2\gamma  \beta^{-1}}}{\sqrt{\ g(R)}}\left[\frac{1}{2 \pi \chi R}\dd W_t^\phi -\dd W_t^z\right]\\
        \dd B^s_t=\frac{\sqrt{2\gamma  \beta^{-1}}}{\sqrt{\ g(R)}}\left[\dd W_t^\phi+\frac{1}{2\pi \chi R} \dd W_t^z\right]
    \end{split}
\end{equation}
Eqs.~\eqref{eq:Weiner_increment_prop_main_text} and~\eqref{eq:dB_n_dB_s_defn} lead to
\begin{equation}
    \begin{split}
        \langle \dd B_t^n\rangle &=0 \\
        \langle \dd B_t^s\rangle&=0 \\
         \langle \dd B_t^n\dd B_{t'}^n \rangle &=2\gamma \beta^{-1}\delta(t-t')\dd t\dd t'\\
        \langle \dd B_t^s\dd B_{t'}^s \rangle &=2\gamma \beta^{-1}\delta(t-t')\dd t\dd t' \\
          \langle \dd B_t^n\dd B_{t'}^s \rangle &=\frac{2\gamma \beta^{-1}}{\ g(R)}\left(\frac{1}{2\pi \chi R}-\frac{1}{2\pi \chi R}\right)\delta(t-t')\dd t\dd t' \\ &= 0
    \end{split}
\end{equation}
Since the cross correlation between the noise terms $dB_t^s$ and $dB_t^n$ is zero, the dynamics in screw and the screw-normal coordinates are uncorrelated, hence these motions proceed independently. This makes the mathematical evaluation simple, still with interesting consequences for observations made along the $z$ and $\phi$ coordinates. This is seen by the following expressions (obtained in Appendix~\ref{app:analytical_soln_isotropic} ) for the stationary time correlation functions,
\begin{subequations}
\label{eq:main_text_isotrpic_time_corr_functions}
\begin{align}
    \langle L_\phi(\tau)L_\phi(0)\rangle&=\frac{\beta^{-1}e^{-\gamma \tau}R^2}{\ g(R)}\left[1+\frac{e^{\frac{\gamma \tau}{2}}}{4\pi^2R^2}\mathcal{K}(\tau)\right],\label{subeq:phiphi_corr_isotropic}\\
    \langle p_z(\tau)p_z(0)\rangle&=\frac{\beta^{-1}e^{-\gamma \tau}}{\ g(R)}\left[\frac{1}{4\pi^2R^2}+e^{\frac{\gamma \tau}{2}}\mathcal{K}(\tau)\right],\label{subeq:zz_corr_isotropic}\\
     \langle L_\phi(\tau) p_z(0)\rangle&=\frac{\beta^{-1}e^{-\gamma \tau}}{2\pi\chi\ g(R)}\left[1-e^{+\frac{\gamma \tau}{2}}\mathcal{K}(\tau)\right],\label{subeq:phiz_corr_isotropic}\\
     \mathcal{K}(\tau)&=\cosh\left(\frac{\Omega \tau}{2}\right)-\frac{\gamma}{\Omega}\sinh\left(\frac{\Omega \tau}{2}\right).
\end{align}
\end{subequations}
where,
\begin{equation}
     \Omega=\sqrt{\gamma^2-4(\omega_0(R))^2}.
\end{equation}
\begin{figure}
    \centering   \includegraphics[width=1.0\linewidth]{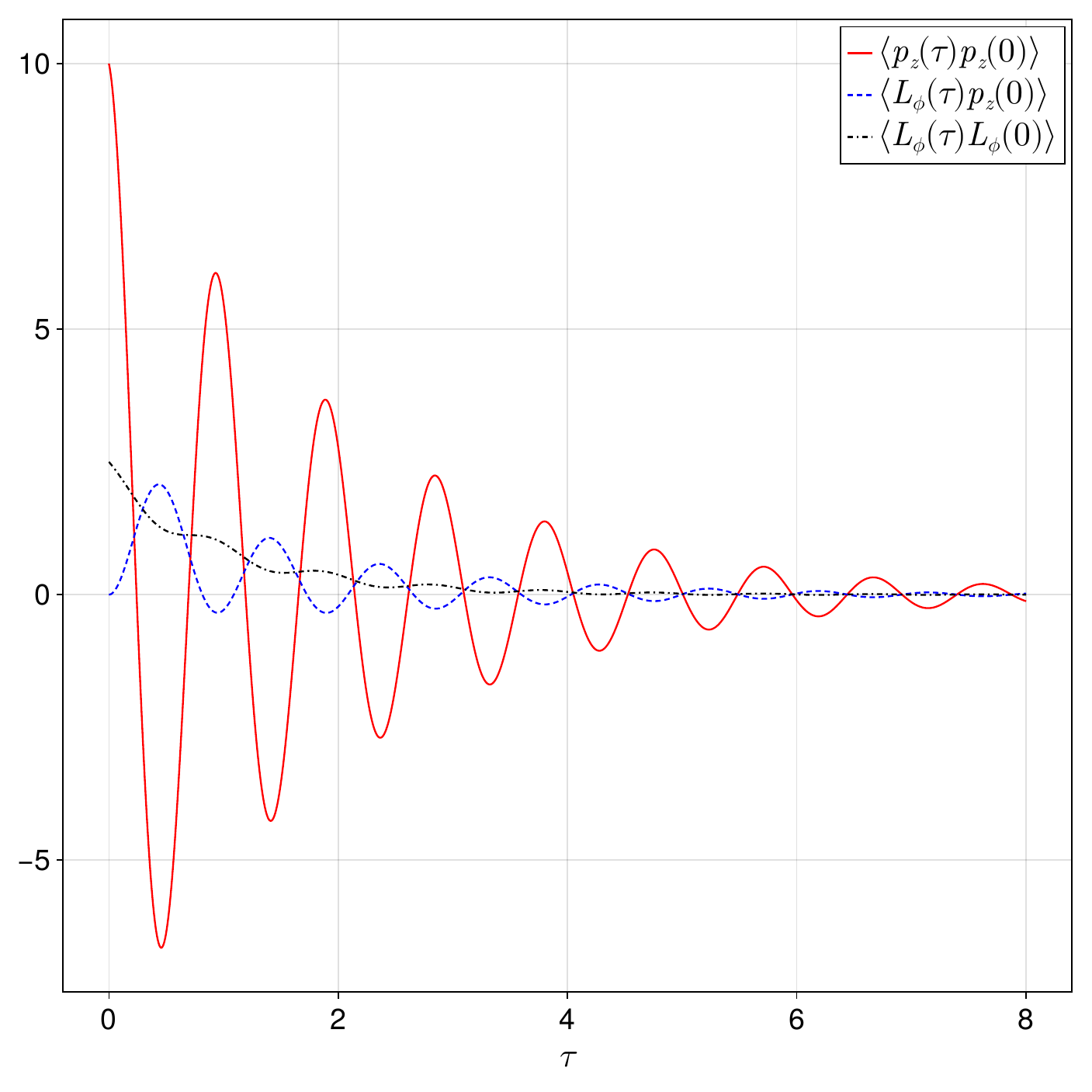}
    \caption{Time correlation functions of the momenta (Eq.~\eqref{eq:main_text_isotrpic_time_corr_functions} or Eq.~\eqref{eqn:S_p_krylov_split_main_text}) for the isotropic case. At $\tau=0$ (equilibrium moments)the diagonal second moments show the expected equipartition, $\langle L_\phi^2\rangle/R^2=\langle p_z^2\rangle=\beta^{-1}$. The cross correlation function vanishes at $\tau=0$, illustrating the dynamic nature of the correlation between the linear and angular momenta. Parameters: $\beta=0.1,R=0.5$ and $\gamma_\phi=\gamma_z=1$. }
    \label{fig:isotropic_time_correlations}
\end{figure}
In particular, even though equal time correlation between the linear and angular momenta , $p_z$  and $L_\phi$, vanishes as seen in Eq.~\eqref{subeq:phiz_corr_isotropic},  they develop a mutual correlation (sometimes referred to as a “locking phenomenon”) at finite times. An example is shown in Fig.~\ref{fig:isotropic_time_correlations}.
Linear response theory implies that these cross-correlation functions are related to cross-mobilities, the way a force in axial ($z$) direction gives rise to angular ($\phi$ direction) motion while an angular driving leads to linear ($z$-direction) motion (see Appendix~\ref{App:generic_linear_response}). 
\subsection{Anisotropic thermal interactions}
\label{sec:anisotropic_model}
Next, we turn to the more general anisotropic case. In this case, the stochastic dynamics described by Eqs.~\eqref{eq:anisotropic_screw},\eqref{eq:anisotropic_transverse} are no longer separable even after transforming to the S/SN coordinates. This is caused by the mutual coupling through the damping ($\Gamma$) terms, and the non-zero cross-correlation in the noise terms. However, the absence of a restoring force that depends on the screw position $q_s$ implies that at infinite time all values of this coordinate are equally probable. This invariance of the dynamics along the translational screw coordinate implies the existence of a zero mode: the (infinite) system does not relax to a stationary distribution at a finite time in the infinite phase space.
It is still possible to describe the linear response behavior in the three-dimensional stable subspace of the other three variables and to obtain stationary time-correlation functions of observables associated with these variables. The detailed mathematical procedure is described in Appendix~\ref{App_sec:analysis_of_general_anisotropic_model_Details} and below we present its main results.
\subsubsection{Time evolution in the stable and zero-mode subspaces}
Since we are primarily interested in measuring responses in the original cylindrical coordinate frame, we analyze the anisotropic system in the $(\phi,L_\phi,z,p
_z)$ phase space and use spectral projectors to separate the zero-mode.
We start with rewriting Eq.~\eqref{eqn:SDE_scaled_cylinder_coordinates} in a matrix stochastic differential equation (SDE) form as 
\begin{equation}
\label{eq:matrix_SDE_anisotropic}
\mathrm{d} \vec{X}(t)
=
\mathbf{C}\ \vec{X}(t)\, \mathrm{d}t
+
\mathbf{\Sigma}\ \mathrm{d}\vec{W}(t),
\end{equation}
with, $\vec{X}(t)= \left(\phi(t) \ \ L_\phi(t) \ \ z(t) \ \ p_z(t)\right)^{T}$ \footnote{the superscript $T$ over a matrix denotes transpose throughout the article}
and,
\begin{equation}
\label{eq:drift_mat}
    \mathbf{C}=
\begin{pmatrix}
0 & \dfrac{1}{R^{2}} & 0 & 0 \\
-1 & -\gamma_{\phi} & 2\pi\chi & 0 \\
0 & 0 & 0 & 1 \\
2\pi\chi & 0 & -4\pi^{2} & -\gamma_{z}
\end{pmatrix},
\end{equation}
\begin{equation}
\label{eq:noise matrix}
\mathbf{\Sigma}=
\begin{pmatrix}
0 & 0 \\
\sqrt{2\gamma_{\phi} R^{2}\beta^{-1}} & 0 \\
0 & 0 \\
0 & \sqrt{2\gamma_{z}\beta^{-1}}
\end{pmatrix}.    
\end{equation}
In Eq.~\eqref{eq:matrix_SDE_anisotropic}$ \mathrm{d}\vec{W}(t)=\left(\mathrm{d}W_t^{\phi} \ \
\mathrm{d}W_t^{z}\right)^T$ is a $2 \times1$ column matrix representing the noise vector with the properties
\begin{equation}
    \begin{split}
    \langle \mathrm{d}\vec{W}(t)\rangle&=\begin{pmatrix}
        0\\
        0
    \end{pmatrix},\\
    \langle \mathrm{d}\vec{W}(t)\mathrm{d}\vec{W}^T(s) \rangle&=\mathrm{d}t \ \mathrm{d}s\ \delta(t-s) \mathbf{I}_2,\\
    \mathbf{I}_2&=\begin{pmatrix}
        1 & 0\\
        0 &1
    \end{pmatrix}.
    \end{split}
\end{equation}
Eq.~\eqref{eq:matrix_SDE_anisotropic} represents the dynamics of the system state $\vec{X}(t)$ as a multivariate Ornstein-Uhlenbeck process\cite{kurtJacobs_2010_book,risken1996fokker} in the $(\phi,L_\phi,z,p_z)$ phase space.\par 
From Eq.~\eqref{eq:matrix_SDE_anisotropic} the mean evolves according to
\begin{equation}
    \langle \vec{X}(t)\rangle=e^{\mathbf{C}t}\langle \vec{X}(0)\rangle,
\end{equation}
while the phase space covariance matrix 
\begin{equation}
    \label{eq:defn_covariance_matrix}
    \mathbf{S}(t)=\langle\delta \vec{X}(t)\delta \vec{X}^T(t)\rangle,
\end{equation}
where $\delta \vec{X}(t)=\vec{X}(t)-\langle\vec{X}(t)\rangle$, evolves under the continuous time Sylvester-Lyapunov equation\cite{behr2019solution} (see Appendix~\ref{App:SylvesterLyapunovEquation_Review}), 
\begin{equation}
\label{eq:Sylvester_eqn}
\begin{split}
    \dv{\mathbf{S}(t)}{t}&= \mathbf{CS}(t)+\mathbf{S}(t)\mathbf{C}^T+\mathbf{\Sigma\Sigma}^T. \\
\end{split}
\end{equation}
The phase space probability distribution $\rho(\phi,L_\phi,z,p_z)$ does not relax to a stationary state due to the existence of the zero mode of the matrix $\mathbf{C}$, which corresponds to the diffusion along the screw coordinate. Thus, to calculate stationary time-correlation functions, we need to exclude this diffusive motion from the otherwise stable dynamics near the stationary state associated with what we refer to as the stable subspace\cite{Ott_2002}. This exclusion is achieved by following a procedure based on spectral projection operators \cite{behr2019solution,nakazima_10.1143/PTP.20.948,zwanzig_10.1063/1.1731409} applied to the continuous-time Sylvester Lyapunov equation ~\eqref{eq:Sylvester_eqn}. It leads (Appendix~\ref{App:Zwanzig-Sylvester}) to the following expression for the covariance matrix in the long time ($t \to \infty$) limit:
\begin{equation}
\label{eq:asymptotic_covariance_matrix_text}
    \mathbf{S}(t)\sim \mathbf{S}_{\text{init}}+\mathbf{D}_0t+\mathbf{S}_\infty
\end{equation}
where
\begin{equation}
    \label{eq:zero_mode_matrices}
    \mathbf{S}_{\text{init}}= \mathbf{\Pi}_0\mathbf{S}(0)\mathbf{\Pi}_0^T, \ \ \mathbf{D}_0= \left(\mathbf{\Pi}_0\mathbf{\Sigma \Sigma}^T\mathbf{\Pi}_0^T\right),
\end{equation}
\begin{equation}
\label{eq:S_infty_expression_main_text}
\begin{split}
\mathbf{S}_\infty&=-\mathbf{C}^{-1}_D\mathbf{\Sigma\Sigma}^T\mathbf{\Pi}_0^T   -\mathbf{\Pi}_0\mathbf{\Sigma\Sigma}^T\left(\mathbf{C}^{-1}_D\right)^T \\
&+ \sum^{2}_{m,n=0}J_{mn}\mathbf{E_m}\mathbf{\Sigma\Sigma}^T\mathbf{E_n}^T,
\end{split}
\end{equation}
\begin{equation}
    \mathbf{I}_4=\mathrm{diag}(1,1,1,1)
\end{equation}
\begin{equation}
    \label{eq:Pi_0_def}
    \mathbf{\Pi}_0=\frac{1}{a_0}\left(\mathbf{C}^3+a_2\mathbf{C}^2+a_1\mathbf{C}+a_0\mathbf{I}_4\right),
\end{equation}
\begin{equation}
    \label{eq:Q_0_mat_maintext_defn}
    \begin{split}
        \mathbf{Q}_0&=\mathbf{I}_4-\mathbf{\Pi}_0\\
        &=-\frac{1}{a_0}\left(\mathbf{C}^3+a_2\mathbf{C}^2+a_1\mathbf{C}\right),
    \end{split}
\end{equation}
\begin{equation}
    \label{eq:Krylov_basis}
    \begin{split}
        \mathbf{E_0}&=\mathbf{Q_0},\mathbf{E_1}=\mathbf{CQ_0},\mathbf{E_2}=\mathbf{C}^2\mathbf{Q_0},
    \end{split}
\end{equation}
\begin{equation}
\label{eq:C_D_inv_expression}
\begin{split}
     \mathbf{C}^{-1}_D&=-\frac{1}{a_0}\left(\mathbf{C}^2+a_2\mathbf{C}+a_1\mathbf{I}_4\right)\mathbf{Q}_0\\
     &=-\frac{1}{a_0}\left(\mathbf{E_2}+a_2\mathbf{E_1}+a_1\mathbf{E_0}\right) ,
\end{split}
\end{equation}
\begin{equation}
\begin{split}
     \label{eq:J_matrix_main_text}
   \mathbf{J}&=[J_{mn}]_{3\times3}\\ &= \left[\begin{matrix}\frac{a_{0} a_{1} - a_{0} a_{2}^{2} - a_{1}^{2} a_{2}}{2 a_{0} \left(a_{0} - a_{1} a_{2}\right)} & - \frac{a_{1} a_{2}^{2}}{2 a_{0} \left(a_{0} - a_{1} a_{2}\right)} & \frac{1}{2 a_{0}}\\- \frac{a_{1} a_{2}^{2}}{2 a_{0} \left(a_{0} - a_{1} a_{2}\right)} & \frac{- a_{0} - a_{2}^{3}}{2 a_{0} \left(a_{0} - a_{1} a_{2}\right)} & - \frac{a_{2}^{2}}{2 a_{0} \left(a_{0} - a_{1} a_{2}\right)}\\\frac{1}{2 a_{0}} & - \frac{a_{2}^{2}}{2 a_{0} \left(a_{0} - a_{1} a_{2}\right)} & - \frac{a_{2}}{2 a_{0} \left(a_{0} - a_{1} a_{2}\right)}\end{matrix}\right],
\end{split}
\end{equation}

\begin{equation}
\label{eq:coefficents_A_text}
    \begin{split}
    a_0&=\left(4\pi^2\gamma_{\phi}+\frac{\gamma_{z}}{R^2}\right)\\ &=\bar{\gamma}4\pi^2\ g(R)+\delta \gamma \left(4\pi^2-\frac{1}{R^2}\right),\\
    a_1&=\left(\gamma_{\phi}\gamma_{z}+4\pi^2 +\frac{1}{R^2}\right)\\ &=\bar{\gamma}^2-\delta\gamma^2 + 4\pi^2\ g(R),\\
    a_2&=\left(\gamma_{\phi}+\gamma_{z}\right)\\&=2\bar{\gamma},\\
    \end{split}
\end{equation}
and
\begin{equation}
\bar{\gamma}=\frac{\gamma_\phi+\gamma_z}{2}, \ \ \delta\gamma=\frac{\gamma_\phi-\gamma_z}{2}.
\end{equation}
In these equations, $\mathbf{\Pi_0}$ is the projector that projects on the zero eigen-mode diffusion along the screw direction, while $\mathbf{Q}_0=\mathbf{I}_4-\mathbf{\Pi}_0$ projects on the complementary stable subspaces\cite{Ott_2002} of the state vector $\vec{X}$, and the evolution matrix $\mathbf{C}$. This is shown in Appendix~\ref{appendix:matrix_solution_sde}, where we further show that the evolution in the stable subspace according to $\mathbf{C}\mathbf{Q}_0$ contains the three decaying dynamical modes. Their decaying nature follows from the negative real parts of the corresponding eigenvalues. The action of the matrix projectors $\mathbf{\Pi}_0$ and $\mathbf{Q}_0$ on a column vector representing the system state, $\vec{X}(t)=\mathbf{\Pi}_0\vec{X}(t)+\mathbf{Q}_0\vec{X}(t)$ splits this state into a component along the zero-eigenmode $\mathbf{\Pi}_0\vec{X}(t)$, and a component in the stable subspace $\mathbf{Q}_0\vec{X}(t)$. For row vectors the same splitting can be represented as $\vec{X}^T(t)=\vec{X}^T(t)\mathbf{\Pi}_0^T+\vec{X}^T(t)\mathbf{Q}_0^T$. Thus in Eq.~\eqref{eq:asymptotic_covariance_matrix_text} $\mathbf{S}_{\text{init}}$ is the initial variance along the zero-eigenmode (the extent of the initial distribution along the screw coordinate), which remains unaffected by the dynamics. The second term $\mathbf{D_0}t$, which grows linearly in time $t$, represents the diffusive growth of the variance along the zero mode . The third term $\mathbf{S_\infty}$ is our primary quantity of interest, the bounded part of the covariance matrix which reaches a stationary value and contains information about correlations of modes with non-zero eigenvalues. Using the stable subspace projection matrix $\mathbf{Q}_0$ (and $\mathbf{Q}_0^T$) we resolve the dynamics generated by $\mathbf{C}$ in the stable subspace by constructing a matrix basis $\{\mathbf{E_0,E_1,E_2}\}$ with $\mathbf{E_n}=\mathbf{C}^n\mathbf{Q_0}$ that act left to right, and $\{\mathbf{E}_0^T,\mathbf{E}_1^T,\mathbf{E}_2^T\}$ that operate from right to left on a vector or a matrix.
 The matrix $\mathbf{C}^{-1}_D$ is the Drazin pseudo-inverse \cite{Mandal2016AnalysisStates,Drazin01081958,landi2024current-6f7} of the drift matrix $\mathbf{C}$, i.e., the inverse when zero mode has been removed and satisfies $\mathbf{C}^{-1}_D\mathbf{C}=\mathbf{C}\mathbf{C}^{-1}_D=\mathbf{Q}_0$ (see Appendix.~\ref{app:S_infty_calc)details}).
 Since, $\mathbf{Q}_0$ can be written as a polynomial in $\mathbf{C}$ (Eq.~\eqref{eq:Q_0_mat_maintext_defn}), we have the commutation relation 
\begin{equation}
    \label{eq:commutation_relation Q_0_C}
    [\mathbf{C},\mathbf{Q}_0]=\mathbf{0}.
\end{equation}
and thus, the dynamics in the stable subspace remains closed (i.e. the time evolution proceeds independently in the zero mode and the stable mode subspaces), and we can write down Markovian equations of motion inside the stable subspace, which we analyze in Appendix~\eqref{app:Markovian_dynamics_stable_subspace}. Note that we recover the isotropic case from the anisotropic in the limit $\bar{\gamma}\to \gamma$ and $\delta\gamma \to 0$; which results in $a_0=4\pi^2\gamma\ g(R), a_1=\gamma^2+4\pi^2\ g(R),a_2=2\gamma$ (see Appendix~\ref{app:anisotropic_to_isotropic_reduction}).
\subsubsection{Time correlation functions in the stable subspace}
\label{Sec:Time_corr_functions}
To quantify the time correlations in the fluctuation of the system state at time $t$: $\delta\vec{X}(t)=\vec{X}(t)-\langle\vec{X}(t)\rangle$ and a future time $t+\tau$, we introduce the time correlation matrix \footnote{ Since $\langle\delta\vec{X}(t+\tau)\delta\vec{X}^T(t)\rangle
        = \langle e^{\mathbf{C}\tau}\delta\vec{X}(t)\delta\vec{X}^T(t)\rangle +\left\langle\left(\int_t^{t+\tau} e^{\mathbf{C}(t+\tau-s)}\Sigma\mathrm{d}\vec{W}(s)\right)\delta\vec{X}^T(t)\right\rangle $. The second term in the r.h.s. containing the integral vanishes due to causality since $\delta\vec{X}^T(t)$ is uncorrelated with the evolution happening due to the future noise $\{\mathrm{d}W(s)|s>t\}$.This leads to $\mathbf{S}(t+\tau|t)=e^{\mathbf{C}\tau}\mathbf{S}(t)$}
\begin{equation}
    \begin{split}
        \mathbf{S}(t+\tau|t)&=\langle\delta\vec{X}(t+\tau)\delta\vec{X}^T(t)\rangle\\
        &=e^{\mathbf{C}\tau}\mathbf{S}(t),
    \end{split}
\end{equation}
where, $\mathbf{S}(t)$ is the phase space covariance matrix defined in Eq.~\eqref{eq:defn_covariance_matrix}.  
In the long time limit, $t\to \infty$, using Eq.~\eqref{eq:asymptotic_covariance_matrix_text} and the property $e^{\mathbf{C}\tau}=\mathbf{\Pi}_0+e^{\mathbf{C}\tau}\mathbf{Q_0}$, we can write the time correlation matrix as
\begin{equation}
\label{eq:asymptotic_time_correlation_matrix}
     \mathbf{S}(t+\tau|t)\sim (\mathbf{S}_{\mathrm{init}} +\mathbf{D}_0t) +\mathbf{\Pi_0}\mathbf{S}_\infty+e^{C\tau}\mathbf{Q}_0\mathbf{S_\infty}
\end{equation}
Here, the matrices $\mathbf{S}_\mathrm{init}$ and $\mathbf{D}_0$ are given by Eq.~\eqref{eq:zero_mode_matrices}, $\mathbf{\Pi}_0$ by Eq.~\eqref{eq:Pi_0_def}, $\mathbf{S}_\infty$ by Eq.~\eqref{eq:S_infty_expression_main_text}. Note that the variance along the diffusive mode does not develop any correlation to the future $t+\tau$ as it remains independent of $\tau$, whereas the bounded part of the covariance matrix $\mathbf{S_\infty}$ develops time correlation through the $e^{\mathbf{C}\tau}\mathbf{Q}_0$ operator. This stable subspace time evolution $e^{\mathbf{C}\tau}\mathbf{Q}_0$ can be evaluated in spirit of the Krylov basis expansion\cite{saad1992analysis-22b,nandy2025quantum-a2d, rabinovici2025krylovcomplexity} (see Appendix~\ref{appendix:matrix_solution_sde})
\begin{equation}
\label{Eq:stable_space_)dynamics_generator_eqn}e^{\mathbf{C}\tau}\mathbf{Q_0}=\left(b_0(\tau)\mathbf{E_0}+b_1(\tau)\mathbf{E_1}+b_2(\tau)\mathbf{E_2}\right),
\end{equation} 
where the matrices$\{\mathbf{E}_n=\mathbf{C}^n\mathbf{Q_0}|n=0,1,2\}$, capture the hierarchy of dynamical correlation generation in stable space time evolution operator in terms of the repeated action of the drift matrix $\mathbf{C}$ on the stable subspace projector, and their corresponding, time dependent coefficients\footnote{not to be confused with Lanczos coefficients. Here $b_n(\tau)$ captures the time dependence in the $e^{\mathbf{C}\tau}\mathbf{Q}_0$generator arising in the order of matrix $\mathbf{C}^n\mathbf{Q}_0$.} $\{b_n(\tau)|n=0,1,2\}$ can be calculated as 
as the matrix exponential
\begin{equation}
    \label{eq:b_vector_matrix_eq_main_text}
\begin{split}
\vec{b}(\tau)&=e^{A \tau}\vec{b}(0),\\
    \vec{b}(\tau)&= \begin{pmatrix}
        b_0(\tau)\\
        b_1(\tau)\\
        b_2(\tau)
    \end{pmatrix}, \ \vec{b}(0)=\begin{pmatrix}
        1\\
        0\\
        0
    \end{pmatrix},\\ \ \mathbf{A}&=\begin{pmatrix}
        0 &0& -a_0\\
        1& 0 &-a_1\\
        0 & 1 & -a_2
    \end{pmatrix},
\end{split}
\end{equation}
where the model parameters $(a_0,a_1,a_2)$ are given in Eq.~\eqref{eq:coefficents_A_text}.
Alternatively, using Laplace-Fourier transform we write
\begin{equation}
\begin{split}
    \tilde{b}_n(\omega)&=\mathcal{F}\left[\theta(\tau)b_n(\tau)\right]\\
    &=\int_0^\infty \mathrm{d}\tau \ e^{-i\omega\tau}b_n(\tau),  
\end{split}
\end{equation}
with  $n=0,1,2$ and $\mathcal{F}$ standing for Fourier transform, $\theta(\tau)$ standing for Heaviside step function. Thus in frequency domain we get,
\begin{equation}
    \label{eq:freq_domain_b_coeffs}
\begin{split}
     \begin{pmatrix}
        \tilde{b}_0(\omega)\\
        \tilde{b}_1(\omega)\\
        \tilde{b}_2(\omega)
    \end{pmatrix}&=-\frac{1}{p(i\omega)}\begin{pmatrix}
            \omega^2 -a_1-ia_2\omega \\
            -a_2-i\omega\\
            -1
        \end{pmatrix}, \\
        p(i\omega)&=-i\omega^3-a_2\omega^2+ia_1\omega+a_0\\
        &=(a_0-a_2\omega^2)+i\omega(a_1-\omega^2)
\end{split}
\end{equation}
where $p(\lambda)$ is the characteristic polynomial in the stable subspace of $\mathbf{C}$.
\par
Next, consider time correlation function of fluctuations between two observables $\vec{A}_Q(t)$ and $\vec{B}_Q(t)$, the observables are taken to be linear functions of the state vector $\vec{X}(t)$ after projection to the stable subspace: $\vec{A}_Q(t)=\mathbf{M_a}\vec{Y}_Q=\mathbf{M_a}\mathbf{Q_0}\vec{X}(t),$ and $\vec{B}_Q(t)=\mathbf{M_b}\vec{Y}_Q=\mathbf{M_b}\mathbf{Q_0}\vec{X}(t)$  where $\mathbf{M}_a$ and $\mathbf{M}_b$ are matrices representing the linear map from the stable subspace component $\vec{Y}_Q(t)=\mathbf{Q}_0\vec{X}(t)$ of the vector $X(t)$ to our observable of interest $\vec{A}_Q(t)$ and $\vec{B}_Q(t)$ respectively. The correlation functions of interest are given by the elements of the matrix $\langle\delta \vec{A}_Q(t+\tau)\delta \vec{B}_Q^T(t) \rangle$, where  
\begin{equation}
    \label{eq:fluctuations_defn}
    \begin{split}
        \delta \vec{A}_Q(t)&=\vec{A}_Q(t) -\langle \vec{A}_Q(t)\rangle \\
        &=\mathbf{M_a}\mathbf{Q_0}\delta\vec{X}(t),\\
        \delta \vec{B}_Q(t)&=\vec{B}_Q(t) -\langle \vec{B}_Q(t)\rangle \\
        &=\mathbf{M_b}\mathbf{Q_0}\delta\vec{X}(t).
    \end{split}
\end{equation}
Thus in the long time limit $(t \to \infty)$ we have
\begin{equation}
\label{eq:time_corr_matrix_formula}
\begin{split}
    \mathbf{S}^Q_{AB}(\tau)&=\lim_{t \to \infty}\langle\delta \vec{A}_Q(t+\tau)\delta \vec{B}_Q^T(t) \rangle \\ &=\lim_{t \to \infty}\mathbf{M}_a\mathbf{Q_0}\mathbf{S}(t+\tau|t)\mathbf{Q_0}^T\mathbf{M_b}^T\\
    &=\mathbf{M_a}\mathbf{Q_0}e^{\mathbf{C}\tau}\mathbf{Q_0}\mathbf{S_\infty} \mathbf{Q_0}^T\mathbf{M_b}^T    \\
    &= \mathbf{M_a}e^{\mathbf{C}\tau}\mathbf{S^\infty_Q} \mathbf{M_b}^T,
\end{split}  
\end{equation} 
where
\begin{equation}
\begin{split}
\mathbf{S^\infty_Q}&=\mathbf{Q_0}\mathbf{S_\infty} \mathbf{Q_0}^T.
\end{split}
\end{equation}
Alternatively in frequency domain we can express the stationary time correlation as 
\begin{equation}
    \begin{split}
    \mathbf{\tilde{S}}^Q_{AB}(\omega)&=\mathcal{F}\left[\theta(\tau)\mathbf{S}^Q_{AB}(\tau)\right]\\
    &=-\mathbf{M_a}(\mathbf{C}-i\omega\mathbf{I}_4)^{-1}_D\mathbf{S}_\infty\mathbf{Q}^T_0\mathbf{M_b}^T
    \end{split}
\end{equation}
where we have defined the resolvent in the stable subspace as 
\begin{equation}
    \label{eq_drazin_resolvant}
    \begin{split}
        (\mathbf{C}-i\omega \mathbf{I}_4)^{-1}_{D}&= -\int_0^\infty \mathrm{d} \tau \ e^{(\mathbf{C}-i\omega \mathbf{I}_4)\tau} \mathbf{Q_0}\\
        &=-(\tilde{b}_0(\omega)\mathbf{E}_0+\tilde{b}_1(\omega)\mathbf{E}_1+\tilde{b}_2(\omega)\mathbf{E}_2)
    \end{split}
\end{equation}
where the matrices$\{\mathbf{E}_i|i=0,1,2\}$ are defined in Eq.~\eqref{eq:Krylov_basis} and $\{\tilde{b}_j|j=0,1,2\}$ are given in Eq.~\eqref{eq:freq_domain_b_coeffs}.
Notice that in $\omega \to 0$ the expression of $(\mathbf{C}-i\omega \mathbf{I}_4)^{-1}_D$ reduces to $\mathbf{C}_D^{-1}$ which can be also evaluated using Eq.~\eqref{eq:C_D_inv_expression}.
Below we illustrate this by providing a matrix expression for the momentum time correlation functions. Notice that the momentum subspace completely exists inside the stable subspace of the dynamics hence action of $\mathbf{Q}_0$ on $\vec{X}(t)$ leaves the momentum sector unchanged. Thus, we can calculate the momentum vector $\vec{P}=(L_\phi \ \ p_z)^T$ directly from $\vec{Y}_Q$ as $\vec{P}=\mathbf{B}^T\vec{Y}_Q=\mathbf{B}^T\vec{X}$, where $\mathbf{B}^T$ is a $2\times4$ matrix that picks up the momentum sector part of the state $\vec{X}(t)$ and can be written as 
\begin{equation}
    \mathbf{B}^T=\begin{pmatrix}
        0 & 1&0&0\\
        0&0&0&1
    \end{pmatrix}
\end{equation}
which satisfies $\mathbf{B}^T\mathbf{Q_0}=\mathbf{B}^T$.
Hence we can write the momentum time correlation function as 
\begin{equation}
    \begin{split}
         \mathbf{S_p^\infty}(\tau)&=\lim_{t\to\infty}\langle\vec{P}(t+\tau)\vec{P}^T(t)\rangle\\&=\mathbf{B}^Te^{\mathbf{C}\tau}\mathbf{Q_0}\mathbf{S}_\infty
 \mathbf{Q}_0^T\mathbf{B}\\
 &=\mathbf{B}^T(e^{\mathbf{C}\tau}\mathbf{Q_0}\mathbf{S}_\infty
 )\mathbf{B},
    \end{split}    
 \end{equation}
where $e^{\mathbf{C}\tau}\mathbf{Q_0}$ is given by Eq.~\eqref{Eq:stable_space_)dynamics_generator_eqn}. Then  $\mathbf{S_p^\infty}(\tau)$ can be expressed 
\begin{equation}
\label{eqn:S_p_krylov_split_main_text}
\begin{split}
\mathbf{S_p^\infty}(\tau)&=  \frac{b_0(\tau)}{\beta}\begin{pmatrix}
        R^2 &0\\
        0 &1
        \end{pmatrix}\\
        &-\frac{b_1(\tau)}{\beta}\begin{pmatrix}
        R^2\gamma_\phi &0\\
        0 &\gamma_z
        \end{pmatrix}\\ &+\frac{b_2(\tau)}{\beta}\begin{pmatrix}
        R^2\gamma_\phi^2 -1  &2\pi\chi\\
        2\pi\chi &\gamma_z^2-4\pi^2
        \end{pmatrix}.   
\end{split}
\end{equation}    
which allows us to resolve the time correlation of momenta in different orders of application of the drift matrix in the stable subspace.

\begin{figure}
    \centering
    \includegraphics[width=1\linewidth]{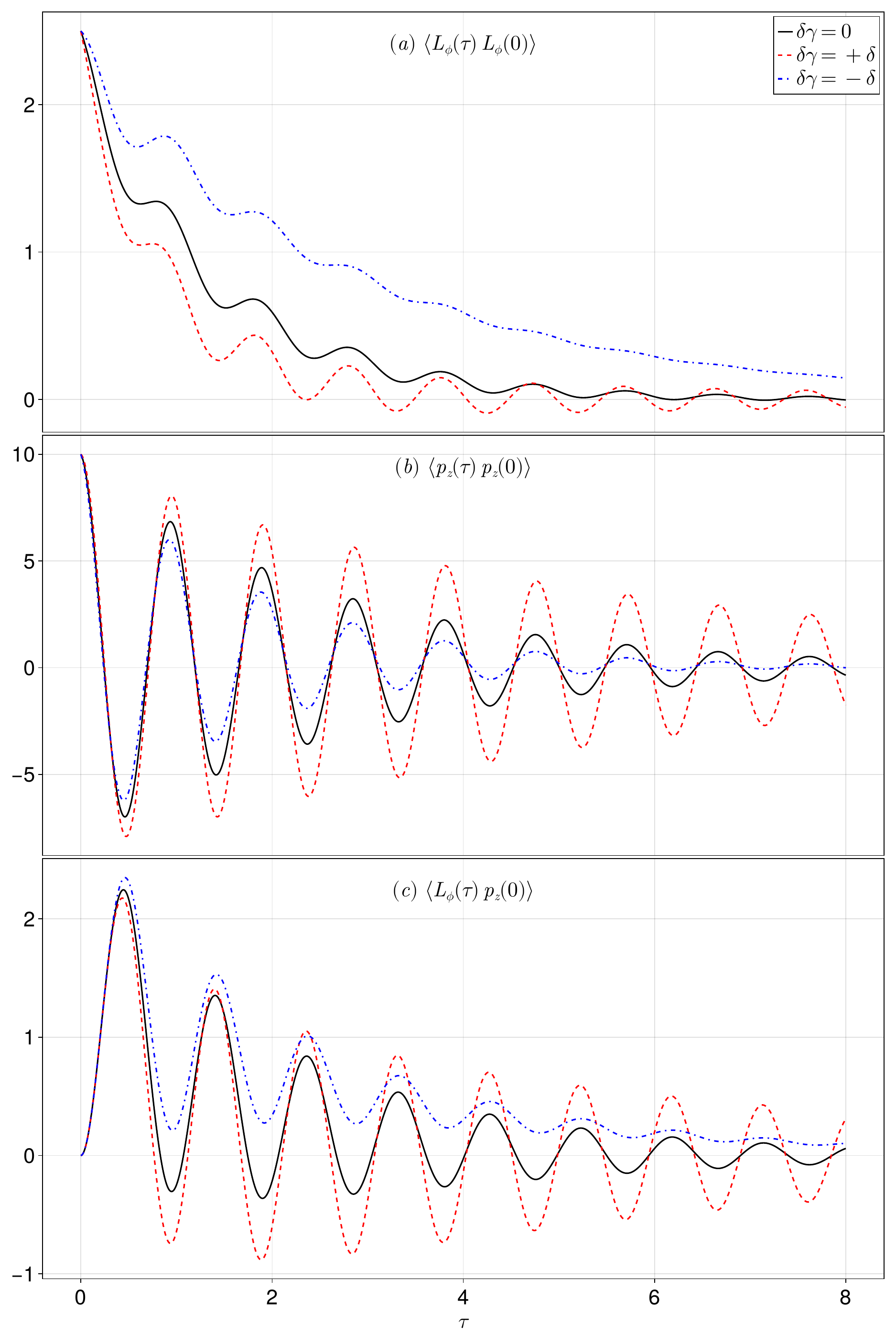}
     \caption{Effect of damping anisotropy in momentum time-correlation functions. Elements of the matrix $\mathbf{S^\infty_p}(\tau)$ are plotted against dimensionless time $\tau$ using Eq.~\eqref{eqn:S_p_krylov_split_main_text}. To illustrate the effect of anisotropy we keep $\bar{\gamma}=(\gamma_\phi+\gamma_z)/2$ same as that of Fig.~\ref{fig:isotropic_time_correlations}, then consider three different values of anisotropy coefficient $\delta\gamma=(\gamma_\phi-\gamma_z)/2$; which correspond to the isotropic case($\delta\gamma=0$, solid black lines), angular damping tilted ($\delta\gamma=+\delta$, red dashed lines), axial damping tilted ($\delta\gamma=-\delta$, blue dash-dot line). Panel(a): $\langle L_\phi(\tau)L_\phi(0)\rangle$ against $\tau$, panel (b): $\langle p_z(\tau)p_z(0)\rangle$ against $\tau$ and panel (c):$\langle L_\phi(\tau)p(0)\rangle$ against $\tau$. Parameters:$\beta=0.1,R=0.5$, $\chi=1$ with $\bar{\gamma}=1.0$, $\delta=0.5$}
     \label{fig:damping_anisotropy_inmom_time_corr}
\end{figure}
The corresponding frequency-domain correlations, $\mathbf{\tilde{S}_p^\infty}(\omega)=\mathcal{F}[\mathbf{S_p^\infty}(\tau)\theta(\tau)]$ can be similarly expressed in terms of $\{\tilde{b}_j(\omega)|j=0,1,2\}$ using Eq.~\eqref{eq:freq_domain_b_coeffs}.
\par 
In Fig.~\ref{fig:damping_anisotropy_inmom_time_corr} we plot the momentum time-correlation functions illustrating the effect of anisotropy in damping coefficients. Several observations can be made:
(a) From Eq.~\eqref{eq:b_vector_matrix_eq_main_text} we see that, at $\tau=0$ we have $b_0(0)=1,b_1(0)=0, b_2(0)=0$. Thus we recover the equipartition result
\begin{equation}
\begin{split}
    \mathbf{S_p^\infty}(0)
 &=\frac{1}{\beta}\begin{pmatrix}
     R^2 &0\\
     0&1
 \end{pmatrix},
\end{split}
\end{equation}
This result implies, as seen in Fig.~\ref{fig:damping_anisotropy_inmom_time_corr} and as already seen in the isotropic case, that there is no equal-time correlation between $p_z$ and $L_\phi$. Any such correlation forms only at $\tau\ne0$, in course of the dynamic evolution.
(b) In Eq.~\eqref{eqn:S_p_krylov_split_main_text}, we see that the first term and the second term in the r.h.s., representing the zeroth and first order contribution respectively, are diagonal for any $\tau \ge 0$, implying no contribution to the cross-correlation between linear and angular momentum. Only the third term in the r.h.s. of Eq.~\eqref{eqn:S_p_krylov_split_main_text} has off diagonal elements which depend on the chirality factor $\chi$, which creates dynamic correlation between linear and angular momentum. The appearance of off-diagonal terms in this term implies that the correlation between linear and angular momentum can only arise due to a second order effect as there is no direct kinetic coupling (e.g. $L_\phi p_z$ type term) present in the Hamiltonian. Correlation $L_\phi$ and $p_z$ arise due to the helical potential $U(\phi,z)$, i.e., through the position variables $\phi$ and $z$.  Finally, in Appendix~\ref{app:anisotropic_to_isotropic_reduction}, we show that the general results for the time correlation functions, Eqs.\eqref{eq:time_corr_matrix_formula}~\eqref{eqn:S_p_krylov_split_main_text} reduce to the results obtained in the isotropic case, Eq.~\eqref{eq:main_text_isotrpic_time_corr_functions}, when $\delta\gamma=0$.
\section{Mobilities and energy dissipation}
\label{Sec:Mobilites}
\subsection{Mobilities}
\label{subsec:mobility_subsection}
\begin{figure*}
    \centering
    \includegraphics[width=1.0\linewidth]{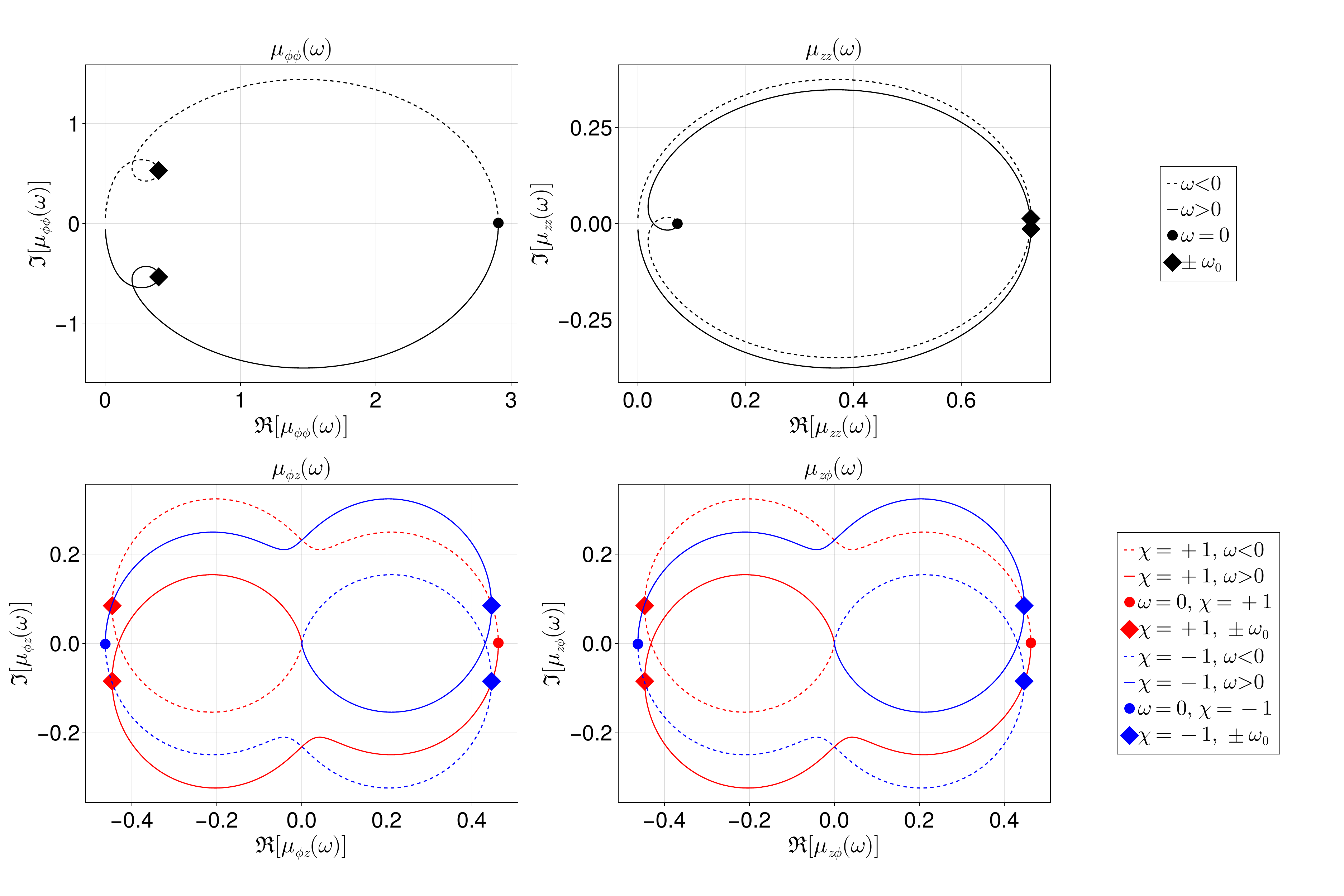}
    \caption{Nyquist plots for visualizing AC mobilities (Eq.~\eqref{eq:AC_mobility_expressions}). Here we plot $(\mathfrak{R}[\mu_{ij}(\omega)],\mathfrak{I}[\mu_{ij}(\omega)])$ parametrically in the drive frequency $\omega$. In these diagrams, at any point parametrized by drive frequency $\omega$, the polar angle with respect to the positive real axis characterizes  the phase response $\Phi_{ij}(\omega)$ and distance from the origin captures the amplitude response $|\mu_{ij}(\omega)|$. $|\omega|$ is varied from $0$ to $10\omega_0$ for all branches of the plots. The $\omega=0$ (solid circular points) in the plots represents the  represent the value of the DC mobility and sits on the real axis. The solid square points represent the values of $\pm \omega_0$.  The dotted lines ($\omega<0$ branch) and solid lines ($\omega>0$ branch) are mirror images of each other when reflected through the real axis. In the cross mobility plots, we also illustrate the effect of changing the handedness of the helix where the red lines correspond to $\chi=1$ and the blue line $\chi=-1$. Parameters for the plots are $\beta=0.1,R=0.5,\gamma_\phi=\gamma_z=1.25$}
    \label{fig:nyquist_1}
\end{figure*}
The natural transport quantity of interest for a driven Brownian particle in the helical system is the dynamical mobility tensor
\begin{equation}
    \boldsymbol{\mu}(\omega)=\begin{pmatrix}
        \mu_{\phi\phi}(\omega)& \mu_{\phi z}(\omega)\\
        \mu_{z\phi}(\omega) & \mu_{zz}(\omega)
    \end{pmatrix}
\end{equation} which connects a generalized force vector $\vec{f}(t)$ in the system containing torque $f_\phi(t)$ and axial force $f_z(t)$ to steady state average velocities $\langle\dot{\phi}\rangle$ and $\langle\dot{z}\rangle$ respectively. Here we first consider a monochromatic force with frequency $\omega$ of the form
\begin{equation}
    \vec{f}(t)=\begin{pmatrix}
        f_\phi(t)\\
        f_z(t)
    \end{pmatrix}=\begin{pmatrix}
        F_\phi(\omega)\\
        F_z(\omega)
    \end{pmatrix}e^{i\omega t}
\end{equation} where $F_\phi$ and $F_z$ represent complex amplitude of the input drive. The average velocity response at the steady state is given as
\begin{equation}
    \begin{pmatrix}
        \langle\dot{\phi}(t)\rangle\\
        \langle \dot{z}(t)\rangle
    \end{pmatrix}
    =\begin{pmatrix}
        \mu_{\phi\phi}(\omega)& \mu_{\phi z}(\omega)\\
        \mu_{z\phi}(\omega) & \mu_{zz}(\omega)
    \end{pmatrix}
    \begin{pmatrix}
        f_\phi(t)\\
        f_z(t)
    \end{pmatrix}
\end{equation}
Here the diagonal terms $\mu_{\phi \phi}(\omega)$ and $\mu_{z z}(\omega)$  represent the response in the direction of the applied generalized force and off-diagonal terms $\mu_{\phi z}(\omega)$ and $\mu_{z \phi}(\omega)$, representing cross responses that arise from motion on the helical landscape. If we consider the output average velocities to be of the form $\langle\dot{\phi}\rangle = v_\phi(\omega)  e^{i\omega t}$ and $\langle\dot{z}\rangle = v_z(\omega)  e^{i\omega t}$, we get the relation
\begin{equation}
     \begin{pmatrix}
        v_\phi(\omega)\\
        v_z(\omega)
    \end{pmatrix}
    =\begin{pmatrix}
        \mu_{\phi\phi}(\omega)& \mu_{\phi z}(\omega)\\
        \mu_{z\phi}(\omega) & \mu_{zz}(\omega)
    \end{pmatrix}
    \begin{pmatrix}
        F_\phi(\omega)\\
        F_z(\omega)
    \end{pmatrix}
\end{equation}
In Appendix~\ref{App:generic_linear_response}, using linear response theory in the stable subspace, we show that the mobility tensor is related to the momentum time correlation matrix via the fluctuation dissipation relation as
\begin{equation}
\label{eq:FDT_main_text}
\begin{split}
      \boldsymbol{\mu}(\omega)&= \beta\mathbf{V}\mathbf{\tilde{S}^\infty_p}(\omega)\mathbf{V}^T, \\ \mathbf{\tilde{S}^\infty_p}(\omega)&=\mathcal{F}\left[\theta(\tau)\mathbf{S_p^\infty}(\tau)\right],\\
      \mathbf{V}&=\begin{pmatrix}
          \frac{1}{R^2} & 0\\
          0 &1
      \end{pmatrix}
\end{split}
\end{equation}
where $\mathbf{S_p^\infty}(\tau)$ is given in Eq.~\eqref{eqn:S_p_krylov_split_main_text}.
After explicit evaluation, the AC (frequency dependent) mobilities are obtained in the forms 
\begin{subequations}
\label{eq:AC_mobility_expressions}
    \begin{align}
        \mu_{\phi\phi}(\omega)&= \frac{
4\pi^2-\omega^2+i\omega\gamma_z }{\zeta(\omega;R,\gamma_z,\gamma_\phi)}\\
 \mu_{zz}(\omega)&= \frac{
1-R^2\omega^2+i\omega R^2\gamma_\phi }{\zeta(\omega;R,\gamma_z,\gamma_\phi)}\\
\mu_{\phi z}(\omega)&=\mu_{z \phi}(\omega)=\frac{2\pi\chi}{\zeta(\omega;R,\gamma_z,\gamma_\phi)}
    \end{align}
\end{subequations}
where the denominator $\zeta(\omega\ ;  R,\gamma_z,\gamma_\phi)\equiv\zeta(\omega)$ is same for all the mobility components and is given by
\begin{equation}
\label{eq:zeta_omega_Expr}
\begin{split}
\zeta(\omega)
&=
\zeta_\mathrm{Re}(\omega)+i\zeta_\mathrm{Im}(\omega).   \\
\zeta_\mathrm{Re}(\omega)&=R^2\left[\gamma_{\mathrm{eff}}(R)\omega_0^2(R)
-(\gamma_\phi+\gamma_z)\omega^2\right]\\
\zeta_\mathrm{Im}(\omega)
&=\omega R^2\left[\left\{
\gamma_\phi \gamma_z+\omega^2_0(R)\right\}-\omega^2
\right].
\end{split}
\end{equation}
 with $\gamma_\mathrm{eff}(R)$ being the effective friction (Eq~\eqref{eq:gamma_effective}) along screw (S) direction and $\omega_0(R)=\sqrt{4\pi^2\ g(R)}$ being the effective oscillation frequency ($g(R)$ given in Eq.~\eqref{eq:geometric_factor}) in the screw-normal (SN) direction. 
 Note that $\mu_{\phi z}=\mu_{z \phi}$ in accordance with the Onsager symmetry relation.
 These results can be visualized using Nyquist diagrams\cite{Bechhoefer_2021}. An example is shown in Fig.~\ref{fig:nyquist_1}. 

\begin{figure}
    \centering
    \includegraphics[width=1\linewidth]{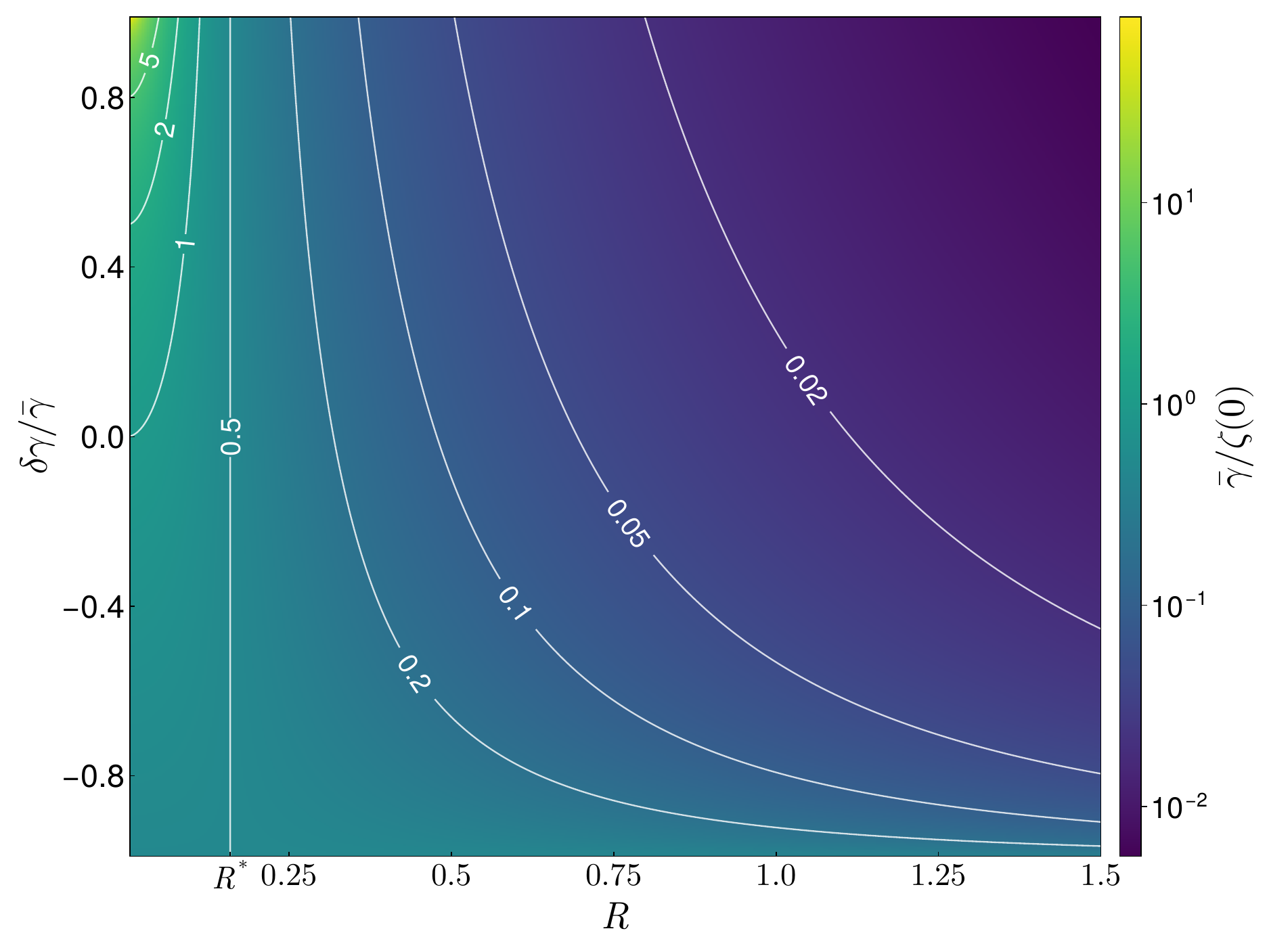}
    \caption{DC mobilities. The scaled mobility prefactor $\bar{\gamma}/\zeta(0)$ is plotted in terms of $R$ and $\delta\gamma/\bar{\gamma}$ , where $\bar{\gamma}=(\gamma_\phi+\gamma_z)/2$, $\delta\gamma=(\gamma_\phi-\gamma_z)/2$. The DC mobilities are given as $\mu^{DC}_{\phi\phi}=4\pi^2/\zeta(0),\mu^{DC}_{zz}=1/\zeta(0)$ and $\mu^{DC}_{z\phi}=\mu^{DC}_{\phi z}=2\pi\chi/\zeta(0)$. The vertical white line corresponds to $R=R^*=1/2\pi$ where dependence on anisotropy vanishes and $\bar{\gamma}/\zeta(0)=1/2$.}
    \label{fig:DC_mobilities}
\end{figure}

 In the DC ($\omega \to 0)$ limit $\zeta_{\mathrm{Im}}(0)$ vanishes, and we get $\zeta(0)=\zeta_{\mathrm{Re}}(0)=R^2 \gamma_{\mathrm{eff}}(R)\omega^2_0(R)=(4\pi^2R^2\gamma_\phi +\gamma_z)$. The DC mobility tensor then becomes
\begin{equation}
\label{DC_mobility_matrix}
\begin{split}
\boldsymbol{\mu}_{DC}\equiv\boldsymbol{\mu}(0)&=\frac{1}{(4\pi^2R^2\gamma_\phi +\gamma_z)}\begin{pmatrix}
    4\pi^2 & 2\pi\chi\\
    2\pi\chi & 1    
    \end{pmatrix}\\
\end{split}
\end{equation}
In the Fig.~\ref{fig:DC_mobilities} we illustrate the parametric variation of the DC mobilities by plotting $\bar{\gamma}/\zeta(0)$, scaled denominator of the mobilities in terms of scaled anisotropy ($\delta\gamma/\bar{\gamma}$) and the radius to pitch ratio of the helix $(R)$, where $\bar{\gamma}=(\gamma_\phi+\gamma_z)/2,\delta\gamma=(\gamma_\phi-\gamma_z)/2$. 
\begin{figure}
    \centering
    \includegraphics[width=1\linewidth]{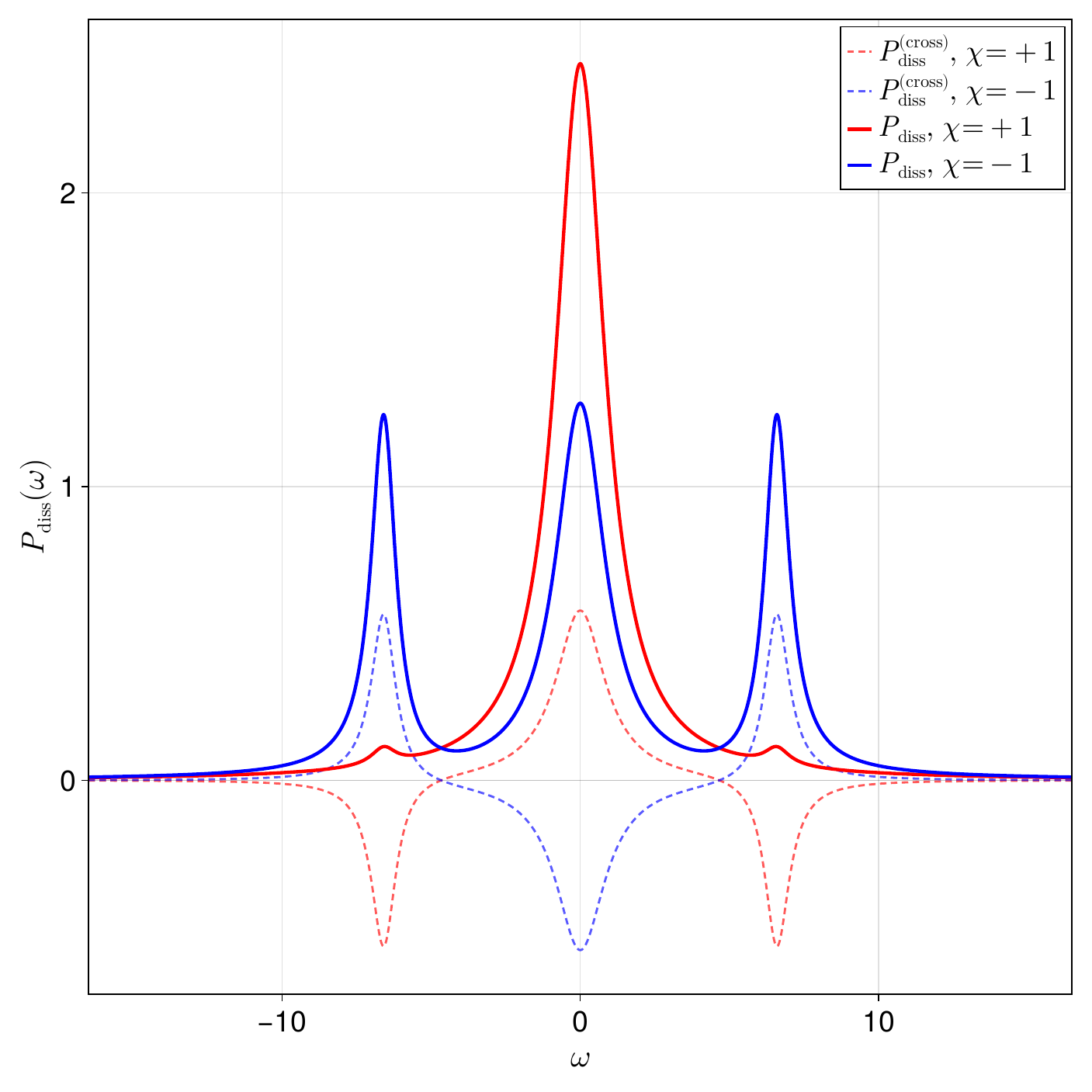}
    \caption{Effect of chirality on steady state energy dissipation rate. The energy dissipation rate $P_\mathrm{diss}(\omega)$ (Eq.~\eqref{eq:energy_diss_rate_defn}) is plotted against the drive frequency $\omega$ is shown for $\chi=+1$ (solid red line) and $\chi=-1$ (solid blue line) for a drive with $F_\phi=F_z=1$. the cross term  $P_\mathrm{diss}^{\mathrm{cross}}(\omega)$ Eq.~\eqref{eq:P_cross_zeta}) is also shown for $\chi=+1$ (dashed red line) and $\chi=-1$ (dashed blue line) . The cross term $P^{\mathrm{cross}}_\mathrm{diss}(\omega)$ depends on $\chi$, leading to asymmetry in the overall energy dissipation for different handedness of the helix. The parameters of the plots are $R=0.5$ and $\gamma_\phi=\gamma_z=\gamma=1.0$, which leads to $\omega_0(R)\approx6.594$. The driving is $F_\phi=F_z=1$.}
    \label{fig:energy_dissipation}
\end{figure}
From Eq.~\eqref{DC_mobility_matrix}, we see that the DC mobility tensor can be factorized into  
\begin{equation}
\label{eq:DC_mobility_factorized}
\begin{split}
    \boldsymbol{\mu}_{DC}&=\frac{1}{(4\pi^2R^2\gamma_\phi +\gamma_z)} \ \vec{v}\vec{v}^T,\\
    \vec{v}&=\begin{pmatrix}
        2\pi\chi \\
        1
    \end{pmatrix},
\end{split}
\end{equation}
 where $\vec{v}$ represents the direction of the helical valley $(q_s)$. Such form could be expected since the average motion along the screw-normal direction becomes zero, i.e., $\langle\dot\phi\rangle/\langle\dot z\rangle=2\pi\chi $, or, $\langle\dot q_n \rangle=0$. It follows that irrespective of the direction of the applied initial force, the steady state current takes place along the screw direction. We further notice that in all the DC mobilities the denominator is $\zeta(0)=R^2 \gamma_{\mathrm{eff}}(R)\omega^2_0(R)$, where $\gamma_{\mathrm{eff}}(R)$ is given by Eq.~\eqref{eq:gamma_effective}. Interestingly, for the special value $R= R^{*}=1/(2\pi),$
 the DC mobilities are independent of the damping-anisotropy $\delta\gamma$. For a given average friction $\bar{\gamma}$ and helix handedness $\chi$, we see that for $R>R^*$, an increase (a decrease) in anisotropy $\delta\gamma$ decreases (increases) the magnitude of each of the DC mobility components and, for $R<R^*$, we get the opposite effect, where an increase ( a decrease) in the anisotropy $\delta\gamma$ increases (decreases) the magnitude of all of the DC mobility components (see Fig.~\ref{fig:DC_mobilities}). 
\subsection{Energy dissipation}
For the drive
\begin{equation}
    \vec{f}(t)=\mathfrak{R}\left[\vec{F}(\omega)e^{i\omega t}\right],\  \vec{F}(\omega)= \begin{pmatrix}
        F_\phi(\omega)\\
        F_z(\omega)
    \end{pmatrix}
\end{equation}
the steady state dissipation can be obtained from the instantaneous input power.
\begin{equation}    
    \begin{split}
P(t)&=\vec{f}(t)\cdot\langle\delta\dot{\vec{Q}}(t)\rangle=\vec{f}^T(t)\langle\delta \dot{\vec{Q}}(t) \rangle\\
&=\mathfrak{R}\left[\vec{F}(\omega)e^{i\omega t}\right]^T\mathfrak{R}\left[\boldsymbol{\mu}(\omega)\vec{F}(\omega)e^{i\omega t}\right]\\
 \end{split}
\end{equation}
Averaging over a period leads to
\begin{equation}
\label{eq:energy_diss_rate_defn}
\begin{split}
P_\mathrm{diss}(\omega)&=\frac{\omega}{2\pi}\int_0^{\frac{2\pi}{\omega}}\mathrm{d}t\ P(t)\\& 
=\frac{1}{2}\mathfrak{R}\left[\vec{F}^{\dagger}(\omega)\boldsymbol{\mu}(\omega)\vec{F}(\omega)\right]\\
   &=P^{(\phi)}_{\mathrm{diss}}(\omega)+P^{(z)}_{\mathrm{diss}}(\omega)+P^{(\mathrm{cross})}_{\mathrm{diss}}(\omega) 
\end{split}
\end{equation}
where
\begin{equation}
    \begin{split}
    P^{(\phi)}_{\mathrm{diss}}(\omega)&=\frac{1}{2}\mathfrak{R}\left[F^*_\phi(\omega) \mu_{\phi\phi}(\omega)F_\phi(\omega)\right]\\P^{(z)}_{\mathrm{diss}}(\omega)&=\frac{1}{2}\mathfrak{R}\left[F^*_z(\omega) \mu_{zz}(\omega)F_z(\omega)\right]\\P^{(\mathrm{cross})}_{\mathrm{diss}}(\omega)&=\frac{1}{2}\mathfrak{R}\left[F^*_\phi(\omega) \mu_{\phi 
        z}(\omega)F_z(\omega) +F^*_z(\omega) \mu_{ z \phi}(\omega)F_\phi(\omega)\right]
    \end{split}
\end{equation}
where AC mobilities are given by Eq.~\eqref{eq:AC_mobility_expressions}. 
Note that $P^{(\mathrm{cross})}_{\mathrm{diss}}(\omega)$ is linear in $\mu_{\phi z}=\mu_{z\phi }$, and hence is proportional to the chirality factor $\chi$. Without loss of generality we can take $\vec{f}(t)=\vec{F}(\omega)\cos(\omega t)$, with real $\vec{F}(\omega)$, which leads to $P_\mathrm{diss}(\omega)$ are given as follows
\begin{equation}
\label{eq:P_phiphi_zeta}
P^{(\phi)}_{\mathrm{diss}}(\omega)
=
\frac{F_\phi^2}{2}
\frac{
\left(4\pi^2-\omega^2\right)\zeta_{\mathrm{Re}}(\omega)
+
\omega\gamma_z\,\zeta_{\mathrm{Im}}(\omega)
}{
\zeta_{\mathrm{Re}}^2(\omega)+\zeta_{\mathrm{Im}}^2(\omega)
}.
\end{equation}
\begin{equation}
\label{eq:P_zz_zeta}
P^{(z)}_{\mathrm{diss}}(\omega)
=
\frac{F_z^2}{2}
\frac{
\left(1-R^2\omega^2\right)\zeta_{\mathrm{Re}}(\omega)
+
\omega R^2\gamma_\phi\,\zeta_{\mathrm{Im}}(\omega)
}{
\zeta_{\mathrm{Re}}^2(\omega)+\zeta_{\mathrm{Im}}^2(\omega)
}.
\end{equation}
\begin{equation}
\label{eq:P_cross_zeta}
P_{\mathrm{diss}}^{\mathrm{(cross)}}(\omega)
=
2\pi\chi F_\phi F_z
\frac{
\zeta_{\mathrm{Re}}(\omega)
}{
\zeta_{\mathrm{Re}}^2(\omega)+\zeta_{\mathrm{Im}}^2(\omega)
}.
\end{equation}
with $\zeta_{\mathrm{Re}}(\omega)$ and $\zeta_{\mathrm{Im}}(\omega)$ given by Eq.~\eqref{eq:zeta_omega_Expr}. 
The cross contribution, Eq.~\eqref{eq:P_cross_zeta} is the source of asymmetry in overall energy dissipation, changing sign with changing handedness $\chi$ or equivalently the sign change of $F_z$ or $F_\phi$. An example is provided in Fig.~\ref{fig:energy_dissipation}, which depicts the absorption spectrum of the same driving applied to the  opposite helical enantiomers.
\section{Conclusion and Outlook}
\label{Sec:Conclusion}
We have studied an analytically solvable model for a Brownian particle with inertia moving in a helical landscape confined in a two dimensional cylindrical manifold, elucidating how the helical symmetry is encoded into the particle dynamics through a model potential. The effect of the thermal environment is included using Langevin dynamics characterized by the environmental temperature T and potentially anisotropic friction. The mathematical problem is separable in the screw/screw-normal coordinates and remains separable for isotropic friction, but this separability is lost when the noises/friction in the axial and angular directions are different. Analytical solution may still be obtained by projecting the system dynamics onto the stable subspace that exclude the zero mode along the helical pathway, making it possible to evaluate the relevant correlation functions in this subspace.
 The linear response behavior characterizing this model clearly shows the consequence of linear-angular momentum correlation, mathematically expressed by the fact that the screw coordinate is the zero-mode of the ensuing stochastic dynamics. Importantly this correlation does not arise from a direct coupling term in the system Hamiltonian, but is a consequence of the system topology expressed by the helical potential. For this reason, equal time linear-angular momentum correlation is zero and correlation only arises over time. This correlation gives rise to cross-mobilities, whereupon driving in the axial (angular) directions lead to motion in the angular (axial) direction. As expected these cross mobilities satisfy the Onsager symmetry relations. Importantly, these cross response coefficients enter into the energy dissipation rate in a way that depends on the helix handedness, leading to dichroism in energy absorption when both axial and angular driving exist simultaneously, that may be attributed to interference between these two drivings. 
We have also found that anisotropy in the system interaction with its thermal environment, where coupling to the thermal environment is different for the axial and angular motions, has quantitative consequences for the relaxation and transport behaviors (in addition to rendering the mathematics considerably more challenging), however with the same qualitative features outlined above. An interesting observation is that in DC response, the dependence on the damping anisotropy disappears namely when the pitch to radius ratio is $2\pi$.

This study is based on a classical, two-dimensional model, however it is able to examine important generic consequences of dynamics in chiral systems: correlation (“locking”) of angular and linear momenta, effect of anisotropy in the interaction with the thermal environment and handedness-asymmetry in dissipation processes involving several (here two) driving fields. Our analytical results are limited to low energy (low temperature) dynamics (for which approximating the helical path as a harmonic well is justified) and should be reexamined with numerical simulations at high energies and temperatures. Numerical simulations will also be needed to study many particle systems with interparticle interactions.  Extending this classical study to the linear and non-linear response of helical quantum systems is an additional major future challenge. Another possible future direction of the work would be to explore the generation of dynamical correlation in stochastic systems due to the (helical) potential landscape in terms of complexity measures like Krylov complexity\cite{rabinovici2025krylovcomplexity}.
Finally, this work suggests that effect of chirality on the thermal transport phenomena can be experimentally observed by impedance spectroscopy, rheological measurements of mobility, measurement of absorption spectra which rely on dynamical manifestation of geometric chirality. Obviously, some or all of the extensions discussed above will be needed for addressing manifestation of chirality on transport in realistic systems.

\begin{acknowledgments}
This research is supported by the National Science Foundation Grant \#2451953. We thank Joel Gersten, Oded Hod, Yun Chen, Xuecheng Tao and Nikhil Gupt for many helpful discussions, and Joel Gersten for suggesting the 2 dimensional helical potential model.
\end{acknowledgments}
\appendix
\begin{widetext}
\section{Derivation for the Langevin equation on the cylinder}
\label{app:derivation_langevin_equations}
We consider an stochastic dynamics of a classical particle which at time $t$ is described by the state vector $(\vec{q}_t,\vec{p}_t)$ where $\vec{q}_t =(x(t) \ y(t) \ z(t))^T$ represents the coordinates, and $\vec{p}_t =(p_x(t) \ p_y(t) \ p_z(t))^T$ represents the corresponding momentum in dimensionless representation. The particle is evolving under a potential $V(\vec{q}) $ in the presence of a weakly coupled thermal bath at an inverse temperature $\beta=\frac{\kappa}{k_B T}$ with $k_B$ being the Boltzmann constant and $T$ being the temperature of the bath. Hamiltonian generates the deterministic part of the dynamics of the particle
\begin{equation}
    H(\vec{q},\vec{p})=\frac{|\vec{p}|^2}{2} +V(\vec{{q}})
\end{equation}
We can describe the phase space stochastic dynamics of the particle by the set of Langevin equations (in dimensionless form) written in Stratonovich representation\cite{chorin_hald2009stochastic,kampen1981it,Seifert_2025,sekimoto_book2010stochastic}: 
\begin{equation}
    \label{eqn:3D-sde}
    \begin{split}
        dx &= p_xdt, \\
   dp_x &= -\,\partial_x V(x,y,z)\,dt 
           \;-\;\gamma_x\,p_x\,dt 
           + \sqrt{2\,\gamma_x \,\beta^{-1}}\;\circ dW_t^{(x)},\\
   dy &= p_y\,dt, \\
   dp_y &= -\,\partial_y V(x,y,z)\,dt 
           \;-\;\gamma_y\,p_y\,dt 
           + \sqrt{2\,\gamma_y\,\beta^{-1}}\;\circ dW_t^{(y)},\\
   dz &= p_z\,dt, \\
   dp_z &= -\,\partial_z V(x,y,z)\,dt 
           \;-\;\gamma_z\,p_z\,dt 
           + \sqrt{2\,\gamma_z\,\beta^{-1}}\;\circ dW_t^{(z)}.
    \end{split}
\end{equation}
where $(\gamma_x, \gamma_y,\gamma_z)$ represent the momentum damping coefficients in $x,y$ and $z$ direction respectively, and the $\mathrm{d}W^{x}_t,\mathrm{d}W^{y}_t$ and $\mathrm{d}W^{z}_t$ represent three independent Wiener increments. For our problem we assume existence of a cylindrical symmetry in the model and impose that the damping rates are isotropic in the $xy$ plane, $\gamma_x=\gamma_y=\gamma_{(xy)}$, where $\gamma_{(xy)}$ represent the damping in the $xy$ plane. In general, $\gamma_{(xy)}$ may not be same as $\gamma_z$. We refer to the case for $\gamma_{(xy)}=\gamma_z$ as the isotropic damping model and  $\gamma_{(xy)}\ne\gamma_z$ as the anisotropic damping model.\par

We now put a constraint on the particle such that it is restricted to move on a cylindrical surface of a constant radius $R$. If we represent the $\vec{q}_t$ in Cartesian coordinates as $(x,y,z)$ and in cylindrical coordinates as $(\rho,\phi,z)$ with $x^2+y^2=\rho^2=R^2$, and $ x=\rho \cos \phi=R\cos \phi$, $y=\rho \sin \phi = R\sin \phi$. Because of this restriction the coordinate space of the accessible to the particle gets constrained to the cylinder $C_R=\{(x,y,z)| x^2+y^2=R^2\}$ where $\phi \in [0,2\pi)$ is an angle variable and $z\in (-\infty,\infty)$, hence we can represent the coordinate of the particle as a 2D vector $\vec{q}=(\phi,z)$. \par 

Since we have $x= R\cos \phi,y=R\sin \phi, z=z$, taking time derivative we get 
\begin{equation}
    \dot{x}= - R \dot{\phi} \sin \phi
\end{equation}
\begin{equation}
    \dot{y}=  R \dot{\phi} \cos \phi
\end{equation}
And we have linear momentum as $p_z= \dot{z}$ and angular momentum as $L_\phi=  R^2 \dot{\phi}=xp_y -yp_x$. Since we are working in Stratonovich we can use the standard chain rule  directly and get:
\begin{equation}
\begin{split}
        \dot{L}_\phi&= \dot{x} p_y + x \dot{p_y} - \dot{y}p_x - y\dot{p}_x \\
        &=  \dot{x}\dot{y} + x \dot{p}_y- \dot{y}\dot{x}- y\dot{p}_x \\
        &= x \dot{p}_y - y\dot{p}_x
\end{split}
\end{equation}
Thus we have 
\begin{equation}
    \begin{split}
            \dd L_\phi = &-\left(x\partial_y - y\partial_x\right)V(x,y,z) \dd t- \gamma_{(xy)}\left(xp_y-yp_x\right) \dd t\\ &+\sqrt{2\,\gamma_{(xy)}\,\beta^{-1}}\;\left(x\circ dW_t^{(y)}-y\circ dW_t^{(x)}\right)
    \end{split}
\end{equation}
Noticing that $\partial_\phi = \left(x\partial_y -y \partial_x\right)$ and, $L_\phi= (xp_y-yp_x)$ and then substituting $x=R \cos \phi, y=R \sin \phi$ with a constant R, we get
\begin{equation}
    \begin{split}
        \dd L_\phi = &-\partial_\phi V(\phi,z) \dd t- \gamma_{(xy)} L_\phi \dd t\\
        &+\sqrt{2\,\gamma_{(xy)}\beta^{-1}}\;\left(R \cos \phi\circ dW_t^{(y)}-R \sin \phi\circ dW_t^{(x)}\right)
    \end{split}
\end{equation}
Which leads to 
\begin{equation}
    \dd L_\phi = -\partial_\phi V(\phi,z) \dd t- \gamma_{(xy)} L_\phi \dd t+\sqrt{2\,\gamma_{(xy)} R^2\,\beta^{-1}}\;\circ dW_t^{(\phi)}
\end{equation}
where,
\begin{equation}
     \mathrm{d}W_t^{(\phi)}= \cos \phi \circ \mathrm{d}W_t^{(y)}-\sin \phi\circ \mathrm{d} W_t^{(x)}
\end{equation}
Clearly we see that $\langle dW_t^{(\phi)}  \rangle = 0$ since both $\langle   dW_t^{(x)}  \rangle =\langle  dW_t^{(y)}  \rangle = 0$. Now we also check
\begin{equation}
\begin{split}
     \ dW_t^{(\phi)}   \  dW_s^{(\phi)}  &= \cos^2 \phi    \ dW_t^{(y)}   \  dW_s^{(y)}  \\ &- \cos \phi \sin \phi   \ dW_t^{(y)}   \  dW_s^{(x)}  \\ &-\sin \phi \cos \phi    \ dW_t^{(x)}  \  dW_s^{(y)}  \\
    &+ \sin^2 \phi  \ dW_t^{(x)}   \  dW_s^{(x)} 
\end{split}
\end{equation}
Using the property $ \ \langle dW^{i}_{s}\ dW^j_t \rangle =\delta_{ij}\delta(t-s) \mathrm{d}t \ \mathrm{d}s$ we get 
\begin{equation}
\begin{split}
      \langle dW_t^{(\phi)}  \  dW_s^{(\phi)}\rangle  &= \left(\cos^2 \phi + \sin^2 \phi\right) \delta(t-s) \mathrm{d}t \ \mathrm{d}s \\ &= \delta(t-s)\mathrm{d}t \ \mathrm{d}s
\end{split}
\end{equation}
and further one can easily verify that $\langle dW_t^{(\phi)}  \  dW_s^{(z)}\rangle  =0$.
Thus we see $dW_t^{(\phi)}$ also acts as a Wiener increment. Hence we can
Thus the equation of motion in the cylindrical coordinate is given as
\begin{equation}
    \begin{split}
     \dd\phi &= \frac{L_\phi}{R^2} \dd t \\
     \dd L_\phi &= -\partial_\phi V(\phi,z) \dd t- \gamma_{(xy)} L_\phi \dd t+\sqrt{2\,\gamma_{(xy)}R^2\,\beta^{-1}}\;\circ dW_t^{(\phi)} \\
         dz &= p_z\,dt, \\
   dp_z &= -\,\partial_z V(\phi,z)\,dt 
           \;-\;\gamma_zp_z\,dt 
           + \sqrt{2\,\gamma_z\,\beta^{-1}}\;\circ dW_t^{(z)}.
    \end{split}
\end{equation}
 Since the pre-factors in the noise terms do not depend on the coordinate or momentum variables,i.e., the noises are additive in nature, Stratonovich and Ito representations of equations of motion will be same. Thus we can write in the Ito form as
\begin{equation}
    \label{eq:app_eq_Ito_general_helical}
    \begin{split}
     \dd\phi &= \frac{L_\phi}{R^2} \dd t \\
     \dd L_\phi &= -\partial_\phi V(\phi,z) \dd t- \gamma_{\phi} L_\phi \dd t+\sqrt{2\,\gamma_{\phi}R^2\,\beta^{-1}} \  dW_t^{(\phi)} \\
         dz &= p_z\,dt, \\
   dp_z &= -\,\partial_z V(\phi,z)\,dt 
           \;-\;\gamma_z p_z\,dt 
           + \sqrt{2\,\gamma_z\,\beta^{-1}} \ dW_t^{(z)}.
    \end{split}
\end{equation}
where we have denoted $\gamma_\phi=\gamma_{(xy)}$ for simplicity. Below we summarize properties of the noises in the equation above:
\begin{equation}
\begin{split}
 \langle  dW_t^{(\phi)} \rangle &=0,\\
 \langle dW_t^{(z)} \rangle &=0,\\
    \langle dW_t^{(\phi)} dW_s^{(\phi)} \rangle  &= \dd t \dd s \delta(t-s),\\
    \langle dW_t^{(z)} dW_s^{(z)} \rangle&= \dd t \dd s\delta(t-s), \\
    \langle \dd W_t^{(\phi)} dW_s^{(z)} \rangle &=0.
\end{split}
\end{equation}
\section{Analytical solution for the isotropic case}
\label{app:analytical_soln_isotropic}
For the transverse direction (Eq.~\eqref{eqn:sde_lambda_coord}), we have a Brownian harmonic oscillator, and the solution to it can be written as
\begin{equation}
    \begin{split}
        q_n(t)&=f_{qq}(t)q_n(0) + f_{qp}(t)p(0) + \int_0^{t}\dd B^{p_n}_{t'}f_{qp}(t-t') \\
        p_n(t)&=f_{pq}(t)q_n(0) + f_{pp}(t)p(0) + \int_0^{t}\dd B^{p_n}_{t'}f_{pp}(t-t')
    \end{split}
\end{equation}
where, 
\begin{equation}
    \label{eq:solution_transverse_direction}
    \begin{split}
        f_{qq}(t)&= e^{-\frac{\gamma t}{2}}\left[\cosh\left(\frac{\Omega t}{2}\right)+\frac{\gamma}{\Omega}\sinh\left(\frac{\Omega t}{2}\right)\right]\\
        f_{qp}(t)&= e^{-\frac{\gamma t}{2}}\left[\frac{2}{\Omega}\sinh\left(\frac{\Omega t}{2}\right)\right]\\
         f_{pq}(t)&= -e^{-\frac{\gamma t}{2}}\left[\frac{2\omega^2_0}{\Omega}\sinh\left(\frac{\Omega t}{2}\right)\right]\\
         f_{pp}(t)&= e^{-\frac{\gamma t}{2}}\left[\cosh\left(\frac{\Omega t}{2}\right)-\frac{\gamma}{\Omega}\sinh\left(\frac{\Omega t}{2}\right)\right]
    \end{split}
\end{equation}
with
\begin{equation}
   \Omega(R)=\sqrt{\gamma^2-4\omega^2_0(R)} \ \ \text{.}
\end{equation}
Hence we get overdamped motion when $\gamma>\gamma_{\mathrm{crit}}(R)=2\omega_0(R)$, i.e., the dependence of $\gamma_{\text{crit}}(R)$ is given as 
\begin{equation}
    \gamma_{\text{crit}}(R) = 4\pi\sqrt{1+\frac{1}{4\pi^2R^2}}
\end{equation}
The expression above interestingly shows that if we decrease the curvature of the cylinder infinitely, i.e.,  $R\to \infty$, we approach the minimum possible value $\gamma_{\mathrm{crit}}\to 4\pi$. Thus, we see that the increase in the curvature (decrease in $R$) of the cylinder effectively expands the under-damped oscillation regime. This reiterates the fact that a geometric parameter $R$ controls the effective inertia of the dynamics, as it can also be seen from the Hamiltonian in Eq.~\eqref{eq:Hamiltonian_scaled}. Now it can be shown that the equilibrium time-correlation of momentum in the transverse direction is given as\cite{risken1996fokker} 
\begin{equation}
    \langle p_n(\tau) p_n(0) \rangle_{\mathrm{eq}}= \beta^{-1} e^{-\frac{\gamma \tau}{2}}\left[\cosh\left(\frac{\Omega\tau}{2}\right)-\frac{\gamma}{\Omega}\sinh\left(\frac{\Omega \tau}{2}\right)\right]
\end{equation}
where we have used the time reversible symmetry to make the equilibrium time-correlation function symmetric. Taking Laplace-Fourier transform we write the this correlation function as 
\begin{equation}
    \begin{split}
        \tilde{C}_{p_n, p_n}(\omega)&=\mathcal{F}\left[\theta(\tau)\langle p_n(\tau) p_n(0) \rangle_{\mathrm{eq}}\right]\\
        &=\int_{0}^\infty \dd \tau e^{-i \omega \tau}\langle p_n(\tau) p_n(0) \rangle_{\mathrm{eq}}
    \end{split}
\end{equation} \par 
For the screw direction (Eq.~\eqref{eq:screw_sde}), the screw momentum evolves as an OU process, and the solution to it can be written as 
\begin{equation}
    \label{eq:solution_screw_direction}
    \begin{split}
          p_s(t) &= p_s(0)e^{-\gamma t} + \int_0^t\dd B_{t'}^se^{-\gamma (t-t')}, \\
          q_s(t)&=q_s(0) +p_s(0)\frac{1-e^{-\gamma t}}{\gamma} + \frac{1}{\gamma}\int_{0}^{t}\dd B_{t'}^s(1-e^{-\gamma (t-t')})
    \end{split}
\end{equation} 
We can also derive the equilibrium time-correlation function for the screw momentum as
\begin{equation}
   \langle p_s(\tau)p_s(0)\rangle_{\mathrm{eq}}= \beta^{-1}e^{-\gamma \tau},
\end{equation}
which we can write in Fourier space by 
\begin{equation}
    \begin{split}
        \tilde{C}_{p_s, p_s}(\omega)&=\int_{0}^\infty \dd \tau e^{-i \omega \tau}\langle p_s(\tau) p_s(0) \rangle_{\mathrm{eq}}
    \end{split}
\end{equation}
We also find that the equilibrium time-correlation between screw momentum and transverse momentum vanishes at equilibrium.
\begin{equation}
    \langle p_s(\tau)p_n(0)\rangle_{\mathrm{eq}}=0
\end{equation}
 Using the solutions from Eqs.~\eqref{eq:solution_transverse_direction},\eqref{eq:solution_screw_direction} and the inverse transforms of Eqs.~\eqref{eq:coord_transform_yz_to_s-lambda},\eqref{eq:momentum_transform_yz_to_s-lambda}, one can easily obtain the full integral solutions in the $(y,z,p_y,p_z)$ space, which we do not show here for conciseness.
 \begin{figure}
    \centering
    \includegraphics[width=\linewidth]{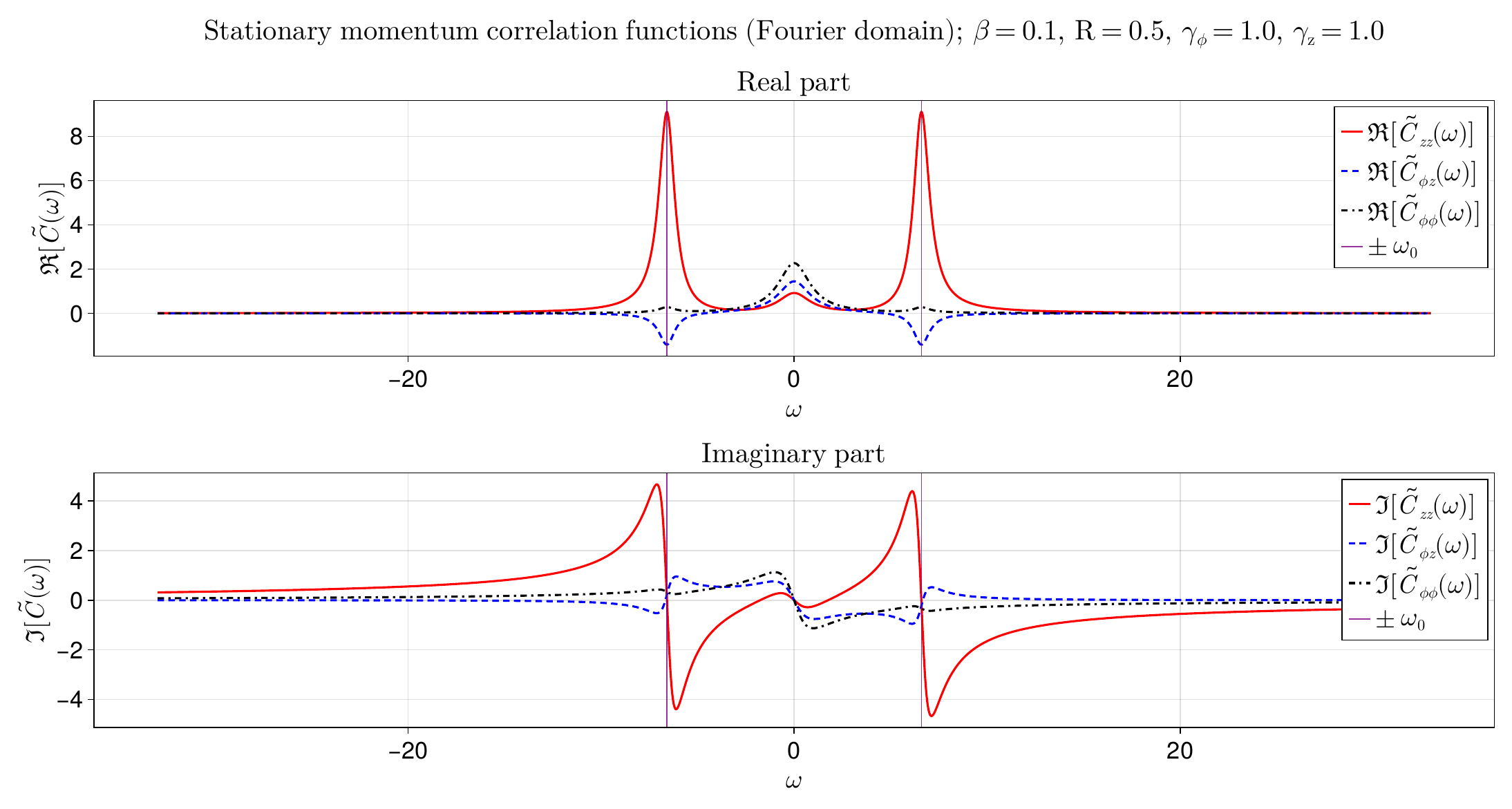}
    \caption{Fourier domain time correlation functions. In this figure we show the real and imaginary parts of the one sided Fourier transform of the time correlation functions shown in Fig.~\ref{fig:isotropic_time_correlations}
    \label{fig:fourier_time corr_isotropic}, which we have denoted as $\tilde{C}_{ab}(\omega)=\mathcal{F}\left[\theta(\tau)\langle p_a(\tau)p_b(0)\rangle\right]$ where $a,b$ represents the coordinate variables and $p_a,p_b$ represent their corresponding conjugate momentum respectively. We also have added vertical lines at $\pm \omega_0$ showing natural frequency of oscillation in the screw normal direction of the helix. The peak at $\omega=0$ corresponds to the purely damping mode with timescale $1/\gamma$ and the peaks near the $\pm \omega$ correspond the modes containing both oscillation and damping to the effective oscillation frequency $\pm\omega_0\sqrt{1-(\gamma^2/4)}$ and damping timescale $2/\gamma$. Parameters of the plot are $\beta=0.1, R=0.5,\gamma_\phi=\gamma_z=1.$}
\end{figure}
Instead, we just present the time equilibrium time correlation functions of momentum in helical coordinates in the Fourier space for simplicity:
 \begin{equation}
     \begin{split}
         \tilde{C}_{L_\phi,L_\phi}(\omega)&= \int_{0}^{\infty}\dd \tau e^{-i \omega \tau}\langle L_\phi(\tau)L_\phi(0)\rangle_{\mathrm{eq}}\\
         &=\frac{R^2}{\ g(R)}\left[\frac{1}{4\pi^2 R^2}\tilde{C}_{p_n, p_n}(\omega)+\tilde{C}_{p_s,p_s}(\omega)\right]\\
          \tilde{C}_{p_z,p_z}(\omega)&=\int_{0}^{\infty}\dd \tau e^{-i \omega \tau}\langle p_z(\tau)p_z(0)\rangle_{\mathrm{eq}}\\ &=\frac{1}{\ g(R)}\left[\frac{1}{4\pi^2 R^2}\tilde{C}_{p_s,p_s}(\omega)+\tilde{C}_{p_n,p_n}(\omega)\right]\\
         \tilde{C}_{L_\phi,p_z}(\omega)&= \int_{0}^{\infty}\dd \tau e^{-i \omega \tau}\langle L_\phi(\tau)p_z(0)\rangle_{\mathrm{eq}}\\ &=\frac{\chi}{2\pi\ g(R)}\left[\tilde{C}_{p_s,p_s}(\omega)-\tilde{C}_{p_n,p_n}(\omega)\right]
     \end{split}
 \end{equation}
 and illustrate them in Fig,~\ref{fig:fourier_time corr_isotropic}.
\section{Ensemble-level conservation law for a noisy helical system with anisotropic damping} 
\label{app:I_s_conservation_and_related_properties}
We provide a  proof for the ensemble-level average conservation of 
\begin{equation}
\begin{split}
    I_s(\phi,L_\phi, z,p_z)
    &=\frac{1}{2\pi\chi R \sqrt{\ g(R)}}\left[(2\pi\chi L_\phi +p_z)+(2\pi\chi R^2 \gamma_\phi \phi+\gamma_zz)\right]\\
    &=p_s(p_z,L_\phi) + \frac{1}{\sqrt{\ g(R)}}\left(R \gamma_\phi\phi+\frac{\gamma_zz}{2\pi\chi R}\right)\\
     &=p_s(p_z,L_\phi) + \bar\gamma q_s (\phi,z)
    +\frac{\delta\gamma}{\sqrt{\ g(R)}}\left(R \phi-\frac{z}{2\pi\chi R}\right)
\end{split} 
\end{equation}
 with  $\bar{\gamma}=\frac{\gamma_\phi+\gamma_z}{2}$ and $\delta\gamma=\frac{\gamma_\phi-\gamma_z}{2}$
for a general stochastic helical system with anisotropic damping given by Eq.~\eqref{eq:app_eq_Ito_general_helical}.
Now we write the equation of motion of $u(\vec{X})=\left(2\pi \chi R\sqrt{\ g(R)}\right)I_s(\vec{X})$ with $\vec{X}=(\phi  \ \ L_\phi \ \ z \ \ p_z)^T$ using Ito Lemma\cite{chorin_hald2009stochastic} as 
\begin{equation}
\begin{split}
    \mathrm{d}u&=\sum_{i}^4\frac{\partial u}{\partial X_i}\mathrm{d}X_i+ \sum_{i}^4\sum_{j}^4\frac{\partial u}{\partial X_i\partial X_j}\mathrm{d}X_i\mathrm{d}X_j\\
    &=\sum_{i}^4\frac{\partial u}{\partial X_i}\mathrm{d}X_i \\
    &=(2\pi\chi \mathrm{d}L_\phi +\mathrm{d}p_z)+(2\pi\chi R^2 \gamma_\phi \mathrm{d}\phi+\gamma_z \mathrm{d}z)
\end{split}
\end{equation}
where the second derivative terms vanish since $u$ (or $I_s$) is linear in $\vec{X}$. Now using Eq.~\eqref{eq:app_eq_Ito_general_helical} from the expression above we get
\begin{equation}
    \begin{split}
        \mathrm{d}u(t) =&-\left[2\pi\chi\partial_\phi V+\partial_z V\right]\mathrm{d}t \\&+\left[2\pi\chi \sqrt{2\gamma_\phi R^2\beta^{-1}} \  \mathrm{d}W^\phi_t +\sqrt{2\gamma_z \beta^{-1}} \  \mathrm{d}W^z_t\right]
    \end{split}
\end{equation}
For helical potential $V(\phi-2\pi\chi z)$, we have $\partial_\phi V=-\frac{1}{2\pi\chi}\partial_z V$, thus the deterministic term vanishes and we get
\begin{equation}
     \mathrm{d}u(t)=\left[2\pi\chi \sqrt{2\gamma_\phi R^2\beta^{-1}} \  \mathrm{d}W^\phi_t +\sqrt{2\gamma_z \beta^{-1}} \  \mathrm{d}W^z_t\right]
\end{equation}
Since $\langle\mathrm{d}W^\phi_t\rangle=\langle\mathrm{d}W^z_t\rangle=0$, we get $\langle\mathrm{d}u(t)\rangle=0$ which implies
\begin{equation}
    \frac{\mathrm{d}\langle I_s(t)\rangle}{\mathrm{d}t}=0
\end{equation}
For the time correlation properties of $I_s$, we calculate
\begin{equation}
\langle\mathrm{d}u(t)\mathrm{d}u(t')\rangle=\left[4\pi^2(2\gamma_\phi R^2\beta^{-1})+2\gamma_z \beta^{-1}\right]\delta(t-t')\mathrm{d}t\mathrm{d}t',
\end{equation}
and by defining an effective friction $\gamma_{\mathrm{eff}}(R)=\frac{4\pi^2 R^2\gamma_\phi +\gamma_z}{4\pi^2R^2+1}$, we get
\begin{equation}
    \langle\mathrm{d}I_s(t)\mathrm{d}I_s(t')\rangle=2\beta^{-1}\gamma_{\mathrm{eff}}(R)\delta(t-t')\mathrm{d}t\mathrm{d}t'
\end{equation}
\section{Analysis of the general anisotropic model}
\label{App_sec:analysis_of_general_anisotropic_model_Details}
\subsection{Continuous time Sylvester-Lyapunov Equation}
\label{App:SylvesterLyapunovEquation_Review}
Consider a generic Ornstein-Uhlenbeck(OU) process is given by the matrix stochastic differential equation (SDE)
\begin{equation}
    \mathrm{d} \vec{X}(t)
=
\mathbf{C}\ \vec{X}(t)\, \mathrm{d}t
+
\mathbf{\Sigma}\ \mathrm{d}\vec{W}(t)
\end{equation}
where $\mathbf{C},\mathbf{\Sigma}$ are given in Eq.\eqref{eq:drift_mat} and Eq.~\eqref{eq:noise matrix} respectively and $\mathrm{d}\vec{W}(t)$ represent a vector Wiener process as column vector with two independent components.
The evolution equation of the mean $\vec{m}(t)=\langle \vec{X}(t)\rangle$ is given as 
\begin{equation}
    \mathrm{d} \vec{m}(t)=\mathbf{C}\vec{m}(t)\mathrm{d}t
\end{equation}
We define the fluctuation as $\delta\vec{X}=\vec{X}(t)-\vec{m}(t)$
. The evolution equation of the fluctuation is given as 
\begin{equation}
\label{eq:fluctuation_eqn_OU_process}
    \mathrm{d}(\delta\vec{X})=\mathbf{C}\delta\vec{X}\mathrm{d}t +
\mathbf{\Sigma}\ \mathrm{d}\vec{W}(t)
\end{equation}
Now we apply Ito rule \cite{kurtJacobs_2010_book,chorin_hald2009stochastic} to calculate the differential of $\delta\vec{X}\delta\vec{X}^T$ by keeping terms that may contain terms second order in $d\vec{W}$, we get 
\begin{equation}
\begin{split}
    \mathrm{d}(\delta\vec{X}\delta\vec{X}^T)= \mathrm{d}(\delta\vec{X})\delta\vec{X}^T+ \delta\vec{X}\mathrm{d}(\delta\vec{X}^T) \\+ \mathrm{d}(\delta\vec{X})\mathrm{d}(\delta\vec{X}^T)  
\end{split}
\end{equation}
Substituting Eq.~\eqref{eq:fluctuation_eqn_OU_process} in the equation above we get
\begin{equation}
    \begin{split}
    \mathrm{d}(\delta\vec{X}\delta\vec{X}^T)&= \mathbf{C}(\delta\vec{X}\delta\vec{X}^T)\mathrm{d}t+(\delta\vec{X}\delta\vec{X}^T)\mathbf{C}^T\mathrm{d}t  \\ &+ \mathbf{\Sigma}\mathrm{d}\vec{W}(t)\mathrm{d}\vec{W}^T(t)\mathbf{\Sigma}^T \\
    &=\mathbf{C}(\delta\vec{X}\delta\vec{X}^T)\mathrm{d}t+(\delta\vec{X}\delta\vec{X}^T)\mathbf{C}^T\mathrm{d}t  \\ &+ \mathbf{\Sigma}(\mathbf{I_2}\mathrm{d}t)\mathbf{\Sigma}^T\\
    &=\mathbf{C}(\delta\vec{X}\delta\vec{X}^T)\mathrm{d}t+(\delta\vec{X}\delta\vec{X}^T)\mathbf{C}^T\mathrm{d}t \\ &+ \mathbf{\Sigma}\mathbf{\Sigma}^T\mathrm{d}t
\end{split}
\end{equation}
Now taking ensemble average of the both sides of the equation and defining the covariance matrix as $S(t)=\langle\delta\vec{X}(t)\delta\vec{X}^T(t)\rangle$ we get
\begin{equation}
    \dfrac{\mathrm{d}\mathbf{S}(t)}{\mathrm{d}t}= \mathbf{CS}(t)+\mathbf{S}(t)\mathbf{C}^T+\mathbf{\Sigma\Sigma}^T
\end{equation}
\subsection{Projection operator approach for solving the continuous time Sylvester-Lyapunov Equation with a zero mode}
\label{App:Zwanzig-Sylvester}
In this appendix we calculate asymptotic ($t \to \infty$) expressions for the covariance matrix by applying projection operator machinery to Sylvester-Lyapunov equation\cite{behr2019solution} to remove the zero-eigenmode.
The SylvesterLyapunov equation is given as
\begin{equation}
\begin{split}
    \dv{\mathbf{S}(t)}{t}&= \mathbf{C}\mathbf{S}(t)+\mathbf{S}(t)\mathbf{C}^T+\mathbf{\Sigma\Sigma}^T \\
        &=\mathcal{L}\left[\mathbf{S}(t)\right] +\mathbf{\Sigma\Sigma}^T
\end{split}
\end{equation}
where we have introduced the  Sylvester superoperator as $\mathcal{L}[\mathbf{M}]=\mathbf{C} \mathbf{M}+\mathbf{M} \mathbf{C}^T$ and the source term matrix is $\mathbf{\Sigma\Sigma}^T=\mathrm{diag}\left(0 \ \ 2\gamma_{\phi}R^2\beta^{-1} \ \ 0 \ \ 2\gamma_{z}\beta^{-1}\right)$
The formal solution covariance matrix $\mathbf{S}(t)$ is given as 
\begin{equation}
    \label{eq:Sylvester_eq_sol}
    \begin{split}
         \mathbf{S}(t)&=e^{t\mathcal{L}}\left[\mathbf{S}(0)\right] + \int_0^t\dd t' e^{(t-t')\mathcal{L}}\left[\mathbf{\Sigma \Sigma}^T\right]\\
         &=e^{\mathbf{C}t}\mathbf{S}(0)e^{\mathbf{C}^Tt}+ \int_0^t\dd t' e^{\mathbf{C}(t-t')}\mathbf{\Sigma\Sigma}^Te^{\mathbf{C}^T(t-t')}\\
    \end{split}
\end{equation}
where $e^{t\mathcal{L}}[\mathbf{M}]=e^{t\mathbf{C}} \mathbf{M} e^{t\mathbf{C}^T}$.\par

To understand the dynamics of the system, it is useful to look at the eigenspectrum of the matrix $\mathbf{C}$. We see that $\vec{r}_0=\left(2\pi\chi \ \ 0 \ \ 1 \ \ 0  \right)^T$, as a column vector satisfies $\mathbf{C}\vec{r}_0=\vec{0}$ becomes the right null-vector of the matrix $\mathbf{C}$ and corresponds to the zero mode of the dynamics, and similarly we take the left null-null-vector as a row-vector as 
$\vec{l}_0^T=\frac{1}{\gamma_z+4\pi^2R^2\gamma_\phi}\left(2\pi\chi\gamma_\phi R^2 \ \ 2\pi\chi \ \ \gamma_z \ \ 1\right)$, which satisfies $\vec{l}^T\mathbf{C}=\vec{0}^T$.
Due to the existence of a zero-eigenvalue of matrix $\mathbf{C}$, the full phase space covariance matrix $\mathbf{S}(t)$ will grow boundlessly with an asymptotic scaling of $\mathbf{S}(t)\sim \mathbf{D}_0t$, where the matrix $\mathbf{D}_0$ captures the diffusive mode. Hence, the probability distribution in full phase space is not normalizable unless some form of confinement is introduced in the dynamics (for example, a periodic boundary condition in both $\phi$ and $z$). However, we can show that other eigenvalues of $\mathbf{C}$ have negative real parts and thus the dynamics lead to a stable subspace (see Appendix~\ref{appendix:matrix_solution_sde}).\par
To analyze this system we introduce a set of projection matrices: $\mathbf{\Pi}_0=\vec{r}_0\vec{l}_0^T$,  $\mathbf{Q}_0=\mathbf{I}-\mathbf{\Pi}_0$ where $\vec{r}_0$ is the right eigen vector corresponding to the zero eigenmode of the drift matrix $\mathbf{C}$. The explicit expressions of this projection matrices have been derived in Appendix~\ref{appendix:matrix_solution_sde}. We also introduce the left eigenvector corresponding to the zero eigen mode as $\vec{l}_0$. The eigenvectors $\vec{r}_0$ and $\vec{l}_0$ satisfy the following properties: $\mathbf{C}\vec{r}_0=\vec{l}_0^T\mathbf{C}=\vec{0}$,and $\vec{l}_0\cdot\vec{r}_0=\vec{l}_0^T\vec{r}_0=1$.
Using the projection matrices $\mathbf{\Pi}_0$ and $\mathbf{Q}_0$, we can write, 
\begin{equation}
    \begin{split}
    e^{t\mathbf{C}}&=\mathbf{\Pi}_0 + e^{t\mathbf{C}}\mathbf{Q}_0\\
    e^{t\mathbf{C}^T}&=\mathbf{\Pi}^T_0+ \mathbf{Q}^T_0e^{t\mathbf{C}^T}
\end{split}
\end{equation}
and thus we can expand the evolution operator for the Sylvester equation as
\begin{equation}
    \begin{split}
        e^{t\mathcal{L}}[\mathbf{M}]&=e^{t\mathbf{C}}\mathbf{M}e^{t\mathbf{C}^T}\\
        &=\mathcal{P}_0\left[\mathbf{M}\right]+e^{t\mathcal{L}}\mathcal{Q}_0\left[\mathbf{M}\right]
    \end{split}
\end{equation}
where we have introduced the following projection superoperators: 
\begin{equation}
    \begin{split}
     \label{eq:projection_superoperator_sylvester}   \mathcal{P}_0[\mathbf{M}]&=\mathbf{\Pi}_0\mathbf{M}\mathbf{\Pi}^T_0\\
        \mathcal{Q}_0[\mathbf{M}]&=(1-\mathcal{P}_0)[M]=\mathbf{M}-\mathbf{\Pi}_0\mathbf{M}\mathbf{\Pi}^T_0
    \end{split}
\end{equation} and written the expression of the Sylvester evolution operator for the source term outside the null-block as
\begin{equation}
    \label{eq:projection_super_Q_evolve}
    \begin{split}   e^{t\mathcal{L}}\mathcal{Q}_0\left[\mathbf{M}\right]&=\mathbf{\Pi}_0\mathbf{M}\mathbf{Q}^T_0e^{t\mathbf{C}^T}+e^{t\mathbf{C}}\mathbf{Q}_0\mathbf{M}\mathbf{\Pi}^T_0\\&+e^{t\mathbf{C}}\mathbf{Q}_0\mathbf{M}\mathbf{Q}^T_0e^{t\mathbf{C}^T}
    \end{split}
\end{equation}
Now rewrite the solution from Eq.~\eqref{eq:Sylvester_eq_sol}  using Eq.~\eqref{eq:projection_superoperator_sylvester} and Eq.~\eqref{eq:projection_super_Q_evolve} as
\begin{equation}
\label{eq:S(t)_soln_operator}
\begin{split}
\mathbf{S}(t)&=\mathcal{P}_0\left[\mathbf{S}(0)\right]+e^{t\mathcal{L}}\mathcal{Q}_0\left[\mathbf{S}(0)\right]+t\mathcal{P}_0\left[\mathbf{\Sigma \Sigma}^T\right]\\
&+\int_0^t\mathrm{d}t'e^{(t-t'))\mathcal{L}}\mathcal{Q}_0\left[\mathbf{\Sigma \Sigma}^T\right]  
    \end{split}
\end{equation}
Here, the first term represents the memory effect of the initial condition of $\mathbf{S}(0)$ that is in the $\mathcal{P}_0$ projection, which stays invariant due to the dynamics. The second term arises  from the part of the initial condition that exists in $\mathcal{Q}_0$ space and it decays with time. The decay of the initial condition in the $\mathcal{Q}_0$ space can be understood by the fact that all non-zero roots of $\mathbf{C}$ has negative real parts (see Appnedix~\ref{appendix:matrix_solution_sde}) and thus both $e^{t\mathbf{C}}\mathbf{Q}_0$ and $\mathbf{Q}_0^Te^{t\mathbf{C}^T}$ creates terms that exponentially decays with time. 
Hence at $t \to \infty$, the second term $e^{t\mathcal{L}}\mathcal{Q}_0\left[\mathbf{S}(0)\right]$ approaches zero. The third term, $t\mathcal{P}_0\left[\mathbf{\Sigma \Sigma}^T\right]$ arises from the part of the source noise matrix $\mathbf{\Sigma\Sigma}^T$ that exists in the $\mathcal{P}_0$ projection space and this represents the variance along the translational screw direction which grows linearly with time and can be attributed as the diffusive part of the covariance matrix. The fourth term arises  from the part of the source noise matrix $\mathbf{\Sigma\Sigma}^T$ that existed in the $\mathcal{Q}_0$. We can write the asymptotic expression of $\mathbf{S}(t)$ as 
\begin{equation}
\label{eq:asymptotic_covariance_matrix}\mathbf{S}(t)\sim\mathcal{P}_0\left[\mathbf{S}(0)\right]+t\mathbf{D}_0+\mathbf{S}_\infty
\end{equation}
where $\mathbf{D}_0=\mathcal{P}_0\left[\mathbf{\Sigma \Sigma}^T\right]$ and
\begin{equation}
    \label{eq:S_infty_defination_drazin_form}
\begin{split}
     \mathbf{S}_\infty&=-\mathcal{L}^{-1}_{D}\left[\mathbf{\Sigma \Sigma}^T\right]
\end{split}
\end{equation}
where we have introduced \begin{equation}
    \label{eq:Drazin_inv_superoperator}
    \mathcal{L}^{-1}_{D}=- \int_0^\infty \dd t e^{t\mathcal{L}}\mathcal{Q}_0,
\end{equation} 
as the Drazin inverse of the Sylvester superoperator $\mathcal{L}$ in spirit of the approach of Drazin inverse\cite{landi2024current-6f7} of Lindbladian superoperator in open quantum systems literature.

\subsection{Spectral projector and construction of the evolution operator of Sylvester-Lyapunov equation }
\label{appendix:matrix_solution_sde}
Starting from the drift matrix $\mathbf{C}$ given in Eq.~\eqref{eq:drift_mat}, we write the characteristic polynomial  as
\begin{equation}
    \det\left(\mathbf{C}-\lambda \mathbf{I}\right)=\prod_{i=0}^3(\lambda-\lambda_i)=\lambda p(\lambda) =\lambda(\lambda^3+a_2\lambda^2 +a_1\lambda+a_0)
\end{equation}
where $\{\lambda_i|i=0,1,2,3\}$ are the eigenvalues of $\mathbf{C}$ with $\lambda_0=0$ and
\begin{equation}
\label{eq:coefficents_A}
    \begin{split}
        a_2&=\left(\gamma_{\phi}+\gamma_{z}\right)\\
        a_1&=\left(\gamma_{\phi}\gamma_{z}+4\pi^2 +\frac{1}{R^2}\right)=\left(\gamma_{\phi}\gamma_{z}+4\pi^2\ g(R) \right)\\
        a_0&=\left(4\pi^2\gamma_{\phi}+\frac{\gamma_{z}}{R^2}\right).
    \end{split}
\end{equation}
Note that, we have $a_0>0,a_1>0,a_2>0$ and $a_1a_2>a_0$ which implies all roots of $p(\lambda)=0$ have negative real part due to Routh-Hurwitz theorem\cite{zwillinger2014table}. This ensures the stability of the subspace of $\mathbf{C}$ when the zero mode is removed.
\par 
Now we calculate $p(\mathbf{C})$ as 
\begin{equation}
    p(\mathbf{C})=\prod_{i=1}^3(\mathbf{C}-\lambda_i\mathbf{I})
\end{equation}
Hence $p(\mathbf{C})$, when operated on a vector, removes all the components outside the null-space of $\mathbf{C}$. Thus we can define the normalized nullspace projector $\mathbf{\Pi}_0$ as 
\begin{equation}
    \label{eq:Pi_0_proj_mat_defn}
    \begin{split}
        \mathbf{\Pi}_0&=\frac{p(\mathbf{C})}{a_0}\\
            &=\frac{1}{a_0}\left(\mathbf{C}^3+a_2\mathbf{C}^2+a_1\mathbf{C}+a_0\mathbf{I}\right).
    \end{split}
\end{equation}
Now we show that the matrix $\Pi_0$ is idempotent $\Pi_0^2=\Pi_0$ and can indeed be used as a projection matrix. To see this we calculate $\mathbf{\Pi}_0^2-\mathbf{\Pi}_0$ as 
\begin{equation}
    \begin{split}
        \mathbf{\Pi}_0^2-\mathbf{\Pi}_0&=\frac{1}{a_0^2}(p(\mathbf{C}))^2-\frac{1}{a_0}p(\mathbf{C})\\
        &=\frac{1}{a_0^2}p(\mathbf{C})\left(p(\mathbf{C})-a_0\mathbf{I}\right)\\
        &=\frac{1}{a_0^2}p(\mathbf{C})\left(\mathbf{C}^3+a_2\mathbf{C}^2+a_1\mathbf{C}\right)\\
         &=\frac{1}{a_0^2}p(\mathbf{C})\mathbf{C}\left(\mathbf{C}^2+a_2\mathbf{C}+a_1\mathbf{I}\right)
    \end{split}
\end{equation}
Now the matrix $\mathbf{C}$ satisfies its own  characteristic equation as $p(\mathbf{C})\mathbf{C}=\mathbf{0}$. This proves
\begin{equation}
    \mathbf{\Pi}_0^2=\mathbf{\Pi_0}.
\end{equation}
Next we define the projector, that is complementary to $\mathbf{\Pi}_0$ as
\begin{equation}
    \label{eq:Q_0_proj_mat_defn}
    \mathbf{Q}_0=\mathbf{I}-\mathbf{\Pi}_0=-\frac{1}{a_0}\left(\mathbf{C}^3+a_2\mathbf{C}^2+a_1\mathbf C\right)
\end{equation}
Now since $\mathbf{Q}_0$ removes all the nullspace components from a vector, we can write
\begin{equation}
    p(\mathbf{C})\mathbf{Q}_0=\mathbf{0}
\end{equation} which simplifies to the relation
\begin{equation}
    \label{eq:matrix_pol_reduction_eqn}
    \mathbf{C}^3Q_0=-(a_2\mathbf{C}^2+a_1\mathbf{C}+a_0\mathbf{I})\mathbf{Q}_0
\end{equation} which will be useful later.
Next we write 
\begin{equation}
\label{eq:time_evolved_splitted}
    \begin{split}
        e^{\mathbf{C}t}
    &=\Pi_0+\mathbf{G}_Q(t)\\
    \end{split}
\end{equation}
where 
\begin{equation}
\begin{split}
    \mathbf{G}_Q(t)&=e^{\mathbf{C}t}\mathbf{Q}_0\\ &=\sum_{j=0}^\infty\frac{(t\mathbf{C})^j}{j!}\mathbf{Q}_0  \\    
\end{split} 
\end{equation}
Now using Eq.~\eqref{eq:matrix_pol_reduction_eqn} we can reduce any power $\mathbf{C}^k\mathbf{Q}_0$ with $k\ge3$ to the form $\eta(\mathbf{C})\mathbf{Q}_0$ where $\eta(\mathbf{C})$ is a quadratic matrix polynomial in $\mathbf{C}$. Hence we can write
\begin{equation}
\label{eq:G_q quadratic}
    \begin{split}
         \mathbf{G}_Q(t)
         &=(b_0(t)\mathbf{I}+b_1(t)\mathbf{C}+b_2(t)\mathbf{C}^2)\mathbf{Q}_0
    \end{split}
\end{equation}
where $b_0(t), b_1(t)$ and $b_2(t)$ is to be calculated.
Now taking time derivative of Eq.~\eqref{eq:time_evolved_splitted} and using $\mathbf{\Pi}_0\mathbf{C}=\mathbf{0}$ we get  $\mathbf{C}\mathbf{G}_Q(t)=\dot{\mathbf{G}}_Q(t)$ and set the initial condition $\mathbf{G}_Q(0)=\mathbf{Q}_0$. Now using Eq.~\eqref{eq:G_q quadratic} and Eq.~\eqref{eq:matrix_pol_reduction_eqn} we can write
\begin{equation}
\begin{split}
   &\left[-b_2a_0\mathbf{I}+(b_0-a_1b_2)\mathbf{C}+(b_1-a_2b_2)\mathbf{C}^2)\right]\mathbf{Q}_0\\=& (\dot{b}_0\mathbf{I}+\dot{b}_1\mathbf{C}+\dot{b}_2\mathbf{C}^2)\mathbf{Q}_0
\end{split}
\end{equation}
 then comparing term by term powers of $\{\mathbf{Q}_0,\mathbf{C}\mathbf{Q}_0,\mathbf{C}^2\mathbf{Q}_0\}$, we get a simple companion ordinary differential equation (ODE) for $\vec{b}=(b_0(t) \ \ b_1(t) \ \ b_2(t) )^T $ as\begin{equation}
\label{eq:companion_ode_b}
    \dot{\vec{b}}(t)=\mathbf{A}\vec{b}(t)
\end{equation} 
where 
\begin{equation}
\label{eq_companion_ode_b_matrices}
    \mathbf{A}=\begin{pmatrix}
        0 & 0 &-a_0 \\
        1 & 0 & -a_1 \\
        0 & 1 & -a_2
    \end{pmatrix}; \ \ \vec{b}(0)=\begin{pmatrix} 1\\0\\ 0
        
    \end{pmatrix}
\end{equation}
Thus the vector
\begin{equation}
    \vec{b}(t)=e^{\mathbf{A}t}\vec{b}(0)
\end{equation} 
can be easily obtained by calculating the matrix exponential directly or by the method of inverse Laplace transform.

\subsection{Markovian dynamics in the stable subspace}
\label{app:Markovian_dynamics_stable_subspace}
 Suppose we project the state vector $\vec{X}(t)$ the stable subspace as $\vec{Y}_Q(t)=\mathbf{Q}_0\vec{X}(t)$, then the equation of motion for $\vec{Y}_Q(t)$ is given as 
\begin{equation}
\label{eq:stableSubspace_eqn}
\begin{split}
   \mathrm{d} \vec{Y}_Q(t)
&=\mathbf{Q_0} \ \mathrm{d}\vec{X}(t)\\
&=\mathbf{Q_0C}\ \vec{X}(t)\, \mathrm{d}t
+
\mathbf{Q_0\Sigma}\ \mathrm{d}\vec{W}(t) \\ 
&=\mathbf{CQ_0}\ \vec{X}(t)\, \mathrm{d}t
+
\mathbf{Q_0\Sigma}\ \mathrm{d}\vec{W}(t)\\ 
&=\mathbf{C}\ \vec{Y}_Q(t)\, \mathrm{d}t
+
\mathbf{Q_0\Sigma}\ \mathrm{d}\vec{W}(t), 
\end{split}  
\end{equation}
which is also a multivariate OU process in the stable subspace of the dynamics. Note we have used the property $[\mathbf{Q}_0,\mathbf{C} ]=\mathbf{0}$ in the equation above, which allows us to close the dynamics completely in the stable subspace without involving the screw translational mode. The covariance matrix of the process in the stable subspace is given as $\mathbf{S_Y}(t)=\mathbf{Q}_0\mathbf{S}(t)\mathbf{Q}^T_0$.
Since the matrix $\mathbf{Q_0C}$ does not have any zero mode, the system relaxes to a unique stationary state with a stationary covariance matrix
\begin{equation}
    \lim_{t \to \infty}\mathbf{S_Y}(t)=\mathbf{S}_\infty^Q=\mathbf{Q}_0\mathbf{S}_\infty\mathbf{Q}^T_0.
\end{equation}
in the Appendix~\ref{app:stationary_state_measure_and_ibp} we further give details about this stationary state.
\subsection{Evaluation of the bounded part of the covariance matrix at long time limit}
\label{app:S_infty_calc)details}
Here we discuss details of the calculating $\mathbf{S}_\infty$. Using Eq.~\eqref{eq:time_evolved_splitted} and Eq.~\eqref{eq:G_q quadratic} we write
\begin{equation}
\begin{split}
    e^{\mathbf{C}t}&= \mathbf{\Pi}_0+(b_0(t)\mathbf{I}+b_1(t)\mathbf{C}+b_2(t)\mathbf{C}^2)\mathbf{Q}_0, \\
    e^{\mathbf{C}^Tt}&= \mathbf{\Pi_0}^T+\mathbf{Q}_0(b_0(t)\mathbf{I}+b_1(t)\mathbf{C}^T+b_2(t)(\mathbf{C}^T)^2)
\end{split}
\end{equation}
where $\vec{b}(t)=(b_0(t) \ \ b_1(t) \ \ b_2(t))^T$ are time-dependent functions and can be written as $\vec{b}(t)=e^{\mathbf{A}t}\vec{b}(0)$ as the solution of \eqref{eq:companion_ode_b} with the matrix $\mathbf{A}$ and $\vec{b}(0)$ can be defined using Eq.~\eqref{eq:coefficents_A} and Eq.~\eqref{eq_companion_ode_b_matrices} 
Thus, using Eq.~\eqref{eq:Drazin_inv_superoperator} and Eq.~\eqref{eq:projection_super_Q_evolve}, we can rewrite  Eq.~\eqref{eq:S_infty_defination_drazin_form} as 
\begin{equation}
\label{eq:S_infty_expanded_form}
\begin{split}
     \mathbf{S}_\infty
&=\int_{0}^{\infty}\dd t e^{\mathbf{C}t}\mathbf{Q_0}\mathbf{\Sigma\Sigma}^T\mathbf{\Pi}_0^T+\int_{0}^{\infty}\dd t \mathbf{\Pi}_0\mathbf{\Sigma\Sigma}^T\mathbf{Q}_0^Te^{\mathbf{C}^{T}t}\\ &+\int_{0}^{\infty}\dd t e^{\mathbf{C}t}\mathbf{Q}_0\mathbf{\Sigma\Sigma}^T\mathbf{Q}_0^Te^{\mathbf{C}^{T}t}
\end{split}
\end{equation}
Now using Eq.~\eqref{Eq:stable_space_)dynamics_generator_eqn},we can express $\mathbf{S}_\infty$ in the basis of $\mathbf{E}_0=\mathbf{Q}_0,\mathbf{E}_1=\mathbf{C}\mathbf{Q}_0,\mathbf{E}_2=\mathbf{C}^2\mathbf{Q}_0$ and get
\begin{equation}
    \begin{split}
        \mathbf{S}_\infty&=\left(\sum_{m=0}^2\alpha_m\mathbf{E_m}\right)\mathbf{\Sigma\Sigma}^T\mathbf{\Pi}^T_0 
    +\mathbf{\Pi}_0\mathbf{\Sigma}\mathbf{\Sigma}^T\left(\sum_{m=0}^2\alpha_m\mathbf{E_m}\right)\\ &+ \sum^{2}_{m,n=0}J_{mn}\mathbf{E_m}\mathbf{\Sigma\Sigma}^T\mathbf{E}_n^T,
    \end{split}
\end{equation}
where we have introduced the constants $\alpha_n$ and $J_{mn}$ as the following integrals
\begin{equation}
    \begin{split}
        \alpha_n&=\int_0^\infty \dd t b_n(t),\ \
        J_{mn}=\int_{0}^\infty \dd t b_m(t)b_n(t)
    \end{split}
\end{equation}
Now we can define the vector $\vec{\alpha}=(\alpha_0 \ \ \alpha_1 \ \ \alpha_2)^T$ which can be written in terms of model parameters $(a_0,a_1,a_2)$ as
\begin{equation}
\vec{\alpha}=\int_0^\infty\dd t  e^{\mathbf{A}t}\vec{b}(0)=-\mathbf{A}^{-1}\vec{b}(0)=\frac{1}{a_0}\begin{pmatrix}
        a_1 \\ a_2 \\ 1
    \end{pmatrix}.
\end{equation}
Hence we get 
\begin{equation}
    \sum_0^{2}\alpha_m \mathbf{E_m}=\frac{1}{a_0}\left(a_1\mathbf{I}+a_2\mathbf{C}+\mathbf{C}^2\right)\mathbf{Q}_0
\end{equation}
For compactness, we can also introduce the pseudo inverse of the matrix $\mathbf{C}$ as  
\begin{equation}
     \mathbf{C}_D^{-1}=-\int_0^{\infty}\mathrm{d}t \ e^{\mathbf{C}t}\mathbf{Q}_0
\end{equation}
where one can show\cite{Mandal2016AnalysisStates} $\mathbf{C}^{-1}_D\mathbf{C}=\mathbf{C}\mathbf{C}^{-1}_D=\mathbf{Q}_0$.
which lets us write the sum involving $\alpha_n$ coefficients as
\begin{equation}
    \begin{split}
    \sum_0^{2}\alpha_m\mathbf{ E_m}=-\mathbf{C}_D^{-1}    
    \end{split}
\end{equation} \par 
Next, we consider $J_{mn}$ to be elements of the matrix $\mathbf{J}$ which is defined as
\begin{equation}
\begin{split}
     \mathbf{J}&=\int_0^\infty \dd t \vec{b}(t)\vec{b}(t)^T \\ 
     &=\int_0^\infty \mathrm{d}t \ e^{\mathbf{A}t}\vec{b}(0)\vec{b}(0)^Te^{\mathbf{A}^Tt}
\end{split}
\end{equation}
which implies $\mathbf{J}$ is the stationary solution to the continuous time Sylvester-Lyapunov equation 
\begin{equation}
    \frac{\mathrm{d}\mathbf{\bar{J}}(t)}{\mathrm{d}t}=\mathbf{A}\mathbf{\bar{J}}(t)+\mathbf{\bar{J}}(t)\mathbf{A^T}+\vec{b}(0)\vec{b}^T(0)
\end{equation}
and $\lim_{t \to \infty}\mathbf{\bar{J}}(t)=\mathbf{J}$ and thus we can solve for $\mathbf{J}$ by setting $\left(\frac{\mathrm{d}\mathbf{\bar{J}}(t)}{\mathrm{d}t}\right)_{\mathbf{\bar{J}}(t)=\mathbf{J}}=0$ as
\begin{equation}
    \mathbf{A}\mathbf{J}+\mathbf{J}\mathbf{A}^T=-\vec{b}(0)\vec{b}(0)^T=-\begin{pmatrix}
        1 & 0 & 0\\
        0 & 0 & 0 \\
        0 & 0 & 0
    \end{pmatrix},
\end{equation}
and using computer algebra system which we can solve as
\begin{equation}
    \mathbf{J}=\left[\begin{matrix}\frac{a_{0} a_{1} - a_{0} a_{2}^{2} - a_{1}^{2} a_{2}}{2 a_{0} \left(a_{0} - a_{1} a_{2}\right)} & - \frac{a_{1} a_{2}^{2}}{2 a_{0} \left(a_{0} - a_{1} a_{2}\right)} & \frac{1}{2 a_{0}}\\- \frac{a_{1} a_{2}^{2}}{2 a_{0} \left(a_{0} - a_{1} a_{2}\right)} & \frac{- a_{0} - a_{2}^{3}}{2 a_{0} \left(a_{0} - a_{1} a_{2}\right)} & - \frac{a_{2}^{2}}{2 a_{0} \left(a_{0} - a_{1} a_{2}\right)}\\\frac{1}{2 a_{0}} & - \frac{a_{2}^{2}}{2 a_{0} \left(a_{0} - a_{1} a_{2}\right)} & - \frac{a_{2}}{2 a_{0} \left(a_{0} - a_{1} a_{2}\right)}\end{matrix}\right].
\end{equation}
\subsection{Reduction of anisotropic case to isotropic case}
\label{app:anisotropic_to_isotropic_reduction}
The Fourier-space coefficients are given as 
\begin{equation}
\begin{split}
     \begin{pmatrix}
        \tilde{b}_0(\omega)\\
        \tilde{b}_1(\omega)\\
        \tilde{b}_2(\omega)
    \end{pmatrix}&=-\frac{1}{p(i\omega)}\begin{pmatrix}
            \omega^2 -a_1-ia_2\omega \\
            -a_2-i\omega\\
            -1
        \end{pmatrix}, \\
        p(i\omega)&=-i\omega^3-a_2\omega^2+ia_1\omega+a_0\\
        &=(a_0-a_2\omega^2)+i\omega(a_1-\omega^2),
\end{split}
\end{equation}
For isotropic case we have $a_0=\gamma\omega_0^2,a_1=\gamma^2+\omega_0^2,a_2=2\gamma$. Thus we get
\begin{equation}
    p(\lambda)=(\lambda+\gamma)(\lambda^2+\gamma\lambda+\omega_0^2).
\end{equation}
Substituting for the isotropic case we get
\begin{equation}
    \begin{split}
           \begin{pmatrix}
        \tilde{b}_0(\omega)\\
        \tilde{b}_1(\omega)\\
        \tilde{b}_2(\omega)
    \end{pmatrix}&=-\frac{1}{p(i\omega)}\begin{pmatrix}
        \omega^2-(\gamma^2+\omega_0^2)-i2\gamma\omega\\-2\gamma-i\omega\\
        -1
    \end{pmatrix},\\
    p(i\omega)&=(i\omega+\gamma)(\omega^2_0-\omega^2+i\gamma \omega)  
    \end{split}
\end{equation}
which can be simplified as 
 \begin{equation}
    \begin{split}
        \tilde{b}_0(\omega)&=G_s(\omega)+\gamma G_n(\omega),\\
        \tilde{b}_1(\omega)&=\frac{\gamma}{\omega_0^2}G_s(\omega)+\left(1-i\frac{\gamma\omega}{\omega_0^2}\right)G_n(\omega),\\
        \tilde{b}_2(\omega)&=\frac{1}{\omega_0^2}G_s(\omega)-\frac{i\omega}{\omega_0^2}G_n(\omega),\\
        G_s(\omega)&=\frac{1}{i\omega+\gamma},\\
        G_n(\omega)&=\frac{1}{\omega_0^2-\omega^2+i\gamma\omega}.
    \end{split}
\end{equation}
Now we use the following Fourier-Laplace transform relation pairs
\begin{equation}
\begin{split}
G_s(\omega)&=\mathcal{F}\left[\theta(\tau)e^{-\gamma \tau}\right],\\
G_n(\omega)&=\mathcal{F}\left[\theta(\tau)\frac{2}{\Omega}\mathcal{S}(\tau)\right],\\
    i\omega G_n(\omega)&=\mathcal{F}\left[\theta(\tau)\left(\mathcal{C}(\tau)-\frac{\gamma}{\Omega}\mathcal{S}(\tau)\right)\right],\\
    \mathcal{S}(\tau)&=e^{-\frac{\gamma\tau}{2}}\sinh\left(\frac{\Omega\tau}{2}\right),\\
    \mathcal{C}(\tau)&=e^{-\frac{\gamma\tau}{2}}\cosh\left(\frac{\Omega\tau}{2}\right),\\
    \Omega &=\sqrt{\gamma^2-4\omega_0^2}
\end{split}
\end{equation}
to calculate the time domain coefficients as
\begin{equation}
\begin{split}
    b_0(\tau)&=e^{-\gamma\tau}+\frac{2\gamma}{\Omega}\mathcal{S}(\tau),\\
    b_1(\tau)&=\frac{\gamma}{\omega_0^2}e^{-\gamma\tau}-\frac{\gamma}{\omega_0^2}\mathcal{C}(\tau)
+\frac{1}{\Omega}\left(2+\frac{\gamma^2}{\omega_0^2}\right)\mathcal{S}(\tau),\\
b_2(\tau)&=\frac{1}{\omega_0^2}e^{-\gamma\tau}-\frac{1}{\omega_0^2}\left(\mathcal{C}(\tau)-\frac{\gamma}{\Omega}\mathcal{S}(\tau)\right).
\end{split}
\end{equation}
Notice that $b_0(0)=1,b_1(0)=0,b_2(0)=0$, implying only $b_0(\tau)$ contributes the stationary correlation at $\tau=0$. Now using the expressions above in Eq.~\eqref{eqn:S_p_krylov_split_main_text} and setting $\gamma_\phi = \gamma_z = \gamma$, we get,
\begin{equation}
    \begin{split}
        \langle L_\phi(0)L_\phi(\tau)\rangle&=\frac{1}{\beta}\left[b_0(\tau)R^2-b_1(\tau) R^2\gamma+b_2(\tau)(R^2\gamma^2-1)\right], \\
        \langle L_\phi(0)p_z(\tau)\rangle&=\langle L_\phi(\tau)p_z(0)\rangle= \frac{1}{\beta}\left[b_2(\tau)2\pi\chi\right],\\
        \langle p_z(0)p_z(\tau)\rangle&=\frac{1}{\beta}\left[b_0(\tau)-b_1(\tau)\gamma+b_2(\tau)(\gamma^2-4\pi^2)\right],
    \end{split}
\end{equation}
which leads to
\begin{equation}
    \begin{split}
         \langle L_\phi(0)L_\phi(\tau)\rangle&=\frac{1}{\beta}\left[e^{-\gamma\tau}\left(R^2-\frac{1}{\omega_0^2(R)}\right)+\frac{1}{\omega_0^2(R)}\left(\mathcal{C}(\tau)-\frac{\gamma}{\Omega}\mathcal{S}(\tau)\right)\right]\\
         &=\frac{1}{\beta}\left[e^{-\gamma\tau}\frac{R^2}{\ g(R)}+\frac{1}{4\pi^2\ g(R)}\left(\mathcal{C}(\tau)-\frac{\gamma}{\Omega}\mathcal{S}(\tau)\right)\right]\\
         &=\frac{R^2}{\beta\ g(R)}\left[e^{-\gamma\tau}+\frac{1}{4\pi^2R^2}\left(\mathcal{C}(\tau)-\frac{\gamma}{\Omega}\mathcal{S}(\tau)\right)\right];
    \end{split}
\end{equation}
\begin{equation}
    \begin{split}
        \langle L_\phi(0)p_z(\tau)\rangle&=\langle L_\phi(\tau)p_z(0)\rangle\\&=\frac{2\pi\chi}{\beta\omega_0^2(R)}\left[e^{-\gamma\tau}-\left(\mathcal{C}(\tau)-\frac{\gamma}{\Omega}\mathcal{S}(\tau)\right)\right]\\
        &=\frac{2\pi\chi}{\beta4\pi^2\ g(R)}\left[e^{-\gamma\tau}-\left(\mathcal{C}(\tau)-\frac{\gamma}{\Omega}\mathcal{S}(\tau)\right)\right];
    \end{split}
\end{equation}
\begin{equation}
    \begin{split}
        \langle p_z(0)p_z(\tau)\rangle&=\frac{1}{\beta}\left[e^{-\gamma\tau}\left(1-\frac{4\pi^2}{\omega^2_0(R)}\right)+\frac{4\pi^2}{\omega_0^2(R)}\left(\mathcal{C}(\tau)-\frac{\gamma}{\Omega}\mathcal{S}(\tau)\right)\right]\\
        &=\frac{1}{\beta}\left[e^{-\gamma\tau}\frac{1}{4\pi^2R^2\ g(R)}+\frac{1}{\ g(R)}\left(\mathcal{C}(\tau)-\frac{\gamma}{\Omega}\mathcal{S}(\tau)\right)\right]\\
        &=\frac{1}{\beta\ g(R)}\left[\frac{e^{-\gamma\tau}}{4\pi^2R^2}+\left(\mathcal{C}(\tau)-\frac{\gamma}{\Omega}\mathcal{S}(\tau)\right)\right].
    \end{split}
\end{equation}
Thus, we recover the results for the isotropic case presented in Eq.~\eqref{eq:main_text_isotrpic_time_corr_functions}.
\section{The stationary Gaussian measure on the stable subspace and integration by parts formula}
\label{app:stationary_state_measure_and_ibp}
\subsection{Stationary measure}
Here we show that the dynamics in the stable subspace relaxes to a Gaussian stationary state, clarify the interpretation of the integration using this stationary measure and derive an integration by parts formula which will be useful for the general derivation of response relations . We choose a orthonormal basis vectors $\{\vec{u}_1,\vec{u}_2,\vec{u}_3\}$ in three dimensional stable subspace existing in the four dimensional phase space and define a $4 \times3$ matrix $\mathbf{U}=(\vec{u}_1 \ \vec{u}_2 \ \vec{u}_3)$ with the property $\mathbf{U}^T\mathbf{U}=\mathbf{I}_3$
This leads to a three dimensional projection of the vector $\vec{Y}_Q$ to a vector $\vec{z}\in \mathbb{R}^3$ as 
\begin{equation}
    \vec{Y}_Q=\mathbf{U}\vec{z}
\end{equation}
Note that, since we have $\mathbf{Q}_0\vec{Y}_Q=\vec{Y}_Q$, we also get the relation
\begin{equation}
    \mathbf{Q}_0\mathbf{U}=\mathbf{U}
\end{equation}
Thus we write the equation of motion for $\vec{z}$ as
\begin{equation}
\begin{split}
    \mathrm{d}\vec{z}&=\mathbf{U}^T\mathrm{d}\vec{Y}_Q\\
    &=\mathbf{U}^T\mathbf{C}\ \vec{Y}_Q(t)\, \mathrm{d}t
+
\mathbf{U}^T\mathbf{Q_0\Sigma}\ \mathrm{d}\vec{W}(t) \\
&=\mathbf{U}^T\mathbf{C}\ \mathbf{U}\vec{z}\, \mathrm{d}t
+
\mathbf{U}^T\mathbf{Q_0\Sigma}\ \mathrm{d}\vec{W}(t)
\end{split}
\end{equation}
which is also an OU process and relaxes to the Gaussian distribution
\begin{equation}
    \pi^z_Q(\vec{z})=\frac{\exp\left[-\frac{1}{2}\vec{z}^T\mathbf{S_z^{-1}}\vec{z}\right]}{\sqrt{(2\pi)^3\det\mathbf{S}_z}}
\end{equation}
where $\mathbf{S_z}$ is the stationary state covariance matrix
\begin{equation}
\begin{split}
     \mathbf{S_z}&=\lim_{t\to \infty}\langle \vec{z}(t)\vec{z}^T(t) \rangle\\ &=\lim_{t\to \infty}\langle \mathbf{U}^T\vec{Y}_Q(t)\vec{Y}_Q^T(t)\mathbf{U}\rangle\\&=\mathbf{U}^T\mathbf{S}^\infty_Q\mathbf{U},
\end{split}
\end{equation}
and similarly we can also show $\mathbf{S}^\infty_Q=\mathbf{U}\mathbf{S}_z\mathbf{U}^T$.
Note that $\mathbf{S_z}$ is an invertible matrix and the inverse is given as $\mathbf{S_z^{-1}}$ which satisfies $\mathbf{S_z^{-1}}\mathbf{S_z}=\mathbf{S_z}\mathbf{S_z^{-1}}=\mathbf{I}_3$. We project this inverse to the full four dimensional space by defining
$\mathbf{\tilde{S}^\infty_Q}=\mathbf{U}\mathbf{S_z^{-1}}\mathbf{U}^T$, a pseudoinverse\cite{Mandal2016AnalysisStates,landi2024current-6f7} of $\mathbf{S}^\infty_Q$, which has the following property:
\begin{equation}
    \begin{split}
        \mathbf{\tilde{S}^\infty_Q}\mathbf{S}^\infty_Q\vec{Y}_Q&=\mathbf{U}\mathbf{S_z^{-1}}\mathbf{U}^T\mathbf{S}^\infty_Q\vec{Y}_Q\\
        &=\mathbf{U}\mathbf{S_z^{-1}}\mathbf{U}^T\mathbf{S}^\infty_Q\mathbf{U}\vec{z}\\        &=\mathbf{U}\mathbf{S_z^{-1}}\mathbf{S_z}\vec{z}\\
        &=\mathbf{U}\vec{z}\\
        &=\vec{Y}_Q
    \end{split}
\end{equation}
and similarly, we can also show $\mathbf{S}^\infty_Q\mathbf{\tilde{S}^\infty_Q}\vec{Y}_Q=\mathbf{U}\mathbf{U}^T\mathbf{U}\vec{z}=\vec{Y}_Q$.
\par 
\subsection{Integration under measure and integration by parts formula}
We express the integration of a test function $F(\vec{y})$ with the stationary measure $\pi_Q(\mathrm{d}\vec{y})$ on the stable subspace (i.e., where $\vec{y}=\mathbf{Q_0}\vec{y}$) in the phase space coordinates  as
\begin{equation}
    \begin{split}
        \int\pi_Q(\mathrm{d}\vec{y}) F(\vec{y})
        &=\int\mathrm{d}\vec{z} \ \pi^z_Q(\vec{z}) F(\mathbf{U}\vec{z})
    \end{split}
\end{equation}
Given the 4 dimensional directional derivative operator projected along the stable subspace is given as $\mathbf{U}\mathbf{U}^T\nabla_y$, suppose we are interested in calculating  integrals of the form
\begin{equation}
\begin{split}
I_{1}&=\int\pi_Q(\mathrm{d}\vec{y})\vec{A}_y\cdot(\mathbf{U}\mathbf{U}^T\nabla_y F(\vec{y}))\\
    &=\int\pi_Q(\mathrm{d}\vec{y})\vec{A}_y^T\mathbf{U}\mathbf{U}^T\nabla_y F(\vec{y})\\
    \end{split}
\end{equation}
where $\vec{A}_y$ is a constant $4\times 1$ vector in the stable subspace. We define a function $f(\vec{z})=F(\mathbf{U}\vec{z})$, and write the vector $\vec{A}_z=\mathbf{U}^T\vec{A}_y$. Using  $\nabla_zf(z)=\mathbf{U}^T\nabla_yF(\mathbf{U}\vec{z})$, we get
\begin{equation}
    \begin{split}
        \vec{A}_y^T\mathbf{U}\mathbf{U}^T\nabla_y F(\vec{y})&=\vec{A}_y^T\mathbf{U}\mathbf{U}^T\nabla_y F(\mathbf{U}\vec{z})\\
        &=\left(\mathbf{U}\vec{A}_z\right)^T\mathbf{U}\mathbf{U}^T\nabla_yF(\mathbf{U}\vec{z})\\
        &=\vec{A}_z^T\mathbf{U}^T\nabla_y F(\mathbf{U}\vec{z})\\
        &=\vec{A}_z^T\nabla_zf(z).
    \end{split}
\end{equation}
Then we can write the integral $I_1$ as 
\begin{equation}
\begin{split}
    I_{1}&=\int\mathrm{d}\vec{z} \ \pi^z_Q(\vec{z}) (\vec{A}_z^T\nabla_zf(z))\\
\end{split}
\end{equation}
Now since $\pi_Q^z(\vec{z})$ is a Gaussian distribution and thus boundary terms in integration by parts dies, we get
\begin{equation}
\begin{split}
    I_{1}&=\int\mathrm{d}\vec{z} \ \pi^z_Q(\vec{z}) (\vec{A}_z^T\nabla_zf(z))\\
    &=-\int\mathrm{d}\vec{z} \  f(z)(\vec{A}_z^T\nabla_z\pi^z_Q(\vec{z}))\\
    &=-\int\mathrm{d}\vec{z} \  f(z)\pi^z_Q(\vec{z})(\vec{A}_z^T\nabla_z\ln\pi^z_Q(\vec{z}))\\
    &=-\int\mathrm{d}\vec{z} \  f(z)\pi^z_Q(\vec{z})(\vec{A}_z^T(-\mathbf{S_z^{-1}}\vec{z}))\\
    &=\int\mathrm{d}\vec{z} \  f(z)\pi^z_Q(\vec{z})\vec{A}_z^T\mathbf{S_z^{-1}}\vec{z}    
\end{split}
\end{equation}
where we have defined $\mathbf{S_z^{-1}}$ as the inverse-covariance matrix and since it is symmetric, we have used $\nabla_z(-\frac{1}{2} \vec{z}^T\mathbf{S_z^{-1}}\vec{z})=-\frac{1}{2}\left(\mathbf{S_z^{-1}}+(\mathbf{S_z^{-1}})^T\right)\vec{z}=-\mathbf{S_z^{-1}}\vec{z}$.\par 
Now rewriting in the original variables using $\vec{A}^T_z=\vec{A}^T_y\mathbf{U}$ and $\vec{z}=\mathbf{U}^T\vec{y}$ we get,
\begin{equation}
    \vec{A}_z^T\mathbf{S_z^{-1}}\vec{z}= \vec{A}^T_y\mathbf{U}\mathbf{S_z^{-1}}\mathbf{U}^T\vec{y}=\vec{A}_y^T\mathbf{\tilde{S}_Q^\infty}\vec{y},
\end{equation}
hence we get,
\begin{equation}
   \begin{split}I_{1}&=\int\pi_Q(\mathrm{d}\vec{y})F(\vec{y})\vec{A}_y^T\mathbf{\tilde{S}_Q^\infty}\vec{y}
   \end{split}
\end{equation}
Thus we have the identity
\begin{equation}
\begin{split}     \int\pi_Q(\mathrm{d}\vec{y})\vec{A}_y^T\mathbf{U}\mathbf{U}^T\nabla_y F(\vec{y})= \int\pi_Q(\mathrm{d}\vec{y})F(\vec{y})\vec{A}_y^T\mathbf{\tilde{S}_Q^\infty}\vec{y}.
\end{split}
\end{equation},
since the observable $\vec{A}_y$ is completely in the stable subspace, we get $\mathbf{U}\mathbf{U}^T\vec{A}_y$, which leads to
\begin{equation}
\label{eq:integration_by_parts_formula_gaussian_zero_mode}
\begin{split}     \int\pi_Q(\mathrm{d}\vec{y})\vec{A}_y^T\nabla_y F(\vec{y})= \int\pi_Q(\mathrm{d}\vec{y})F(\vec{y})\vec{A}_y^T\mathbf{\tilde{S}_Q^\infty}\vec{y},
\end{split}
\end{equation}
which we have used in Appendix ~\ref{App:generic_linear_response}.

\section{Derivation of the Mobility tensor via linear response theory}
\label{App:generic_linear_response}

Since we are dealing with existence of zero mode and non-stationary distribution phase space, we approach the response theory in the operator formalism for stochastic dynamics\cite{risken1996fokker,Seifert_2025} \cite{baiesi2013update-6b6}\cite{Marconi2008Fluctuation-Dissipation:Physics}. We first provide a review of the general response calculation using Kubo-Agarwal conjugate approach and then focus on the special case of mobility calculation.

We consider a general non-conservative driving of force $\mathbf{B}\vec{f}(t)$ in the momentum equations in the state vector $\vec{X}$ as 
\begin{equation}
\label{eq:full_space_perturbation_equation}
    \mathrm{d} \vec{X}(t)
=
\mathbf{C}\ \vec{X}(t)\ \mathrm{d}t
+\epsilon \mathbf{B}\vec{f}(t) \ \mathrm{d}t+
\mathbf{\Sigma}\ \mathrm{d}\vec{W}(t)
\end{equation}
where 
\begin{equation}
    \mathbf{B}=\begin{pmatrix}
        0 & 0 \\
        1 & 0\\
        0 & 0\\
        0 & 1
    \end{pmatrix}, \ \ \vec{f}(t)=\begin{pmatrix}
        f_\phi(t)\\
        f_z(t)
    \end{pmatrix}
\end{equation}
The projected equation in the stable space can be written as 
\begin{equation}
    \label{eq:perturbed_eqn_sde}
       \mathrm{d} \vec{Y}_Q(t)
=
\mathbf{C}\ \vec{Y}_Q(t)\ \mathrm{d}t
+\epsilon\mathbf{Q}_0 \mathbf{B}\vec{f}(t) \ \mathrm{d}t+
\mathbf{Q_0}\mathbf{\Sigma}\ \mathrm{d}\vec{W}(t)
\end{equation}
\subsection{Review of response relation calculations}
When $\epsilon=0$, for a test function $\phi(\vec{y})$  we write\cite{chorin_hald2009stochastic} 
\begin{equation}
    \hat{L}\phi(\vec{y})= (\mathbf{Q_0C}\vec{y})\cdot\nabla\phi(\vec{y}) +\frac{1}{2} \left(\mathbf{\mathbf{Q_0}\Sigma \Sigma}^T\mathbf{Q_0}^T\right) :\nabla\nabla\phi(\vec{y}),
\end{equation}
where $\hat{L}$ is the infinitesimal generator \cite{chorin_hald2009stochastic} of the unperturbed process in the stable subspace projection which describes the time evolution of a function $u_t(\vec{y})=\mathbb{E}\left[O(\vec{Y}(T))|\vec{Y}(t)=\vec{y}\right]$, i.e., the conditional expectation value of the observable $O(\vec{Y})$ at time $T\ge t\ge 0$ 
We write the backward time evolution of $u_t$ at any time $T\ge t\ge 0$ as the backward Kolmogorov equation
\begin{equation}
    \label{eq:unperturbed_backward_Kolmogorov}
    \partial_t u_t(\vec{y}) = -\hat{L}u_t(\vec{y}), \ \  u_T(\vec{y})=O(\vec{y}) 
\end{equation}
For any two time $t,s \in [0,T]$ we write the solution to this equation as 
\begin{equation}
    u_s(\vec{y})=\hat{P}_{(t,s)}   u_t(\vec{y})
\end{equation}
where 
\begin{equation}
\hat{P}_{(t,s)}=e^{(t-s)\hat{L}}    
\end{equation}
 is the propagator under the under the unperturbed dynamics  for $T\ge t\ge s \ge 0$. 
We can obtain the  expected value of $O(\vec{Y})$ at time $T$ by averaging over the initial conditions sampled from some probability measure $\mu\left(\mathrm{d}\vec{y}\right)$ in the sable subspace. Thus similar to the Heisenberg evolution form in quantum mechanics here we can write 
\begin{equation}
  \langle O(T)\rangle_{\mu} = \int \mathrm{\mu}(\mathrm{d}\vec{y})u_0(\vec{y}) = \int \mathrm{\mu}(\mathrm{d}\vec{y})\hat{P}_{(T,0)}  u_T(\vec{y})
\end{equation}
When we perturb the system $(\epsilon \ne 0)$ with the non-conservative force, we get the modified generator of the process as $\hat{L}^\epsilon(t)=\hat{L}+ \epsilon\hat{L}_1(t)$ where $\hat{L}_1(t)$
is the infinitesimal generator arising due to the perturbation. Under the perturbed dynamics we write the conditional expected value of the observable as $u_t^\epsilon(\vec{y})=\mathbb{E}^{\epsilon}\left[O(\vec {Y_Q}(T))|\vec{Y_Q}(t)=\vec{y}\right]$ which evolves under the evolution equation  
\begin{equation}
    \label{eq:perturbed_backward_Kolmogorov}\partial_tu_t^\epsilon(\vec{y})=-\left[\hat{L}+\epsilon \hat{L}_1(t)\right]u_t^\epsilon(\vec{y}), \ \  u_T^{\epsilon}(\vec{y})=O(\vec{y}) 
\end{equation}
We express the evolution under the perturbed dynamics as $u_s^\epsilon(\vec{y}_s)=\hat{P}_{T,s}^\epsilon O(\vec{y}))$, where 
\begin{equation}
    \hat{P}_{t,s}^\epsilon= \mathcal{T}_{+}\exp \left[\int_{s}^t\mathrm{d}s' \ (\hat{L}+\epsilon \hat{L_1}(s'))\right]
\end{equation} is the propagator under the perturbed dynamics with appropriate time ordering. Hence the response to the expected value observable $B(Y_Q)$ at time $T$ due to the application of the perturbation can be expressed as\cite{baiesi2013update-6b6} 
\begin{equation}
    \label{eq:general_response_formula}
    \begin{split}
        \delta\langle O(T)\rangle_\mu &=\int\mu( \mathrm{d}\vec{y}) \left(u^\epsilon_0(\vec{y})-u_0(\vec{y})\right)  \\ &=\int\mu( \mathrm{d}\vec{y}) \ (\hat{P}_{(T,0)}^\epsilon-\hat{P}_{(T,0)})u_T(\vec{y}) 
    \end{split}
\end{equation}
Now we assume the effect of perturbation can be expressed as the perturbation series in the powers of $\epsilon$ as
\begin{equation}
    \begin{split}
        \delta\langle O(T)\rangle_\mu&=\sum_{j=0}^\infty\epsilon^j \ \delta\langle O(T)\rangle_\mu^{(j)},\\ \hat{P}_{(T,0)}^\epsilon&=\hat{P}_{(T,0)}+\sum_{j=1}^\infty\epsilon^j \hat{P}_{(T,0)}^{(j)}
    \end{split}
\end{equation}
Now substituting the perturbation series in Eq.~\eqref{eq:general_response_formula} and comparing powers of $\epsilon$ we get
\begin{equation}
    \begin{split}
        \delta\langle O(T)\rangle_\mu^{(0)}&=0,\\
        \delta\langle O(T)\rangle_\mu^{(j)}&=\int\mu( \mathrm{d}\vec{y}) \ \hat{P}_{(T,0)}^{(j)}u_T(\vec{x}). 
    \end{split}
\end{equation}
and substituting the perturbation series in Eq.~\eqref{eq:perturbed_backward_Kolmogorov} and comparing powers of $\epsilon$ gives us the relation
\begin{equation}
    \begin{split}
         \partial_t \hat{P}_{(T,t)}^{(0)}&=-\hat{L}\hat{P}_{(T,t)}^{(0)},\\
         \partial_t \hat{P}_{(T,t)}^{(j)}&=-\left[\hat{L}\hat{P}_{(T,t)}^{(j)}+\hat{L}_1(t)\hat{P}_{(T,t)}^{(j-1)}\right], \ j\ge 1
    \end{split}
\end{equation}
Integrating this we get,
\begin{equation}
    \begin{split}
        \hat{P}_{(T,0)}^{(0)}&=e^{T\hat{L}},\\
        \hat{P}_{(T,0)}^{(j)}&=\int_0^{T}\mathrm{d}se^{s\hat{L}}\hat{L}_1(s)\hat{P}_{(T,s)}^{(j-1)}, \ j\ge 1
    \end{split}
\end{equation}

Hence we have the response relations 
\begin{equation}
    \delta\langle O(T)\rangle_\mu=\int_0^{T}\mathrm{d}s\sum_{j=1}^\infty\epsilon^j R^{(\mu)}_j(T,s)
\end{equation}
where
\begin{equation}
    \begin{split}
        R_j^{(\mu)}(T,s)&=\int\mathrm{\mu}(\mathrm{d}\vec{y})e^{s\hat{L}}\hat{L}_1(s)\hat{P}_{(T,s)}^{(j-1)} O(\vec{y})\\
    \end{split}
\end{equation}
is the $j$th order response kernel to the perturbation generated by $\hat{L}_1(t)$ to  $\langle B(t)\rangle $ when the initial conditions have been sampled from the probability measure $\mu$, and in the linear response regime for $j=1$ we have
\begin{equation}
\label{eq:linear_response_eqn_general_nonstationary}
     R_1^{(\mu)}(T,s)=\int\mathrm{\mu}(\mathrm{d}\vec{y})e^{s\hat{L}}\hat{L}_1(s)e^{(T-s)\hat{L}} O(\vec{y})
\end{equation} 
Writing $T=s+\tau$ we get
\begin{equation}
     R_1^{(\mu)}(s+\tau,s)=\int\mathrm{\mu}(\mathrm{d}\vec{y})e^{s\hat{L}}\hat{L}_1(s)e^{\tau\hat{L}} O(\vec{y})
\end{equation} 
Now we choose the initial conditions to be sampled from a probability measure $\mu(\mathrm{d}\vec{y})=\pi_Q(\mathrm{d} \vec{y})$ such that the dynamics in the $\mathbf{Q_0}$ projected subspace has reached a stationary state (see Appendix~\ref{app:stationary_state_measure_and_ibp}).\par 
\subsection{Special case: forcing perturbation in momentum equations and response in average velocities }
For the forcing perturbation presented in Eq.~\eqref{eq:perturbed_eqn_sde} the generator due to the perturbation can be written as 
\begin{equation}    \hat{L}_1(t)=\left(\mathbf{Q}_0\mathbf{B}\vec{f}(t)\right)\cdot \nabla
\end{equation}
We set $s=0$ to represent the represent a time where the system has reached a stationary state in the stable subspace; then we have $R_1^{(\pi_Q)}(\tau,0)\equiv R_1^{(\pi_Q)}(\tau)$ as 
\begin{equation}
    \begin{split}
         R_1^{(\pi_Q)}(\tau)&=\int\pi_Q(\mathrm{d}\vec{y})\hat{L}_1(0)e^{\tau\hat{L}} O(\vec{y})\\
         &=\int\pi_Q(\mathrm{d}\vec{y})\left(\mathbf{Q_0B}\vec{f}(0)\right)\cdot \left( \nabla e^{\tau\hat{L}}O(\vec{y})\right)\\
         &=\int\pi_Q(\mathrm{d}\vec{y})\left(\mathbf{Q_0B}\vec{f}(0)\right)^T\left( \nabla e^{\tau\hat{L}}O(\vec{y})\right)\\
    \end{split}
\end{equation}
We now choose $O(\vec{y})=\dot{\vec{Q}}=\mathbf{V}\mathbf{B}^T\vec{y}$ as our choice of observable and we get
\begin{equation}
     R_1^{(\pi_Q)}(\tau)=\int\pi_Q(\mathrm{d}\vec{y})\left(\mathbf{Q_0B}\vec{f}(0)\right)^T\left( \nabla e^{\tau\hat{L}}\mathbf{V}\mathbf{B}^T\vec{y}\right)
\end{equation}
Now, given $ \mathbf{Q_0B}\vec{f}(0)$ is a vector that exists completely in the stable subspace, we use the integration by parts formula given in Eq.~\eqref{eq:integration_by_parts_formula_gaussian_zero_mode} to get
\begin{equation}
    \begin{split}
        R_1^{(\pi_Q)}(\tau) &=\int\pi_Q(\mathrm{d}\vec{y})\left( e^{\tau\hat{L}}\mathbf{V}\mathbf{B}^T\vec{y}\right)\left(\mathbf{Q_0B}\vec{f}(0)\right)^T\mathbf{\tilde{S}_Q^{\infty}}\vec{y}\\
        &=\langle\dot{\vec{Q}}(\tau)g(0)\rangle_{\pi_Q}
    \end{split}
    \end{equation}
    where  
    \begin{equation}
    \label{eq:Agarwal-conjugate}
        \begin{split}
             g(s)&=\left(\vec{f}(s)^T\mathbf{B}^T\mathbf{Q}_0^T\mathbf{\tilde{S}}^{\infty}_Q\vec{y}\right)\\
             &=\left(\mathbf{\tilde{S}}^{\infty}_Q\vec{y}\right)\cdot \left(\mathbf{Q_0}\mathbf{B}\vec{f}(s)\right)
        \end{split}
    \end{equation}
is the corresponding Agarwal conjugate variable\cite{baiesi2013update-6b6} to the observable $\dot{\vec{Q}}$. Thus we write the first order response kernel for observable $\dot{\vec{Q}}$ as
\begin{equation}
    \begin{split}
         \langle\delta\dot{\vec{Q}}(t)\rangle_{\pi_Q}
         &=\epsilon\int_0^t \mathrm{d}s \int \pi_Q(\mathrm{d}\vec{y})\dot{\vec{Q}}(t-s) \vec{y}^T\left(\mathbf{\tilde{S}}^{\infty}_Q\right)\left(\mathbf{Q_0}\mathbf{B}\vec{f}(s)\right)\\
         &=\epsilon\int_0^t \mathrm{d}s \int \pi_Q(\mathrm{d}\vec{y})\mathbf{V}\mathbf{B}^Te^{\mathbf{C}(t-s)}\vec{y} \vec{y}^T\left(\mathbf{\tilde{S}}^{\infty}_Q\right)\left(\mathbf{Q_0}\mathbf{B}\vec{f}(s)\right)\\
         &=\epsilon\int_0^t \mathrm{d}s\mathbf{V}\mathbf{B}^Te^{\mathbf{C}(t-s)}\mathbf{S}^{\infty}_Q\mathbf{\tilde{S}}^{\infty}_Q\left(\mathbf{Q_0}\mathbf{B}\vec{f}(s)\right)\\
         &=\epsilon\int_0^t \mathrm{d}s\mathbf{V}\mathbf{B}^Te^{\mathbf{C}(t-s)}\mathbf{Q_0}\mathbf{B}\vec{f}(s)
    \end{split}  
\end{equation}
where we have used the following properties: (i) $\mathbf{\tilde{S}}^{\infty}_Q$ is symmetric, (ii) at the stationery state we choose an initial condition such that $\langle\vec{y}\rangle_{\pi_Q}=\vec{0}$ and we can write $\langle\vec{y}\vec{y}^T\rangle=\mathbf{S}^{\infty}_Q$ and, (iii) $\tilde{\mathbf{S}}^{\infty}_Q \mathbf{S}^{\infty}_Q\mathbf{Q}_0=\mathbf{Q_0}$. \par
Next, in Appendix~\ref{app_subsec:matrix_proof_FDT} we prove the matrix relation
\begin{equation}
     \mathbf{Q}_0\mathbf{B}=\beta \mathbf{S^{\infty}_Q}\mathbf{B}\mathbf{V}^T,
\end{equation}
which directly implies,
\begin{equation}
    \begin{split}
         \langle\delta\dot{\vec{Q}}(t)\rangle_{\pi_Q}
          &=\epsilon\int_0^t \mathrm{d}s\mathbf{V}\mathbf{B}^Te^{\mathbf{C}(t-s)}\mathbf{Q_0}\mathbf{B}\vec{f}(s)\\
          &=\epsilon\beta\int_0^t \mathrm{d}s\mathbf{V}\mathbf{B}^Te^{\mathbf{C}(t-s)} \mathbf{S^{\infty}_Q}\mathbf{B}\mathbf{V}^T\vec{f}(s)\\
          &=\epsilon\beta\int_0^t \mathrm{d}t'\mathbf{V}\mathbf{B}^Te^{\mathbf{C}t'} \mathbf{S^{\infty}_Q}\mathbf{B}\mathbf{V}^T\vec{f}(t-t')\\
    \end{split}
\end{equation}
where we identify the time domain response kernel to average velocity as
\begin{equation}
    \label{eq_app:time_domain_FDT}
    \begin{split}
         \boldsymbol{\mu}(\tau)&=\beta\mathbf{V}\mathbf{B}^Te^{\mathbf{C}\tau} \mathbf{S^{\infty}_Q}\mathbf{B}\mathbf{V}^T\\
         &=\beta\mathbf{V} \mathbf{S^{\infty}_p}(\tau)\mathbf{V}^T.
    \end{split}
\end{equation}
The splitting of $\mathbf{S_p^\infty}(\tau)$ from Eq.~\eqref{eqn:S_p_krylov_split_main_text} allows us to split the response kernel into similar components which are characterized by $b_0(\tau),b_1(\tau)$ and $b_2(\tau)$. Expressing Eq.~\eqref{eq_app:time_domain_FDT} in the frequency domain as $\boldsymbol{\mu}(\omega)=\mathcal{F}\left[\theta(\tau)\boldsymbol{\mu}(\tau)\right]$, we get 
\begin{equation}
    \label{eq_app:freq_domain_FDT}
\begin{split}
      \boldsymbol{\mu}(\omega)&= \beta\mathbf{V}\mathbf{\tilde{S}^\infty_p}(\omega)\mathbf{V}^T, \\ \mathbf{\tilde{S}^\infty_p}(\omega)&=\mathcal{F}\left[\theta(\tau)\mathbf{S_p^\infty}(\tau)\right]
\end{split}
\end{equation}
which is the fluctuation-dissipation relation (Eq.~\eqref{eq:FDT_main_text} in Sec.~\ref{Sec:Time_corr_functions}) that connects the dynamic mobility to the stationary momentum time-correlation functions.
\subsection{Proof of $ \mathbf{Q}_0\mathbf{B}=\beta \mathbf{S^{\infty}_Q}\mathbf{B}\mathbf{V}^T$}
\label{app_subsec:matrix_proof_FDT}
Given $\vec{Y}_Q=\mathbf{Q}_0\vec{X}$, the stationary state in the stable space relaxes to a stationary state, and in the orthogonal representation $\vec{Y}_Q=\mathbf{U}\vec{z}$, it relaxes to a the distribution
\begin{equation}
    \begin{split}
    \pi_Q^z &\propto \exp\left[-\frac{1}{2}\vec{z}^T\mathbf{S}_z^{-1}\vec{z}\right]\\
    \end{split}
    \end{equation}
Also, the harmonic Hamiltonian Eq.~\eqref{eq:Hamiltonian_scaled} can be written as the matrix relation
\begin{equation}
    H=\frac{1}{2}\vec{X}^T\mathbf{K}\vec{X}
\end{equation}
with the Hessian matrix $\mathbf{K}$ defined as 
\begin{equation}
   \mathbf{K}= \begin{pmatrix}
1 & 0 & -2\pi \chi & 0 \\
0 & \dfrac{1}{R^{2}} & 0 & 0 \\
-2\pi \chi & 0 & 4\pi^2 & 0 \\
0 & 0 & 0 & 1
\end{pmatrix}
\end{equation}
and since the noise model is taken satisfying detailed balance we also have the Gibbs distribution (restricted on the stable subspace) as
\begin{equation}
\begin{split}
    \pi_Q &\propto \exp\left[-\frac{\beta}{2}\vec{Y}^T_Q\mathbf{K}\vec{Y}_Q\right]\\
    &= \exp \left[-\frac{1}{2} \vec{z}^T(\beta\mathbf{U}^T\mathbf{K}\mathbf{U})\vec{z}\right],
\end{split}
\end{equation}
which implies
\begin{equation}\begin{split}
    \mathbf{S_z^{-1}}&= \beta\mathbf{U}^T\mathbf{K}\mathbf{U} \\
\end{split}
\end{equation}
Thus we have 
\begin{equation}
    \begin{split}
        &\ \ \mathbf{U}\mathbf{S_z^{-1}}\mathbf{U}^T= \beta\mathbf{U}\mathbf{U}^T\mathbf{K}\mathbf{U}\mathbf{U}^T\\
        \implies &\mathbf{\tilde{S}_Q^\infty}=\beta\mathbf{P_U}\mathbf{K} \mathbf{P_U} \\
        \implies &\mathbf{\tilde{S}_Q^\infty}\mathbf{Q}_0\mathbf{B}=\beta\mathbf{P_U}\mathbf{K} \mathbf{P_U}\mathbf{Q}_0\mathbf{B} \\  
    \end{split}
\end{equation}
where $\mathbf{P_U}=\mathbf{U}\mathbf{U}^T$ is the orthogonal projector to the stable subspace. Note that $\mathbf{P_U}\ne \mathbf{Q_0}$, but satisfies $\mathbf{P_U}\mathbf{Q_0=\mathbf{Q}_0}$ and $\mathbf{P_U}\mathbf{\Pi}_0=\mathbf{0}$. Thus we can write
\begin{equation}
    \begin{split}
        \mathbf{\tilde{S}_Q^\infty}\mathbf{Q}_0\mathbf{B}&=\beta\mathbf{P_U}\mathbf{K} \mathbf{Q}_0\mathbf{B} 
        \\&=\beta\mathbf{P_U}\mathbf{K} (\mathbf{I}_4-\mathbf{\Pi}_0)\mathbf{B}\\
        &=\beta\mathbf{P_U}\mathbf{K}\mathbf{B}
    \end{split}
\end{equation}
since the right eigenvector $\vec{r}_0=(2\pi\chi \ \ 0 \ \ 1 \ \ 0)^T$  of $\mathbf{C}$, is also a right eigen vector of $\mathbf{K}$. Thus $\mathbf{K}\vec{r}_0=\vec{0} \implies\mathbf{K}\vec{r}_0\vec{l}^T=\mathbf{K}\mathbf{\Pi}_0=\mathbf{0}$. Thus multiplying both sides of the equation above with $\mathbf{S_Q^\infty}$ from left we get,
\begin{equation}
    \mathbf{S_Q^\infty}\mathbf{\tilde{S}_Q^\infty}\mathbf{Q}_0\mathbf{B}        =\beta\mathbf{S_Q^\infty}\mathbf{P_U}\mathbf{K}\mathbf{B}
\end{equation}
Since $\mathbf{S_Q^\infty}\mathbf{\tilde{S}_Q^\infty}\vec{y}=\vec{y}$ for any $\vec{y}$ in the stable subspace, and $\mathbf{Q}_0\mathbf{P_U}=\mathbf{Q}_0$, we get
\begin{equation}
    \mathbf{Q}_0\mathbf{B}        =\beta\mathbf{S_Q^\infty}\mathbf{K}\mathbf{B}
\end{equation}
Next we have directly verifiable relation $\mathbf{K}\mathbf{B}=\mathbf{B}\mathbf{V}^T$, which leads to the final result
\begin{equation}
     \mathbf{Q}_0\mathbf{B}        =\beta\mathbf{S_Q^\infty}\mathbf{B}\mathbf{V}^T
\end{equation}
\end{widetext}

\providecommand{\noopsort}[1]{}\providecommand{\singleletter}[1]{#1}%
%


\end{document}